\documentclass[a4paper,11pt,onecolumn]{article}

\usepackage[hyperindex=true, hidelinks]{hyperref}
\usepackage{bbm}
\usepackage{geometry}
\usepackage{fancyhdr}
\usepackage{fancyvrb}
\usepackage{fancybox}
\usepackage{amsmath,amsfonts,amssymb,amscd,graphics,graphicx,latexsym,shuffle,bookmark}
\usepackage{subfigure}
\usepackage{indentfirst}
\usepackage{bm,tikz} \usetikzlibrary{cd}
\usepackage{multicol}
\usepackage{abstract}
\usepackage{anysize}
\usepackage{listings}
\usepackage{color}
\usepackage{units}
\usepackage{tabularx}
\usepackage{mathrsfs}
\usepackage[all]{xy}
\usepackage{authblk}
\usepackage{enumerate}

\usepackage[normalem]{ulem}
\usepackage{datetime2,soul,amsthm,mathtools}

       \font\tenmsb=msbm10
       \font\sevenmsb=msbm7
       \font\fivemsb=msbm5
       \catcode`\@=11
       \ifx\amstexloaded@\relax\catcode`\@=\active
       \endinput\else\let\amstexloaded@\relax\fi
       \def\spaces@{\space\space\space\space\space}
       \def\spaces@@{\spaces@\spaces@\spaces@\spaces@\spaces@}
       \def\space@.{\futurelet\space@\relax}
       \space@. %
       \def\Err@#1{\errhelp\defaulthelp@\errmessage{AmS-TeX error: #1}}
       \def\relaxnext@{\let\next\relax}
       \def\accentfam@{7}
       \def\noaccents@{\def\accentfam@{0}}

       \def\Cal{\relaxnext@\ifmmode\let\next\Cal@\else
       \def\next{\Err@{Use \string\Cal\space only in math mode}}\fi\next}
       \def\Cal@#1{{\Cal@@{#1}}}
       \def\Cal@@#1{\noaccents@\fam\tw@#1}

       \def\Bbb{\relaxnext@\ifmmode\let\next\Bbb@\else
       \def\next{\Err@{Use \string\Bbb\space only in math mode}}\fi\next}
       \def\Bbb@#1{{\Bbb@@{#1}}}
       \def\Bbb@@#1{\noaccents@\fam\msbfam#1}

       \newfam\msbfam
       \textfont\msbfam=\tenmsb
       \scriptfont\msbfam=\sevenmsb
       \scriptscriptfont\msbfam=\fivemsb

\def\Z{\mathbb{Z}}
\def\R{\mathbb{R}}
\def\C{\mathbb{C}}
\def\Q{\mathbb{Q}}

\def\D{\mathbb{D}}

\mathchardef\mx="2D

\newtheorem{Theorem}{Theorem}[section]
\newtheorem{Proposition}[Theorem]{Proposition}
\newtheorem{Lemma}[Theorem]{Lemma}
\newtheorem{Corollary}[Theorem]{Corollary}
\newtheorem*{thmA}{Theorem A}
\newtheorem*{thmB}{Theorem B}
\newtheorem*{thmC}{Theorem C}
\newtheorem*{thmA1}{Theorem A'}
\newtheorem*{thmA2}{Theorem A''}
\newtheorem*{thmB1}{Theorem B'}
\newtheorem*{thmC1}{Theorem C'}

\theoremstyle{definition}

\newtheorem{Remark}[Theorem]{Remark}
\newtheorem{Definition}[Theorem]{Definition}
\newtheorem{Example}[Theorem]{Example}

\@addtoreset{equation}{section}

\newcommand{\la}{\langle  }

\newcommand{\bq}{\begin{equation}}
\newcommand{\eq}{\end{equation}}
\newcommand{\bp}{\begin{Proposition}}
\newcommand{\ep}{\end{Proposition}}
\newcommand{\bdf}{\begin{Definition}}
\newcommand{\edf}{\end{Definition}}
\newcommand{\bl}{\begin{Lemma}}
\newcommand{\el}{\end{Lemma}}
\newcommand{\ba}{\begin{array}}
\newcommand{\ea}{\end{array}}
\newcommand{\bea}{\begin{eqnarray}}
\newcommand{\eea}{\end{eqnarray}}

\newcommand{\BB}{\mathscr{B}}
\newcommand{\LL}{\mathscr{L}}
\newcommand{\AS}{\mathscr{S}}
\newcommand{\ZA}{\mathscr{A}}
\newcommand{\DD}{\mathscr{D}}
\newcommand{\DDo}{\ti{\DD}}
\newcommand{\DDu}{\ud{\DD}}
\newcommand{\NN}{\mathscr{N}}
\newcommand{\RT}{\mathscr{R}}
\newcommand{\cA}{\mathcal{A}}
\newcommand{\FF}{\mathcal{F}}
\newcommand{\HHu}{\mathcal{H}^\mathrm{u}}
\newcommand{\GG}{\mathcal{G}}
\newcommand{\FFs}{\FF^{\hspace{.075em}\mathrm{s}}}
\newcommand{\FFg}{\FF_{\hspace{-.15em}g}}
\newcommand{\FZ}{\mathcal{Z}}
\newcommand{\FZtop}{\FZ_{\mathrm{top}}}
\newcommand{\FZtops}{\FZ_{\mathrm{top}, \mathrm{s}}}

\def\llra{\longmapsto}
\def\ra{\rightarrow}
\newcommand{\Td}{\widetilde}
\newcommand{\ti}{\widetilde}
\def\wh{\widehat}
\def\al{\alpha}
\def\be{\beta}
\def\th{\theta}

\def\de{\delta}
\def\De{\Delta}

\def\ga{\gamma}
\def\Ga{\Gamma}
\newcommand{\ze}{\zeta}
\def\ka{\kappa}
\def\la{\lambda}

\def\si{\sigma}
\def\Si{\Sigma}

\def\eps{\epsilon}
\def\ve{\varepsilon}
\newcommand{\vp}{\varphi}
\newcommand{\ph}{\varphi}
\def\om{\omega}
\def\Om{\Omega}

\def\eq{\equiv}

\def\ov{\overline}
\def\ud{\underline}

\def\pt{\partial}
\def\ta{\tau}

\def\sls{\sum\limits}

\def\Da(\Downarrow)
\def\bup{\bigcup}
\def\lqw{\leqslant}
\def\gqw{\geqslant}
\newcommand{\f}{\frac}
\newcommand{\tf}{\tfrac}

\newcommand{\datestampa}{{\small{File:\ens\hbox{\tt\jobname.tex}
\ens \today}}} % \;(Beijing time)}}}
\newcommand{\datestamp}{\blue{\small{File:\ens\hbox{\tt\jobname.tex}
\ens \DTMnow}}} % \;(Beijing time)}}}
\newcommand{\ens}{\enspace}
\newcommand{\blue}[1]{\textcolor{blue}{#1}}
\newcommand{\newc}[1]{\textcolor{violet}{#1}}

\newcommand{\defeq}{\coloneqq}           % \newcommand{\defeq}{:=}
\newcommand{\col}{\colon\thinspace}          %% for a map f \col A \to B
\newcommand{\DDE}{%
  {{\Delta}} \hspace{-.6em}
  \raisebox{.37ex}{\resizebox{.6ex}{!}{$\mathrm{/}$}}
  \hspace{.4em} {}
}
\newcounter{parag}[subsection]
\renewcommand{\theparag}{{\thesubsection.\arabic{parag}}}
\newcommand{\parag}{\medskip \refstepcounter{parag} \noindent{{\bf\theparag}\ } }
\newcounter{parage}[section]
\renewcommand{\theparage}{{\thesection.\arabic{parage}}}
\newcommand{\parage}{\medskip \refstepcounter{parage} % \addtocounter{parage}{1}
\noindent{\bf\theparage\ } }
\newcommand{\zu}{^{(0),\mathrm{u}}}
\newcommand{\ozu}{^{(1),\mathrm{u}}}
\newcommand{\nuu}{^{(n),\mathrm{u}}}
\newcommand{\uu}{^{\mathrm{u}}}
\newcommand{\conv}{_{\mathrm{conv}}}
\newcommand{\isequiv}[1]{\underset{#1}{\sim}}

\newcommand{\tNNe}{{\ti\NN_{\text{ext}}}}

\newcommand{\wrt}{with respect to}
\newcommand{\lhs}{left-hand side}
\newcommand{\rhs}{right-hand side}

\newcommand{\IM}{\mathop{\Im m}\nolimits}
\newcommand{\RE}{\mathop{\Re e}\nolimits}

\newcommand{\ie}{i.e.\ }
\newcommand{\ii}{^{-1}}

\newcommand{\simp}{{\operatorname{simp}}}
\newcommand{\ID}{\mathop{\hbox{{\rm Id}}}\nolimits}

\DeclareMathOperator{\sing}{sing}
\DeclareMathOperator{\minor}{min}
\DeclareMathOperator{\Ai}{Ai}
\DeclareMathOperator{\cont}{cont}

\newcommand{\Gadiff}{\Ga_{\!\textrm{diff}}}

\newcommand{\dst}{\displaystyle}
\newcommand{\pa}{\partial}

\newcommand{\iimp}{\,\Longrightarrow\,}
\newcommand{\imp}{\ens\Longrightarrow\ens}
 \newcommand{\Imp}{\quad\Longrightarrow\quad}

\newcommand{\gO}{\mathscr{O}}
\newcommand{\gP}{\mathscr{P}}
 
\newcommand{\cS}{\mathcal{S}}
\newcommand{\SOM}{\cS_\Om}
\newcommand{\OOM}{0_\Om}
\newcommand{\tom}{{\ti\om}}
\newcommand{\TOM}{{\ti\Om}}

\newcommand{\ASm}{\AS_{\operatorname{med}}}
 
\newcommand{\Pialth}{\Pi^\th_{\al(\th)}}
\newcommand{\Pialtho}{\ti{\Pi}\,\!^\th_{\al(\th)}}
\newcommand{\Pialthp}{\Pi^{\th'}_{\al(\th')}}

\newcommand{\begla}{\begin{equation}}
\newcommand{\beglab}[1]{\begin{equation}	\label{#1}}
\newcommand{\edla}{\end{equation}}

\newcommand{\goa}{{\mathfrak{A}}}
\newcommand{\tpha}{{ \ti\ph_{\hspace{-.1em}\raisebox{-.7ex}{$\scriptstyle \goa$}} }}
\newcommand{\nua}{{ \nu_{\hspace{-.05em}\raisebox{-.4ex}{$\scriptstyle \goa$}} }}
\newcommand{\snua}{{
    \nu_{\hspace{-.07em}\raisebox{-.4ex}{$\scriptscriptstyle \goa$}}
  }}

\newcommand{\nuap}[1]{{
    \nu_{\hspace{-.05em}\raisebox{-.5ex}{$\scriptstyle
        \goa$}}^{(#1)} }}

  \newcommand{\snuap}[1]{{
    \nu_{\hspace{-.07em}\raisebox{-.5ex}{$\scriptscriptstyle \goa$}}^{(#1)}
  }}

\newcommand{\ch}[1]{{\stackrel{\raisebox{-.23ex}{$\scriptscriptstyle\vee$}}{#1}}}
\newcommand{\trR}{\raisebox{.1ex}{${\stackrel{
            \raisebox{-.23ex}{$\scriptscriptstyle\bigtriangledown$}
          }{\RT}}$}}

\newcommand{\bem}{\mbox{}^\flat\hspace{-.8pt}}    

\newcommand{\chb}[1]{\raisebox{-.23ex}{${\stackrel{
            \raisebox{-.23ex}{$\scriptscriptstyle\vee$}
          }{#1}
     }$}}

\newcommand{\trb}[1]{\raisebox{-.23ex}{${\stackrel{
             \raisebox{-.23ex}{$\scriptscriptstyle\triangledown$}
           }{#1}
      }$}}
\newcommand{\chn}[1]{\raisebox{.20ex}{${\stackrel{
            \raisebox{-.20ex}{$\scriptscriptstyle\vee$}
          }{#1}
     }$}}
\newcommand{\trn}[1]{\raisebox{.20ex}{${\stackrel{
            \raisebox{-.20ex}{$\scriptscriptstyle\triangledown$}
          }{#1}
     }$}}
\begin{document}
\setlength{\columnsep}{5pt}
%\title{Resurgent treatment of all-genus differential equation}
\title{\vspace{-1.25cm} Resurgence in the Universal Structures in
  B-model Topological String Theory}

%\author{ author\footnote{Partially supported by project No. 12345678 of qq.} }

\author[1]{Yong LI\thanks{Email: liyongmath@qq.com}}
\author[1]{David SAUZIN \thanks{On leave from CNRS -- Observatoire de
    Paris, PSL Research University, 75014 Paris, France. Email: David.Sauzin@obspm.fr}}
\author[1, 2]{Shanzhong SUN\thanks{Email: sunsz@cnu.edu.cn \\ \datestamp}}

\affil[1]{Department of Mathematics, Capital Normal University, Beijing 100048 P. R. China}
\affil[2]{Academy for Multidisciplinary Studies, Capital Normal University, Beijing 100048, P. R. China}

\date{}

%%%%%%%%%%%%%%%%%%%%%%%%%%%%%%%%%%%%%%%%%%%%%%%%
%%%%%%%%%%%%%%%%%%%%%%%%%%%%%%%%%%%%%%%%%%%%%%%%

\maketitle
%\vspace{-.75cm}

\begin{abstract}
  We propose a systematic analysis of
    Alim-Yau-Zhou's double scaling limit
    and Couso-Santamar\'{i}a's large radius limit
    for the perturbative free energies in B-model
  topological string theory based on \'Ecalle's Resurgence Theory.
  {Taking advantage of the known resurgent properties of the
    formal solutions to the Airy equation and of the stability of
    resurgent series under exponential/logarithm and nonlinear changes
    of variable,} we show how to rigorously derive the
  non-perturbative information from the perturbative one
  by means of alien calculus in this context,
spelling out the notions of formal integral and Bridge Equation,
  typical of the resurgent approach to ordinary differential
  equations.
  We also discuss the Borel-Laplace summation of the obtained
  resurgent transseries, including a study of real analyticity
  based on the connection formulas stemming from the resummation of the
  Bridge Equation.
%
  % \blue{I eliminated the following sentence:} As a byproduct, we also show that the Borel singularities in the
  % large radius limit can be derived from the simple singularities in
  % the Borel plane for the double scaling limit.
  %
\end{abstract}

\bigskip

\bigskip

%\smallskip

{\small Key Words: topological string theory; free energy; partition
  function; resurgence theory; asymptotic expansion; alien calculus;
  formal integral; resurgent transseries; bridge equation

}

\newpage

\tableofcontents

\bigskip

%%%%%%%%%%%%%%%%%%%%%%%%%%%%%%%%%%%%%%%%%%%%%%%%
%%%%%%%%%%%%%%%%%%%%%%%%%%%%%%%%%%%%%%%%%%%%%%%%

\section {Introduction}\label{sec-Intro}

%%%%%%%%%%%%%%%%%%%%%%%%%%%%%%%%%%%%%%%%%%%%%%%%
%%%%%%%%%%%%%%%%%%%%%%%%%%%%%%%%%%%%%%%%%%%%%%%%

\parage
Topological string theory is a supersymmetric conformal field theory
built up through a nonlinear sigma model which studies the maps from
the string world-sheet Riemann surfaces to a target Calabi-Yau (CY)
threefold (\cite{Alim12}).
Mathematically, there are various ways to formulate it.  It can be
defined through the generating functions of Gromov-Witten invariants
(A-model side) or in terms of deformation theory of complex structures
(B-model side) of CY manifolds which are related by mirror symmetry
(\cite{[K.C.]}).
Topological string theory is in general only defined perturbatively
and, for computational and conceptual reasons, the structure of
  the non-perturbative completions is a delicate matter, still largely
  conjectural.
 %
 % and the structure of the non-perturbative completions is \ul{hard to
 %   derive} due to computational and conceptual reasons, and is
 % \ul{largely conjectural and mysterious.}\marge{English to be improved...}

Using string/gauge theory duality, many spectacular progresses have
already been made by taking advantage of Chern-Simons theory and matrix
models (see e.g.\ \cite{[M1],[M2],[M3]}), and these fruitful
interactions are still actively investigated.
%
% Now \ul{it is still a fruitful and active interaction for mutual benefit}.

It is generally believed that the perturbative free energy is an
asymptotic divergent series in the string coupling constant.  If the
partition functions or correlation function% at hand
---or rather the series supposed to represent these functions---%
is resurgent in the sense of \'{E}calle
(\cite{[J.E.],[M1],[M.S.],primerABS,EncyclDS}), as is conjectured in
most QFT theories, then one may use the Stokes structures to guess or
define what the non-perturbative contributions should be.
In fact, a resurgent transseries ansatz has already been used to check
against the large-order perturbative information in a series of works
(see \cite{[M.]} and references there in, especially
\cite{CESV1,CESV2}).

For a given CY threefold, the topological string partition function is $\FZtop=\exp \FF$, with
the free energy~$\FF$ defined schematically as
 \begin{equation}   \label{eqschematic}
 \FF(g_s, z, \bar{z})=\sum_{g=0}^\infty g_s^{2g-2}\FFg(z,
 \bar{z})
\end{equation}
 where $g_s$ is the topological string coupling constant and the
 summation index~$g$ is
 the genus of the world-sheet Riemann surface. The amplitudes
 $\FFg(z, \bar{z})$ are functions of a set of
   variables, that we collectively denote by~$z$, representing
   \begin{itemize}
     \item   either, on the A-model side, the moduli of
       complexified K\"{a}hler structures of the CY manifold,
       \item  or, on the B-model side, the moduli of complex structures of its
         mirror manifold,
       \end{itemize}
       together with other parameters of the theory.
     The notation $\FFg(z, \bar{z})$ is meant to emphasize that
     $\FFg$ is not holomorphic in~$z$. % \marge{OK?...}
 It is notoriously difficult %  well known that it is very complicated
 to determine the functions $\FFg(z, \bar{z})$.

 %However,
 On the B-model side, which is our main concern in this paper, following Bershadsky-Cecotti-Ooguri-Vafa (\cite{[BCOV], [BCOV2]}), the $\FFg(z, \bar{z})$ recursively solve the {\it holomorphic anomaly equation} (HAE):
 \[
   \partial_{\bar z}\FFg=\frac{1}{2}\overline{C}_{\bar
     z}^{zz}\left (\sum_{h=1}^{g-1} D_z \FF_h
     D_z\FF_{\hspace{-.15em}g-h}+D_zD_z \FF_{\hspace{-.15em}g-1}\right )
   \quad  \text{for}\enspace g\gqw 2,
 \]
 with similar equations for $g=0, 1$. For more details
 concerning the coefficients and the covariant derivative
 $D_z$, the reader is referred to the aforementioned references. This
 way one can study $\FFg$ up to very high values of $g$, but a
 closed formula for $\FFg$ is still missing---see however
   the promising closed-form formulas recently proposed in \cite{IM}
   for the Stokes structure.
%
% however recently there is exciting conjectural transseries ansatz
% proposed in \cite{IM} which is very promising.

%%%%%%%%%%%%%%%%%%%%%%%%%%%%%%%%%%%%%%%%%%%%%%%%
%%%%%%%%%%%%%%%%%%%%%%%%%%%%%%%%%%%%%%%%%%%%%%%%

 \parage   \label{paragonetwo}
\textbf{Alim-Yau-Zhou's double scaling limit.}
 One efficient approach to compute the free energies by means of
the HAE is to observe that they depend on the complex conjugate~$\bar{z}$
  of the complex structure modulus~$z$ as polynomials in
  the ``nonholomorphic generators''
  $S^{zz},  S^z, S$ and~$K_z$, also known as propagators
%
% {\ul{polynomials in a finite set of nonholomorphic
%     generators $(S^{zz}, S^z, S, K_z)$ instead of the complex
%     conjugate $\bar{z}$ of the complex structure modulus $z$. These
%     generators are also known as propagators}
%
(\cite{YY04}, \cite{Alim12}).

Using the polynomial structure of the free energy in terms
of the nonholomorphic  propagators, Alim, Yau and Zhou (\cite{[AYZ]})
studied the leading order terms \wrt\ the
propagator~$S^{zz}$;
they got rational coefficients~$a_g$ such that
\begin{equation}   \label{eqFCzzzSzz}
  \FFg=a_g C_{zzz}^{2g-2} (S^{zz})^{3g-3}
  +\,\text{lower order terms in~$S^{zz}$},
  \end{equation}
where $C_{zzz}$ is the holomorphic Yukawa coupling.
The generating series % of the highest degree terms
%
% \blue{I CHANGE $\FF_{\mathrm{s}}$ \chg $\FFs$ EVERYWHERE (to avoid confusion with $\FFg$)}
  %
  \begin{equation}   \label{eqdefFFs}
    \FFs(\lambda_s) \defeq \sum_{g\gqw 2}\, a_g \lambda_s^{2g-2}
    = \f{5}{24}\la_s^2 + \f{5}{16}\la_s^4 + \f{1105}{1152}\la_s^6 + \cdots
  \end{equation}
  (where we omit $g=0,1$ and the superscript~``s'' stands for ``scaling'')
  %
% \blue{DO WE START AT $g=0$ OR $g=2$?} ANSWER: we start at $g=2$ (2024/01/19)
  %
  turns out to be universal, \ie independent of the CY geometry
  under consideration.
  Following~\cite{[AYZ]}, one may obtain it by scaling two of the
    variables in
    $\FF(g_s, z, S^{zz}, S^z, S, K_z)$:
%
  % introducing the double scaling limit
%
\begin{equation}\label{dsl}
  g_s = \varepsilon \,C_{zzz}^{-1} \,\Si^{-3/2} \la_s,
    \qquad S^{zz}= \varepsilon^{-\frac{2}{3}} \Si
    %
%    \qquad g_s = \varepsilon \td g_s, \qquad \lambda_s^2 = \td g_s^2 C_{zzz}^2 \Si^3}
%
\end{equation}
and taking the limit $\varepsilon\rightarrow 0$. {Thus, the new
  indeterminate~$\la_s$ in~\eqref{eqdefFFs} is essentially $C_{zzz} (S^{zz})^{3/2}g_s$}.

A key observation is that, as a consequence of HAE, $\FFs$ satisfies the ODE
\begin{equation}\label{eq1}   % \label{eq_ageq}
  \th^2_{\la_s}\FF+(\th_{\la_s}\FF)^2+2\left(1-\f{2}{3\la_s^2}\right)\th_{\la_s}\FF+\f{5}{9}=0,
  \qquad \th_{\la_s}\defeq\la_s\f{\pt}{\pt \la_s}
\end{equation}
%
% an equation that will be the starting point of our resurgent
% analysis in Section~\ref{resurgentfreeenergy}.
%
and, for the corresponding partition function
$\FZtops=\exp \FFs$, this amounts to the
modified Bessel equation with parameter $\frac{1}{3}$
in the variable $z=\f{1}{3}\la_s^{-2}$ % in the appropriate variable
(\cite[Prop.~3.2]{[AYZ]});
  hence, up to an elementary factor, $\exp \FFs$ solves the Airy equation
  \begin{equation}\label{eqairy}
\f{d^2y}{dw^2}=wy, \qquad w=(2\la_s^2)^{-2/3},
\end{equation}
which allows one to obtain the coefficients~$a_g$ from the well-known
asymptotics of the solutions of~\eqref{eqairy} and also suggests a
non-perturbative completion of~$\FFs$.

The solutions to the Airy equation~\eqref{eqairy} are known to
  be resurgent \wrt\ the appropriate variable---see e.g.\
  \cite[Sec.~6.14]{[M.S.]} (reviewed in Section~\ref{sec-linear}).
  In this paper, we will revisit \cite{[AYZ]}'s double scaling limit
  in the light of Resurgence Theory and show in
  Section~\ref{resurgentfreeenergy} how \'Ecalle's alien calculus
  leads to the non-perturbative completion of~$\FFs$.

%%%%%%%%%%%%%%%%%%%%%%%%%%%%%%%%%%%%%%%%%%%%%%%%
%%%%%%%%%%%%%%%%%%%%%%%%%%%%%%%%%%%%%%%%%%%%%%%%

\parage
\textbf{Couso-Santamar\'{i}a's large radius limit.}
The physical interpretation of the parameter~$\varepsilon$ in the
double scaling~\eqref{dsl} was left open in \cite{[AYZ]}.
This issue was touched upon in \cite{[RCS]} by Couso-Santamar\'{i}a,
who took a different route.
He demonstrated on several examples of CY threefolds a mechanism by
which a sequence of polynomials
\[
  H_g\zu(u) = a_g u^{3g-3} +\,\text{lower order terms in~$u$}, \qquad g\gqw2,
\]
with the same leading coefficients $a_g$'s as in~\eqref{eqFCzzzSzz}, % \cite{[AYZ]},
appears in a ``large radius limit''.
The latter phrase refers to a generic feature of CY geometries:
  the presence of a special point in the $z$-space ($z=0$ in
  \cite{[RCS]}'s examples with one-dimensional~$z$), at which the
  Yukawa coupling $C_{zzz}$ is singular:
$C_{zzz} \isequiv{z\to0} \ka z^{-3}$ with a topological factor~$\ka$, at
least in the cases with a single modulus~$z$. Couso-Santamar\'{i}a then finds
\begin{equation} \label{eqFFzdSi}
\lim_{z\to0} \FFg(z,S^{zz}=z^2\Si) = a_g \ka^{2g-2} \Si^{3g-3}
  +\,\text{lower order terms in~$\Si$},
\end{equation}
whence the coefficients~$a_g$ can be retrieved by considering the
large radius limit $z\to0$ and then extracting the asymptotic behaviour as
$\Si\to\infty$.

If, after taking the large radius limit $z\to0$, one
keeps~$\Si$ finite and replaces it by a certain geometry-dependent
affine function~$u$ of~$\Si$, then,
up to a topological factor,
the \rhs\ of~\eqref{eqFFzdSi} becomes\footnote{{More precisely,
    $\lim_{z\to0} \FFg(z,S^{zz}=z^2\Si) = \Big(\Td
    b^3\kappa^{-1}\Big)^{g-1} H_g\zu(u)$
    with $\Si=\Td b\kappa^{-1}u+\si_{\mathrm{hol}}$,
    where the constants $\kappa$, $\Td b$ and~$\si_{\mathrm{hol}}$ are
    defined in~\cite{[RCS]}.
    Consequently, in~\eqref{eqdefHzugsu}, the indeterminate~$g_s$ is
    not the original string coupling constant but a rescaled one,
    namely
    $\Td b^{3/2}\kappa^{-1/2} g_s$.
  }
}
the aforementioned polynomial~$H_g\zu(u)$, which thus contains contributions from the
non-holomorphic lower order terms of~\eqref{eqFCzzzSzz}.
It is argued in~\cite{[RCS]} that these polynomials are universal,
hence the superscript~``u'' (not to be confused with the
variable~$u$ stemming from the rescaled propagator~$S^{zz}$).
%
% proposed a relation between~$\varepsilon$ and the complex structure modulus~$z$:
%
% In this way the double scaling $\varepsilon$-limit in \cite{[AYZ]}
% was geometrically interpreted as a large radius limit %, what's more
% and, moreover, some nonholomorphic information could be retrieved. % are restored.
% %
% When $z$ goes
% to zero, it still recovers the series $\FFs$. To make the lower order
% terms visible, the author proposed to consider the large radius limit
% which does NOT rescale $g_s$ as in \cite{[AYZ]} while doing the
% same rescaling for the propagator $S^{zz}$:
% $S^{zz}\mapsto \varepsilon^{\frac{2}{3}} S^{zz}$. This rescaling of
% the propagator by $z^2$ is natural as explained in \cite{[RCS]}
% (the end of Section $2$).  From this new perspective, the
% $\varepsilon$-limit is interpreted geometrically as the large radius
% limit $z\rightarrow 0$ (eqn.~(24) of~\cite{[RCS]}), and at the same time
% one can capture some nonholomorphic terms. Up to some topological
% factors, the rescaled propagator corresponds to a new variable $u$,
% and the large radius limit becomes $u\rightarrow \infty$ with rescaled
% free energy $H^{(0), u}$ incorporating the complex structure moduli
% nonholomorphically. The author argued that $H^{(0),u}$ is also
% universally valid for any CY geometry at least for the case of \blue{a single}
% complex structure modulus.
%
% \emph{\blue{I will modify what follows... [David]}}

The analysis in~\cite{[RCS]} is based on the fact that the all-genus large radius limit free energy
  \begin{equation} \label{eqdefHzugsu}
    H\zu(g_s,u) \defeq \sum_{g\gqw1} g_s^{2g-2} H_g\zu(u),
  \end{equation}
  with a suitably defined contribution~$H_1\zu$ from genus~$1$,
  satisfies a rescaled version of the HAE
in the antiholomorphic modulus~$u$
(\cite[eqn.~(45)]{[RCS]}, referred to as the $u$-equation later on):
\begin{equation}\label{uE}
\partial_u H-\frac{3}{2} g_s^2 u^3\left( \partial_u H+\frac{u}{3}\partial_u^2 H+\frac{u}{3}(\partial_u H)^2\right)=\frac{1}{2u}+\frac{1}{u^2}.
\end{equation}
Solving the ODE~\eqref{uE} % which
leads to a universal non-perturbative completion in the form of a transseries
  \begin{equation} \label{eqdefHuugsusi}
  H\uu(g_s,u,\sigma) = \sum_{n\gqw0} \si^n e^{-\f{2n}{3u^3g_s^2}} H\nuu(g_s,u),
\end{equation}
where each $H\nuu(g_s,u)$, like $H\zu(g_s,u)$, is a series in~$g_s^2$.
The method in~\cite{[RCS]} is quite empirical, with a guess taken at
the coefficients of $H\zu_g(u)$, leading indirectly to a so-called $\tau_s$-equation, namely
an ODE for $H\zu$ written in the variables $\tau_s\defeq u^3 g_s^2$ and~$u$
  (see \cite[eqn.~(49)]{[RCS]}), for
  which a transseries ansatz is introduced, eventually resulting
  in~\eqref{eqdefHuugsusi};
  the resurgent character of each~$H\nuu(g_s,u)$
  % \wrt\ the variable~$g_s^{-2}$
  is derived from this $\tau_s$-equation,
  using a fact that amounts to saying that the $\tau_s$-equation
    is amenable to the Airy equation by an appropriate change of
    variable, however this is quite implicit in~\cite{[RCS]} and it is
    not clear how rigorous that is as a mathematical proof.
    {Moreover, \cite{[RCS]} deals with resurgence \wrt~$\tau_s^{-1}$
    but with $\ta_s/u$ kept fixed.}

In Section~\ref{LRL} of this paper, we will provide a fully
  rigorous treatment of the $u$-equation~\eqref{uE} from the viewpoint
  of Resurgence Theory, clarifying certain passages of~\cite{[RCS]}
  and expressing the non-perturbative transseries
  completion~\eqref{eqdefHuugsusi} in terms of \'Ecalle's alien
  calculus applied to the perturbative series~\eqref{eqdefHzugsu}.

% The author % described the large-order growth ($u\rightarrow \infty$
% % and $\tau_s$ fixed),
% proposed a transseries ansatz % (\cite{[RCS]}, (50))
% for the all-genus large limit free energy based on~\eqref{uE} and
% initiated its resurgent analysis. % with the help of those of Airy equations.

%%%%%%%%%%%%%%%%%%%%%%%%%%%%%%%%%%%%%%%%%%%%%%%%
%%%%%%%%%%%%%%%%%%%%%%%%%%%%%%%%%%%%%%%%%%%%%%%%

\parage
\textbf{Results on the resurgent character and
  resurgence relations of the perturbative series.}   \label{paragResultsRes}
%
% As is well-known, the Airy equation~\eqref{eqairy} is ubiquitous in
% mathematical physics. % For our purpose,
% \blue{In this context, it plays a key
%   role % in~\cite{[AYZ]} and~\cite{[RCS]}
% % the above works studying the universal
% %
% to identify nonperturbative instanton actions and suggest a
% transseries completion by means of resurgence
% %and transseries where
% (the instanton actions appearing as the singular points of the
% Borel transform of the perturbative series for the free energy).}
%
This paper aims to illustrate the power of the resurgent tools on
\cite{[AYZ]}'s double scaling limit and
\cite{[RCS]}'s large radius limit.
We will elucidate the resurgent structure of the perturbative all-genus
rescaled free energies in both cases
and extract their non-perturbative content, \ie the exponentially
ambiguities inherently attached to them, by means of alien calculus.
%
% and the corresponding partition function by a rigorous analysis based
% on the resurgence theory for singular ordinary differential equations.
% %
% We also discuss % establish
% the realness of their formal integral obtained by resummation of the
% full transseries,
% %
% based on the associated Bridge Equation
%
% One can see the incarnations of the parametric or coequational resurgence.

We now give our first main theorem, with explanations on the resurgent
terminology right after the statement:

%%%%%%%%%%%%%%%%%%%%%%%%%%%%%%%%%%%%%%%%%%%%%%%%
%%%%%%%%%%%%%%%%%%%%%%%%%%%%%%%%%%%%%%%%%%%%%%%%

\begin{thmA}
  \emph{(i)}  The all-genus double scaling limit free energy $\FFs(\la_s)$
  in~\eqref{eqdefFFs} and the corresponding partition function
  $\FZtops=\exp \FFs$ of the B-model topological string theory are
  simple $2\Z$-resurgent divergent series \wrt\ the variable
  $z=\f{1}{3}\la_s^{-2}$.
  % \blue{WHY DO WE USE $\f{1}{3}\la_s^{-2}$ AND NOT $\la_s^{-2}$?}
  % \textcolor{red}{ Because using $\f{1}{3}\la_s^{-2}$ as the
  % coordinate transformation can give a new series $\Td{g}(z)$ and
  % $\Td{\psi}(z)$ satisfy the logarithmic relationship, and the
  % resurgent property of $\Td{\psi}(z)$ is clear If we use the
  % variable $\la_s^{-2}$ then the point at which the singularity
  % occurs goes from 2 to $\f{2}{3}$.  }.

  \smallskip

    \emph{(ii)} A more general formal solution of the rescaled
    HAE~\eqref{eq1} is the so-called ``formal integral''
    \begin{equation}\label{tssss}
      \GG(\la_s,\si_1,\si_2) % = \si_1 + \FFs(\la_s,\si_2)
      = \si_1 + \sum_{n\gqw0} \si_2^n e^{-\f{2}{3}n\la_s^{-2}} \GG_n(\la_s),
    \end{equation}
      with arbitrary constants $\si_1,\si_2$ and $\GG_0=\FFs(\la_s)$,
      where the $\GG_n$'s for $n\gqw1$ are simple
      $2\Z$-resurgent divergent series \wrt\ $z=\f{1}{3}\la_s^{-2}$
      that can be obtained from the formula
  \begin{equation}   \label{eqdefFFslasi}
    %
    %
    % \FFs(\la_s,\si) \defeq
    %
      \sum_{n\gqw0} \si^n e^{-\f{2}{3}n\la_s^{-2}} \GG_n(\la_s) =
    \exp\!\Big( i \si e^{-\f{2}{3}\la_s^{-2}}\De_2 \Big) \FFs(\la_s),
    %
    % \qquad\blue{\text{IS IT $i\si$ OR $-i\si$?}}
    %
  \end{equation}
  where the operator~$\De_2$ is \'Ecalle's alien derivation at index~$2$.

  \smallskip

    \emph{(iii)} The action of all alien derivations on the~$\GG_n$'s
    can be compactly written in terms of the formal integral~\eqref{tssss} as the ``Bridge Equation''
  \begin{equation}   \label{eqBEFFslasi}
    \De_{2}\GG = - i e^{\f{2}{3}\la_s^{-2}} \f{\pt}{\pt\si_2}\GG,
    \qquad
    \De_{-2}\GG = - i e^{-\f{2}{3}\la_s^{-2}}
    \Big( \si_2\f{\pt}{\pt\si_1} - \si_2^2\f{\pt}{\pt\si_2} \Big)\GG
      \end{equation}
      and $\De_{\om}\GG = 0$ for $\om\in 2\Z^*\setminus\{-2,2\}$.

  \medskip

  \emph{(iv)} The action of the symbolic Stokes automorphisms on~$\GG$ is given by
%  with Stokes automorphisms~\eqref{eq_nts} and~\eqref{eq_22}.
  %
  \begin{align}
    \DDE_{\R_{\gqw 0}}^+ \GG(\la_s,\si_1,\si_2) &= \GG(\la_s,\si_1,\si_2-i),
    \\[.5ex]
    \label{DDlqwonGG}
    \DDE_{\R_{\lqw 0}}^+ \GG(\la_s,\si_1,\si_2) &
                                                  =
    \GG\Big(\la_s,\si_1+\log(1-i\si_2),\f{\si_2}{1-i\si_2}\Big).
  \end{align}
\end{thmA}

%%%%%%%%%%%%%%%%%%%%%%%%%%%%%%%%%%%%%%%%%%%%%%%%
%%%%%%%%%%%%%%%%%%%%%%%%%%%%%%%%%%%%%%%%%%%%%%%%
\smallskip

\noindent\textit{Explanation of the terminology:}
\smallskip

%%%%%%%%%%%%%%%%%%%%%%%%%%%%%%%%%%%%%%%%%%%%%%%%

\textbf{(i)} Given a lattice~$\Om$ of~$\C$, e.g.\ $\Om=2\Z$, a formal series % of the form
$\Td{\vp}(z)=\sls_{k\gqw0} c_k z^{-k}$ is said to be
``$\Om$-resurgent'' if the Borel transform of $\ti\ph(z)-c_0$, defined as
$\wh{\vp}(\ze)=\sls_{k\gqw1} c_k \f{\ze^{k-1}}{(k-1)!}$, satisfies a certain
property:
\begin{equation}
\label{eqdefOmcont}
\parbox{11cm}{$\wh{\vp}(\ze)$ has positive radius of convergence and defines a holomorphic function
that admits analytic continuation along all the paths in the complex plane that start
near~$0$ and avoid the points of~$\Om$.}
\end{equation}
Notice that the analytic continuation of $\wh{\vp}(\ze)$ may be multivalued. If at
least one of the branches is singular somewhere, then
the radius of convergence of $\wh{\vp}(z)$ is finite, hence
the radius of convergence of $\Td{\vp}(z)$ is zero: the original
series is a divergent one.

Beware that, in this section, the resurgence variable~$z$ represents
$\f{1}{3}\la_s^{-2}$ but, with a slight abuse of notation, we keep on
expressing our series in terms of~$\la_s$ instead of introducing new
notations like $\Td g(z)\defeq \FFs\big( (3z)^{-1/2}\big)$ or
$\Td G_n(z)\defeq\GG_n\big( (3z)^{-1/2}\big)$.

\smallskip

%%%%%%%%%%%%%%%%%%%%%%%%%%%%%%%%%%%%%%%%%%%%%%%%

\textbf{(ii)} We say that $\Td{\vp}(z)$ is a ``simple $\Om$-resurgent series'' if,
moreover,
\begin{equation}
\label{eqdefsimpleness}
\parbox{11cm}{the singularities of the analytic continuation of
$\wh{\vp}(\ze)$ are all of the form simple pole $+$ logarithmic
singularity with regular monodromy}
\end{equation}
(see Definition~\ref{defsimpsing}). % (see Section~\ref{secAlCalsimp} and especially formula~\eqref{eq_ancon}).
For such series, an operator~$\De_\om$ can be defined for each
$\om\in\Om^*\defeq \Om-\{0\}$, that acts on~$\Td\vp$ according to
Definition~\ref{DefDeompDeom}. %\eqref{eq_ancon}--\eqref{eqdefDeom}.
%
% \ens \blue{CHECK REFERENCES TO FORMULAS} \ens
%
The operators~$\De_\om$ are called ``alien derivations'' because they
are derivations (they satisfy the Leibniz rule) but of a very
different nature than the usual differential operators.

In particular,
any convergent series is a simple $\Om$-resurgent series annihilated by all
operators~$\De_\om$ (because its Borel transform is an entire
function).
It is thus because the series~$\FFs(\la_s)$ is divergent that, when
expanding~\eqref{eqdefFFslasi} as
\begin{equation}    \label{eqexpandGGn}
  \GG_1=i\De_2\FFs(\la_s), \quad
  \GG_2=-\f{1}{2!}\De_2^2\FFs(\la_s), \quad
  %
%  \GG_3=-\f{i}{3!}\De_2^3\FFs(\la_s), \quad
  %
  \ldots\ , \quad
  \GG_n=\f{i^n}{n!}\De_2^n\FFs(\la_s), \quad
  \end{equation}
  we get non-trivial series.
It so happens that the operator $e^{-\om z}\De_\om$ is a derivation
that commutes with $\f{\pt}{\pt z}$,
hence % for each $\si\in\C$,
  $\exp\!\big( i \si e^{-2z}\De_2 \big)$ is an algebra automorphism\footnote{%
\label{ftnrestr}
Notice that the space $\Td\RT_{2\Z}^\simp$ of simple
  $2\Z$-resurgent series is an algebra, on which~$\De_\om$ acts as a
  derivation for each
$\om\in2\Z^*$,
but we must go to the algebra $\Td\RT_{2\Z}^\simp[[e^{-2z}]]$ of simple
$2\Z$-resurgent \emph{transseries} to get meaningful automorphisms like
$\exp\!\big( i \si e^{-2z}\De_2 \big)$.
Indeed, we can view $\Td\RT_{2\Z}^\simp[[e^{-2z}]]$ as a completed
graded algebra, with the grading induced by the powers of $e^{-2z}$, and
thus compute the exponential of any operator that increases the
grading; in such a context, the exponential of a derivation is always
an automorphism---see Section~\ref{secResTransseries}.}
that commutes with $\f{\pt}{\pt z}$ and acts trivially on every convergent series,
and thus
formula~\eqref{eqdefFFslasi} necessarily produces a solution to any
  analytic differential equation that~$\FFs(\la_s)$ satisfies.
This corresponds to the Galoisian aspect of Resurgence Theory (by
way of analogy: an
algebraic number over~$\Q$, when it is not rational, has non-trivial
conjugates and they can be obtained by letting the Galois group act on it).

The name ``formal integral''~(\cite{[J.E.]}) is meant to indicate a formal object more
general than a formal series of $\C[[z^{-1}]]$, namely a transseries, here belonging to
$\C[[z^{-1}]][[e^{-2z}]]$, that satisfies the ODE at hand and depends
on the appropriate number of free parameters (or ``constants of integration''), here~$2$ since the HAE~\eqref{eq1}
is a second-order ODE.

\smallskip

%%%%%%%%%%%%%%%%%%%%%%%%%%%%%%%%%%%%%%%%%%%%%%%%

\textbf{(iii)} The terminology ``Bridge Equation'' % for~\eqref{eqBEFFslasi}
too comes from~\cite{[J.E.]};
it brings out the fact that, for an ODE like~\eqref{eq1}, the action
of the alien derivations~$\De_\omega$ on the formal integral coincide
with the action of a certain differential operator in the usual sense,
here a differential operator \wrt\ the free parameters~$\si_1$
and~$\si_2$, thus establishing a connection, or bridge, between alien
calculus and ordinary differential calculus \emph{when acting on the
  formal integral}.
%
% \textcolor{red}{The Bridge equation by \'{E}calle is the connection
%   between the normal derivative (involving $\f{\pt}{\pt\si_1},~\f{\pt}{\pt\si_2}$)
%   and the Alien derivative (the operator $\De_\om$ act on transseries).}
%
The Bridge Equation~\eqref{eqBEFFslasi} is the compact writing of
infinitely many resurgence relations
\begin{gather}
  \De_2 \GG_n = - (n+1) \,i\, \GG_{n+1}, \qquad n\gqw0,
  \\[1ex]
  \De_{-2} \GG_0 = 0, \qquad
  \De_{-2} \GG_1 = -i, \qquad
  \De_{-2} \GG_n = (n-1) \,i\, \GG_{n-1}, \quad n\gqw2.
  \end{gather}

\smallskip

%%%%%%%%%%%%%%%%%%%%%%%%%%%%%%%%%%%%%%%%%%%%%%%%

\textbf{(iv)} The symbolic Stokes automorphism
$\DDE_{\R_{\gqw 0}}^+$ and $\DDE_{\R_{\lqw 0}}^+$ are defined by
\begin{equation}    \label{eqdefDDERlqwzp}
  \DDE_{\R_{\gqw 0}}^+ \defeq \exp\!\bigg( \sum_{k=1}^{\infty}
  e^{-2kz}\De_{2k} \bigg), \qquad
  \DDE_{\R_{\lqw 0}}^+ \defeq \exp\!\bigg( \sum_{k=1}^{\infty}
  e^{2kz}\De_{-2k} \bigg)
  \end{equation}
  or, equivalently, by~\eqref{eqdefDeom} and~\eqref{eqDDdDDdp} (see Theorem~\ref{thm1}).
  These are algebra automorphisms that commute with $\f{\pt}{\pt
    z}$. The first one is defined on the algebra of simple
  $2\Z$-resurgent transseries $\Td\RT_{2\Z}^\simp[[e^{-2z}]]$ of
  footnote~\ref{ftnrestr}, the second one on
  $\Td\RT_{2\Z}^\simp[[e^{2z}]]$.
One can also define the action of~$\DDE_{\R_{\lqw 0}}^+$ on a subalgebra\footnote{%
  \label{ftnsubalgforDDElqw}
  e.g.\ $\big\{\,\Td\vp\in \Td\RT_{2\Z}^\simp[[\si_2,e^{-2z}]] \mid
  \big(\DDE_{\R_{\lqw 0}}\big)^r\Td\vp \in
  \si_2^r \, \Td\RT_{2\Z}^\simp[[\si_2,e^{-2z}]]
  \enspace\text{for each $r\gqw0$} \,\big\}$,
  where $\DDE_{\R_{\lqw 0}} \defeq \sum_{k=1}^{\infty}
  e^{2kz}\De_{-2k}$---see Lemma~\ref{lemcAsubalgforDDElqw}.
}
of $\Td\RT_{2\Z}^\simp[[\si_2,e^{-2z}]]$ that contains~$\GG$ for
each~$\si_1$, thus the \lhs\ of~\eqref{DDlqwonGG} is well-defined.
The definition of $\DDE_d^+$ with direction $d=\R_{\gqw 0}$ or $\R_{\lqw 0}$ is
such that, after Borel-Laplace resummation, it allows one to measure
the Stokes phenomenon associated to direction~$d$
(the general theory is recalled in Section~\ref{secResTransseries}); %{sec-linear}
in the case of the formal integral~$\GG$, this will give rise to
connection formulas between the analytic solutions of the HAE obtained
by Borel-Laplace summation---see Theorem~B below.
%
%
%     \textcolor{red}{Symbolic Stokes automorphism in the space of transseries
%   $\Td{\RT}_\Om^\simp[[e^{\mp2z}]]$ is the automorphism operator given by theorem \ref{thm1}
%   which means that mapping $\Td{\Phi}\cdot\Td{\Psi}\mapsto \DDE_d^+(\Td{\Phi})\cdot\DDE_d^+(\Td{\Psi})$
%   where $\Td{\Phi}$ and $\Td{\Psi}$ belong to $\Td{\RT}_\Om^\simp[[e^{\mp2z}]]$.
% }

\begin{Remark}
  The notation $\underline{\mathfrak{S}}_d$ is used e.g.\ in
    \cite{primerABS} for the inverse operator of~$\DDE_d^+$. We follow
    \'Ecalle's convention~\cite{[J.E.]}.
  \end{Remark}

  \smallskip

  Corresponding to the results that we just gave for the resurgent structure of
    \cite{[AYZ]}'s double scaling limit free energy,
    there are parallel results for $H\zu(g_s,u)$, \cite{[RCS]}'s large radius
    limit free energy~\eqref{eqdefHzugsu}.
    {We will see in Section~\ref{secresurLR} below that, \wrt\
      the variable $z=\f{1}{3}g_s^{-2}$, this formal series is $2
      u^{-3}\Z$-resurgent for any $u\in\C^*$,
      due to the explicit nonlinear change of variable~\eqref{eqdefphigs}
      which allows one to pass from $\FFs(\la_s)$ to $H\zu(g_s,u)$,
    and alien calculus produces a transseries completion that formally
    solves the $u$-equation~\eqref{uE}.}  \label{secannumthreeZresur}
  See Theorem~A' and, for the corresponding Bridge Equation, Theorem~A''.

    \smallskip

%%%%%%%%%%%%%%%%%%%%%%%%%%%%%%%%%%%%%%%%%%%%%%%%
%%%%%%%%%%%%%%%%%%%%%%%%%%%%%%%%%%%%%%%%%%%%%%%%

  \parage
  \textbf{Results on Borel-Laplace summation, connection
    formulas and real solutions.}   \label{paragResultsBLconnreal}
  %
  % \color{blue}
  %
  We now state summability results that allow one to get analytic
  functions out of the perturbative series and even the formal
  integral, and connection formulas linking the various resummations thus
  obtained.

\begin{figure}[ht]
\centering
\includegraphics[scale=0.19]{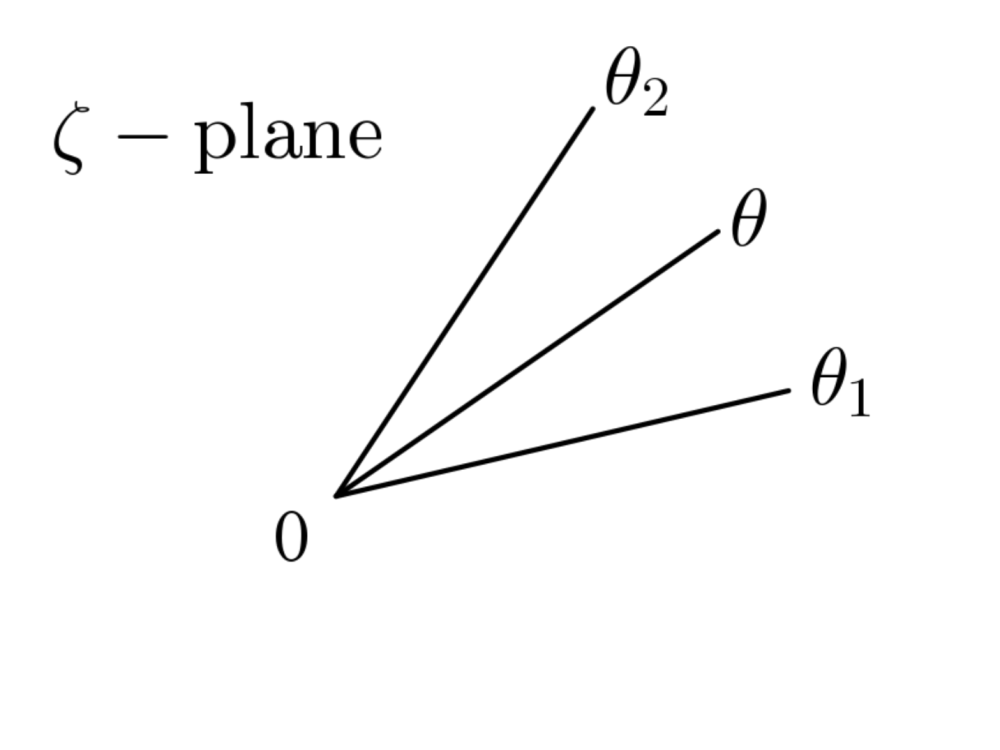}
\hspace{3em}
\includegraphics[scale=0.19]{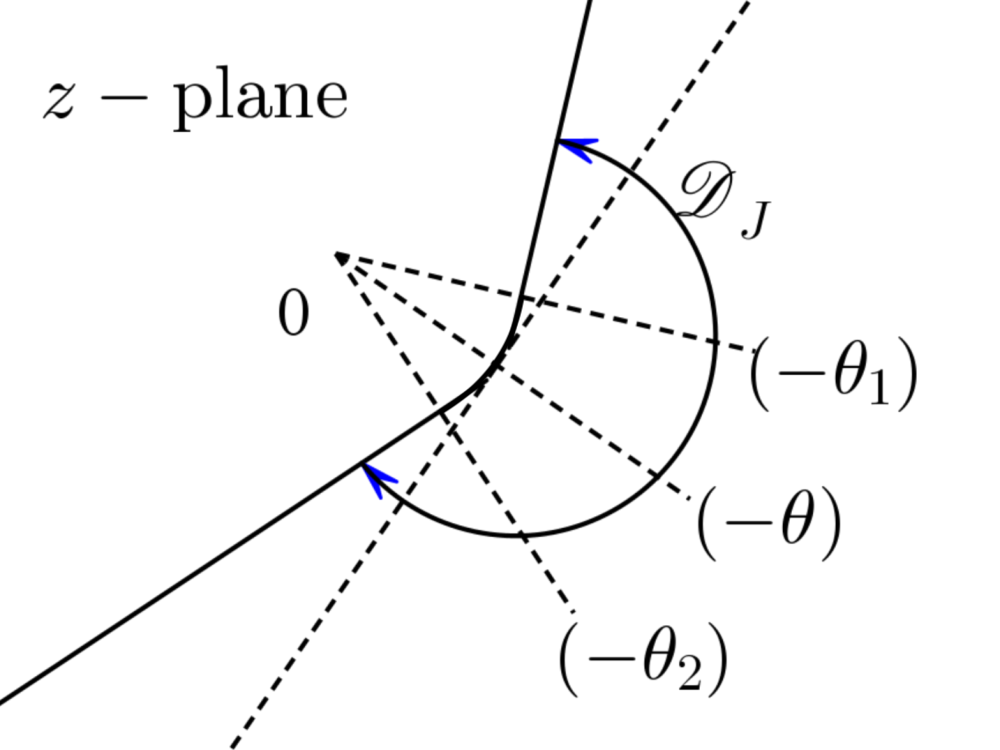}
\caption{Left: Directions for Laplace integration with
  $\th \in J=[\th_1,\th_2]$. Right: The union of half-planes $\DD_J$.}
\label{figdomDJ}
\end{figure}

  We first set our notations. Given an open interval~$I$, a formal series
  $\Td\vp(z)=\sum_{k\gqw0} c_k z^{-k}$ is said to
  be $1$-summable in the directions
  of~$I$ if the Borel transform $\wh\vp(\ze)$ of $\ti\ph(z)-c_0$ has positive radius of
  convergence and extends analytically to the sector $\{\arg\ze\in
  I\}$, with uniform bounds
  \begin{equation}   \label{eqexpboundsJ}
    |\wh\vp(\ze)| \lqw \be_J \, e^{\al_J|\ze|}
    \quad\text{for $\arg\ze\in J$},
  \end{equation}
  for every compact subinterval~$J$, with suitable constants $\al_J,\be_J\in\R$.
  Then, the Laplace transforms
  \begin{equation}   \label{eqdefLLth}
    \LL^\th\wh{\vp}(z) \defeq \int_{0}^{e^{i\th}\infty}\wh{\vp}(\ze)e^{-z\ze}\,d\ze
    \end{equation}
  associated with the various $\th\in J$ can be glued together so as to define one function
  \begin{equation}   \label{eqdefDDJ}
    \text{$\LL^J\wh\vp(z)$
  analytic in
  $\DD_J\defeq \bigcup_{\th\in J} \{ \RE(z\,e^{i\th})>\al_J \}$,}
\end{equation}
where the union of half-planes~$\DD_J$
(see Figure~\ref{figdomDJ})
is to be considered as a subset
of the Riemann surface of the logarithm\footnote{%
  \label{ftnshiftsheet}
  With the convention
  $\arg(z\,e^{i\th})\in\big(-\frac\pi2,\frac\pi2\big)$
  in~\eqref{eqdefDDJ} if $\al_J\gqw0$---see Section~\ref{secBLsum} for
  the general case.
  Notice that shifting~$J$ by $2\pi$ does not change anything in the
  Borel plane but amounts to shifting $\arg z$ by $-2\pi$, thus
  changing sheet on the Riemann surface of the logarithm:
  $\LL^{J+2\pi}\wh\vp(z)=\LL^J\wh\vp(e^{2\pi i}z)$.
}
\wrt\ the variable~$z$.

We then use the notation
\begin{equation}   \label{eqfirstdefASJ}
  \AS^J\Td\vp(z) \defeq c_0 + \LL^J\wh\vp(z)
\end{equation}
(recall that the constant term~$c_0$ of~$\ti\ph(z)$ had been discarded
when defining the Borel transform~$\wh\ph$)
and this function is uniformly $1$-Gevrey asymptotic to~$\Td\vp(z)$
in~$\DD_J$.
The Borel-Laplace sum of~$\Td\vp(z)$ is then the function
$\AS^I\Td\vp(z)$ obtained by glueing together the functions $\AS^J\Td\vp(z)$;
it is analytic in
  $\DD_I\defeq \bigcup_{J\subset\joinrel\subset I} \DD_J$,
 a set to be viewed as a sectorial neighbourhood of infinity of
opening $|I|+\pi$
(see Section~\ref{secBLsum}).

%%%%%%%%%%%%%%%%%%%%%%%%%%%%%%%%%%%%%%%%%%%%%%%%

\begin{figure}[ht]
  % \centering
  \begin{minipage}{.48\linewidth}
%    \begin{flushleft}
      \includegraphics[scale=0.18]{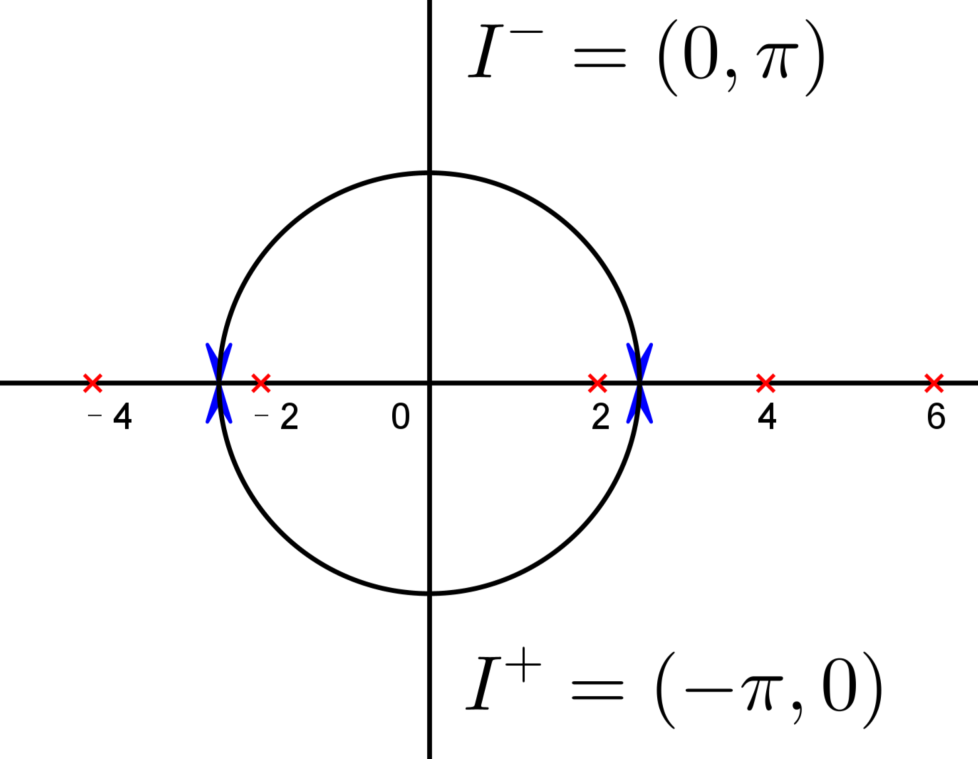}
%    \end{flushleft}
  \end{minipage}
  \hfill
  \begin{minipage}{.48\linewidth}
%    \begin{flushright}
      \includegraphics[scale=0.245]{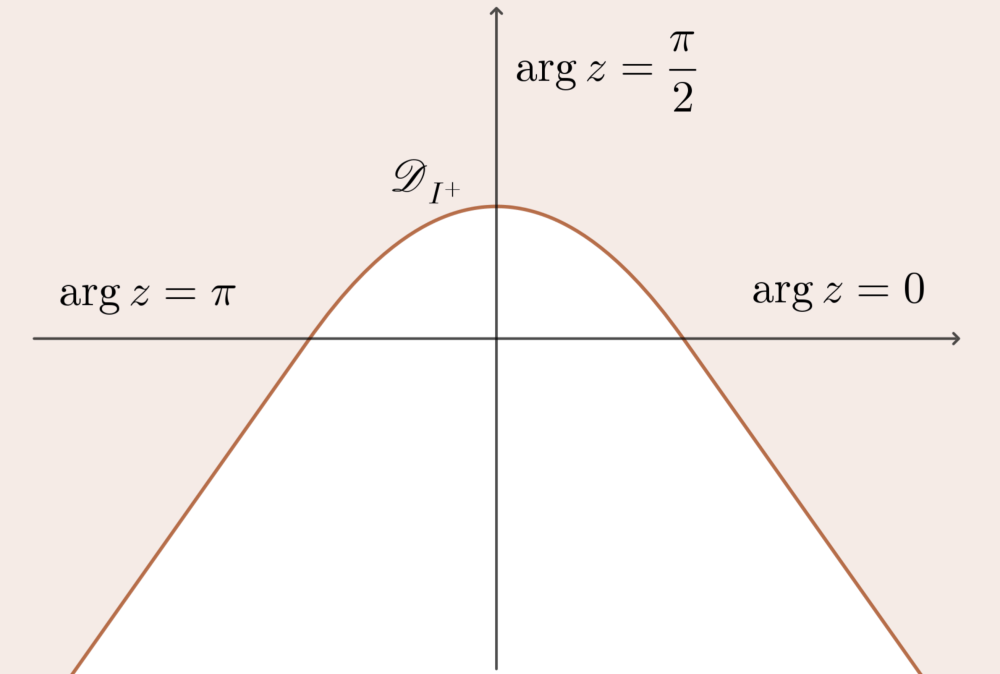}
      \\[5ex]
\includegraphics[scale=0.245]{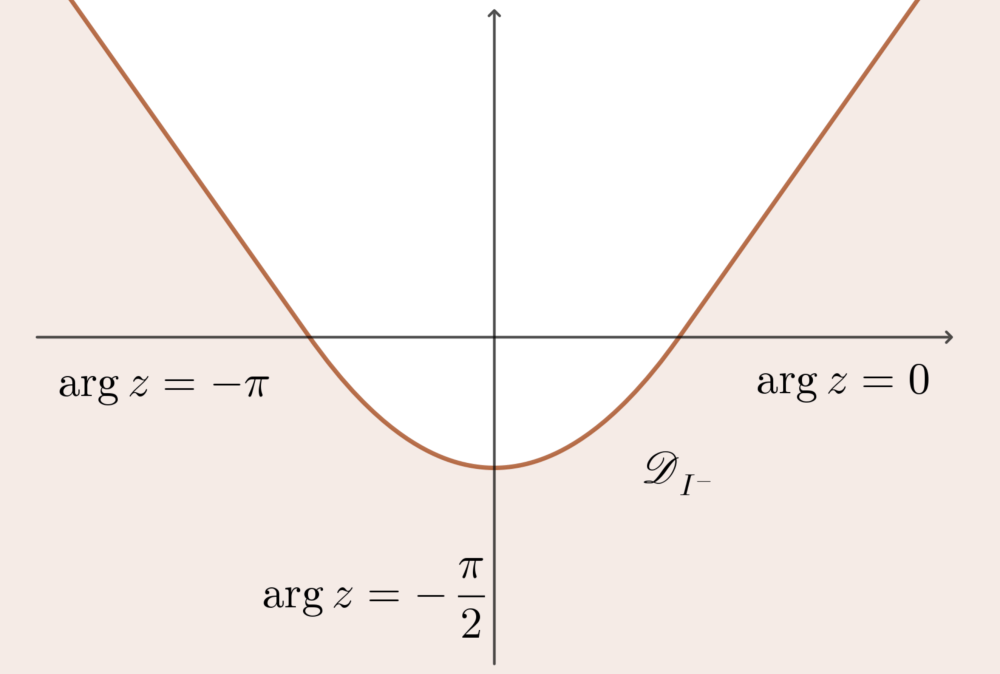}
%    \end{flushright}
  \end{minipage}
  \caption{Left: The arcs of directions $I^+$ and $I^-$ in the Borel plane. Right: The domains
    $\DD_{I^+}$ and $\DD_{I^-}$ in the plane of the variable $z=\f{1}{3}\la_s^{-2}$.}
\label{figdomain1}
\end{figure}

%%%%%%%%%%%%%%%%%%%%%%%%%%%%%%%%%%%%%%%%%%%%%%%%
%%%%%%%%%%%%%%%%%%%%%%%%%%%%%%%%%%%%%%%%%%%%%%%%

\begin{thmB}
  \emph{(i)}
  The perturbative series $\GG_0=\FFs(\la_s)$ is $1$-summable in the
  directions of $(-2\pi,0)$ (as well as in those of $(0,2\pi)$---cf.\
  footnote~\ref{ftnshiftsheet}) \wrt\ the variable
  $z=\f{1}{3}\la_s^{-2}$.
  Each $\GG_n$, $n\gqw1$, is $1$-summable \wrt~$z$ in the directions of both
  \beglab{eqdefIpm}
    I^+ \defeq (-\pi,0) \enspace\text{and}\enspace
    I^- \defeq (0,\pi).
  \edla
  There exist sectorial neighbourhoods of infinity $\DD_{I^+}$
  and~$\DD_{I^-}$ of opening~$2\pi$, with $\DD_{I^\pm}$ centred on
  $\arg z = \pm\f\pi2$, such that, for each choice of sign and each $(\si_1,\si_2)\in\C^2$, the series of functions
\begin{equation}\label{eqbspart}
\AS^{I^\pm}\GG(\la_s,\si_1,\si_2)=\si_1+\sls_{n\gqw 0}\si_2^n\, e^{-\f{2}{3}n\la_s^{-2}}\AS^{I^\pm} \GG_n(\la_s)
\end{equation}
is convergent in the domain
\begin{equation}   \label{eqdefDDpmsi2}
  \DD^\pm(\si_2)\defeq \big\{ z \in \DD_{I^\pm} \mid \Re e z>
  \tf12\ln|2\si_2| \big\}
  \end{equation}
with $z=\f{1}{3}\la_s^{-2}$
and defines an analytic solution\footnote{%
  For each choice of sign, the condition $\f{1}{3}\la_s^{-2} \in
  \DD_{I^\pm}$ defines one sectorial neighbourhood of~$0$ of
  opening~$\pi$ in the Riemann surface of the logarithm \wrt\ the
  variable~$\la_s$, centred on the ray $\arg\la_s=\mp\f\pi4$.
}
to the HAE~\eqref{eq1}.
\smallskip

\emph{(ii)} Near the direction $\arg z=0$ (\ie $\arg\la_s=0$), the connection between the
two families of solutions is given by
\begin{equation}\label{eqbsr}
  \AS^{I^+}\GG(\la_s,\si_1,\si_2)=\AS^{I^-}\GG(\la_s,\si_1,\si_2-i)
\end{equation}
for $z=\f{1}{3}\la_s^{-2}\in \DD^+(\si_2)\cap\DD^-(\si_2-i).$
\smallskip

\emph{(iii)} Near the direction $\arg z=-\pi$ (\ie $\arg\la_s=\f\pi2$),
when $|\si_2|<1$ is small enough,
there is a connection formula
\begin{equation}\label{eqbsl}
  \AS^{I^+}\GG(e^{-i\pi}\la_s,\si_1,\si_2)=
  \AS^{I^-}\GG\Big(\la_s,\si_1+\log(1+i\si_2),\f{\si_2}{1+i\si_2}\Big)
\end{equation}
in the domain
$\big\{\, \la_s \mid z=\f{1}{3}\la_s^{-2}
\in \DD^-(\f{\si_2}{1+i\si_2})\cap(e^{-2\pi
  i}\DD^+(\si_2))\big\}$.
\end{thmB}

\begin{figure}[ht]
\centering
\includegraphics[scale=0.245]{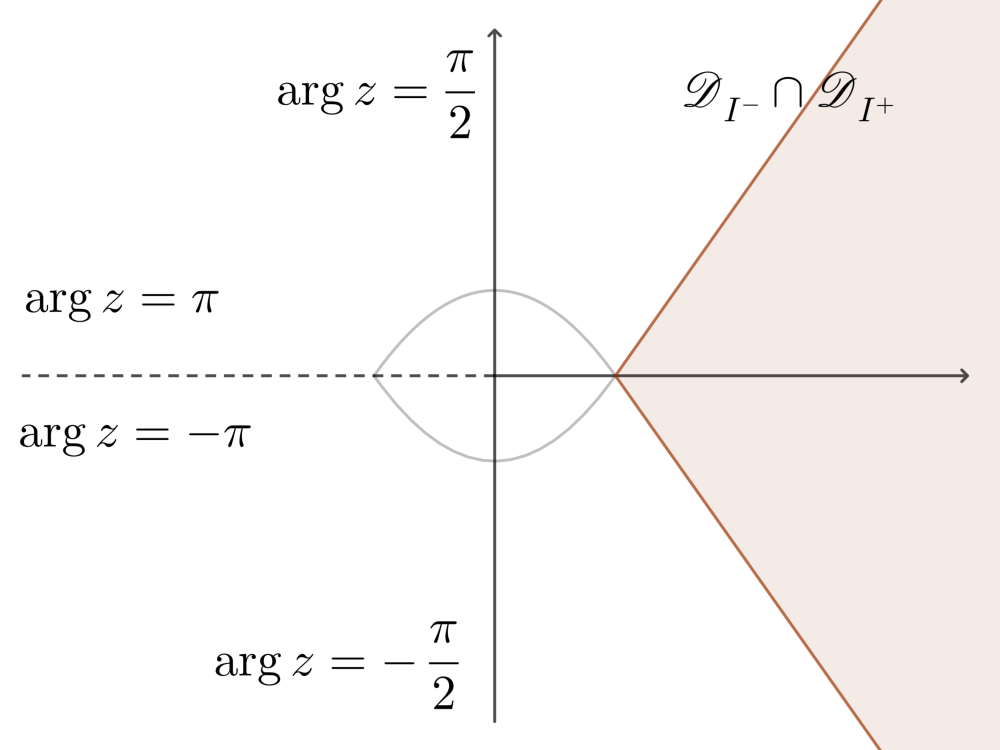}
\hspace{1em}
\includegraphics[scale=0.245]{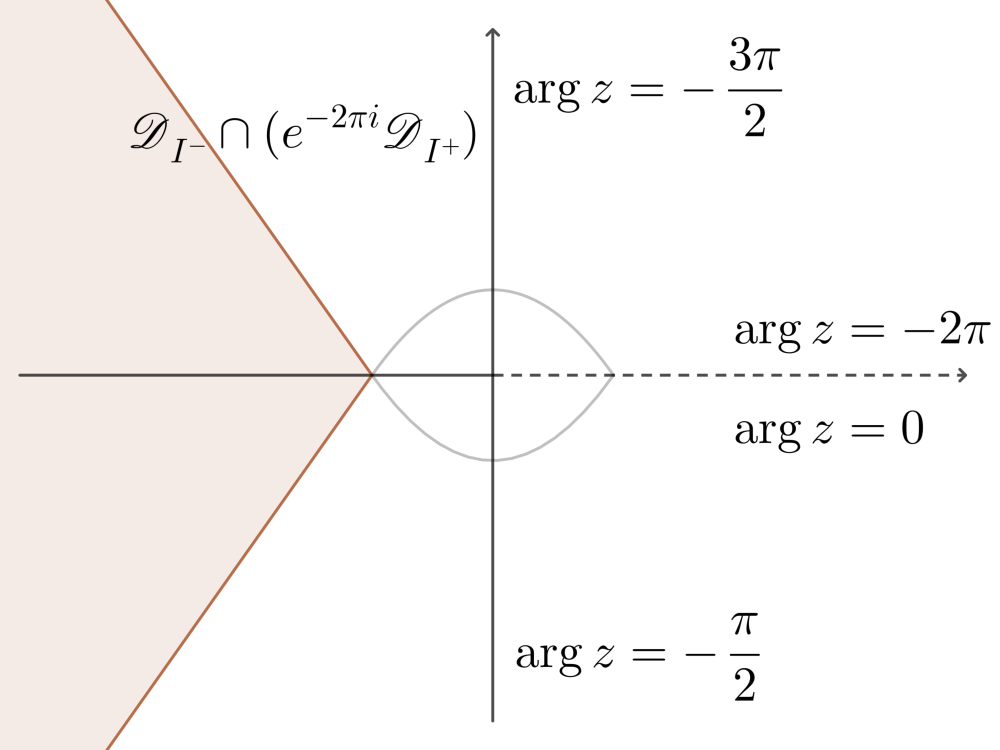}
\caption{The domains $\DD_{I^+}\cap\DD_{I^-}$ and $\DD_{I^-}\cap(e^{-2\pi
  i}\DD_{I^+})$.}
\label{figdomain2}
\end{figure}

Theorem~B(i) gives two families of solutions,
  $\AS^{I^+}\GG$ and $\AS^{I^-}\GG$, parametrized by $\si=(\si_1,\si_2)$.
  For a given parameter~$\si$, the corresponding solutions are a
  priori defined for $z=\f{1}{3}\la_s^{-2} \in \DD^\pm(\si_2)$, which is a sectorial
  neighbourhood of infinity of opening~$\pi$ only (due to the
  necessity of taking the intersection of~$\DD_{I^\pm}$ with a
  half-plane $\Re e z> \text{constant}$---see
    Figures~\ref{figdomain1} and~\ref{figdomainLEFT}). % \f12\ln|2\si_2|$).
  The connection formula~\eqref{eqbsr} stems from the Stokes
  phenomenon across the ray $\arg\ze=0$;
  it is valid for $z\in \DD^+(\si_2)\cap\DD^-(\si_2-i)$, which is
  always non-empty (see the left part of Figure~\ref{figdomain2}
  and Figure~\ref{figdomainINTERSEC}), and thus implies that $\AS^{I^+}\GG$ extends
  analytically to $\DD^+(\si_2)\cup\DD^-(\si_2-i)$.

The connection formula~\eqref{eqbsl} stems from the Stokes
  phenomenon across the ray $\arg\ze=\pi$, which is why it involves
  $\AS^{I^+}\GG(e^{-i\pi}\la_s,\si_1,\si_2)=\AS^{I^++2\pi}\GG(\la_s,\si_1,\si_2)$
  (because footnote~\ref{ftnshiftsheet} implies that
  $\AS^{I+2\pi}\Td\vp(z)=\AS^I\Td\vp(e^{2\pi i}z)$ and
  $e^{2\pi i}z$ corresponds to $e^{-i\pi}\la_s$).
  It is valid for $z\in \DD^-(\f{\si_2}{1+i\si_2}) \cap (e^{-2\pi
    i}\DD^+(\si_2))$;
  we need to require that $|\si_2|$ is sufficiently small to ensure
  that this intersection is non-empty (see the right part of
    Figure~\ref{figdomain2} and Figure~\ref{figdomainINTERSEC}),
  in which case the solution thus extends to
 $z\in \DD^-(\f{\si_2}{1+i\si_2}) \cup (e^{-2\pi
   i}\DD^+(\si_2))$.
 %
% \red{GIVE MORE DETAILED EXPLANATIONS SOMEWHERE IN Section~\ref{resurgentfreeenergy} WITH PICTURES...}

Finally, we use the connection formulas to distinguish real analytic functions among
the solutions $\AS^{I^\pm}\GG(\la_s,\si_1,\si_2)$ (compare with~\cite{AS}).

\begin{thmC}
\emph{(i)}
For any $a,b\in \R$,
the particular solution
\begin{equation}   \label{eqASGGab}
  \AS^{I^+}\GG\big(\la_s,a,b+\tfrac{i}{2} \big)=\AS^{I^-}\GG\big(\la_s,a,b-\tfrac{i}{2}\big)
\end{equation}
is analytic in $\big\{\, z=\f{1}{3}\la_s^{-2}\in
\DD_{I^+}\cup\DD_{I^-}
\text{\ and\, \ }
\Re e z > \f14\ln(1+4 b^2) \,\big\}$, and it is real-valued along the ray $\{\arg z=0\}$.
\smallskip

\emph{(ii)}
There exists $0<\th_*<\f\pi4$ such that, for any $a\in \R$ and
$\th\in(-\th_*,\th_*)$,
the particular solution
\begin{equation}   \label{eqASGGath}
  \AS^{I^+}\GG\big(e^{-i\pi} \la_s,a+i\tfrac{\th}{2}, i(1-e^{-i\th})\big)\\
  =\AS^{I^-}\GG\big(\la_s,a-i\tfrac{\th}{2}, -i(1-e^{i\th})\big)
  \end{equation}
is analytic in $\big\{\, z=\f{1}{3}\la_s^{-2}\in
\DD_{I^-}\cup(e^{-2\pi i}\DD_{I^+})
\text{\ and\, \ }
\Re e z > \f12\ln(2|1-e^{i\th}|) \,\big\}$, and it is real-valued along the ray $\{\arg z=-\pi\}$.
\end{thmC}

%%%%%%%%%%%%%%%%%%%%%%%%%%%%%%%%%%%%%%%%%%%%%%%%
%%%%%%%%%%%%%%%%%%%%%%%%%%%%%%%%%%%%%%%%%%%%%%%%

Notice that the identities~\eqref{eqASGGab}--\eqref{eqASGGath}
  are particular cases of the connection
  formulas~\eqref{eqbsr}--\eqref{eqbsl}.
  The condition $\arg z=0$ in Theorem~C(i) is
  equivalent to $\arg\la_s=0$ and thus has natural physical meaning of
  a positive real rescaled coupling constant
  $\la_s=C_{zzz}(S^{zz})^{3/2}g_s$.
  However this is not the case for the condition $\arg z=-\pi$ in~(ii),
  amounting to $\arg\la_s=\f\pi2$, for which physical implications are yet to be found.

\smallskip

Corresponding to Theorems~B and~C,
there are parallel results in the case of \cite{[RCS]}'s large radius
limit free energy {for the Borel-Laplace sums of the perturbative
  series~\eqref{eqdefHzugsu} and its transseries completion}---see
Sections~\ref{secresurLR} and~\ref{realLR} below,
particularly Theorems~B' and~C'.

%%%%%%%%%%%%%%%%%%%%%%%%%%%%%%%%%%%%%%%%%%%%%%%%
%%%%%%%%%%%%%%%%%%%%%%%%%%%%%%%%%%%%%%%%%%%%%%%%

%%%%%%%%%%%%%%%
%%%%%%%%%%%%%%%
\iffalse
%%%%%%%%%%%%%%%
%%%%%%%%%%%%%%%

\begin{thmC} The formal series solution $H^{(0)}(\tau_s, u)$ to the
  HAE in the large radius limit is resurgent with transseries
  completion \blue{$H(\tau_s,u;\si)$}  \blue{which has nice polynomial genus
  expansion} \blue{MEANING?...}.
\end{thmC}

\begin{thmD}
When $a,b\in \R,~u\in \R\setminus \{0\}$ and $b^2<\f{1}{4}(e^{4\Re
  e((3\tau_s)^{-1})}-1)$, the Borel sum of the formal integrals $H(z,u;\si_1,\si_2)$
$$\AS^{I_0^+}H\left((3\tau_s)^{-1},u;a,b+\f{i}{2} \right)=\AS^{I_0^-}H\left((3\tau_s)^{-1},u;a,b-\f{i}{2}\right)$$
to the HAE in the large radius limit is real analytic. When
$a\in\R,~k\in\Z,~u\in \R\setminus \{0\}$ and $\th$ is in an
appropriate sector, the Borel sum of the formal integrals $H(z,u;\si_1,\si_2)$
%
\begin{align*}
  &\AS^{I_\pi^+}H\left((3\tau_s)^{-1},u;a+i(\th+2k\pi),e^{i(\th+2k\pi)}-i\right)\\
=&\AS^{I_\pi^-}H\left((3\tau_s)^{-1},u;a-i(\th+2k\pi),e^{-i(\th+2k\pi)}+i\right)
\end{align*}
 to the HAE in the large radius is real analytic.
\end{thmD}

%%%%%%%%%%%%%%%
%%%%%%%%%%%%%%%
\fi
%%%%%%%%%%%%%%%
%%%%%%%%%%%%%%%

%%%%%%%%%%%%%%%%%%%%%%%%%%%%%%%%%%%%%%%%%%%%%%%%
%%%%%%%%%%%%%%%%%%%%%%%%%%%%%%%%%%%%%%%%%%%%%%%%

\parage
\textbf{Organization of the paper and outlook.}
 %
  % Theorem A will be proven in Theorem \ref{th_mbs} and Theorem
  % \ref{th_omrs} (for (i)), Theorem \ref{propkey0} (for (ii)) in \S
  % \ref{DSL}, Theorem \ref{thmbridge} (for (iii)) and Theorem
  % \ref{thmstk} (for (iv)) in \S \ref{bridgetransseries}. Theorem B
  % will be proven in Theorem \ref{th_abs} (for (i)), Theorem
  % \ref{sph2} (for (ii)) in \S \ref{bridgetransseries} and Theorem
  % \ref{realness} (for (iii)) in \S \ref{real}. The corresponding
  % Theorem A' for large radius limit will be proven in Theorem
  % \ref{pro_cp}, Theorem \ref{thmbridge2} and Theorem \ref{thmsph3},
  % and Theorem B' in Theorem \ref{bs3}, Theorem \ref{thmsph3} and
  % Theorem \ref{ess} in \S\ref{LRL} .
%
% \medskip
\\[-1.5ex]
  
\noindent -- In Section~\ref{sec-linear} we review all the essential
notions and structures in resurgence theory such as Borel-Laplace
summation and alien derivatives following \cite{[M.S.]}, and we make
it self-contained for the reader's convenience.
We seize the opportunity to add some explanations on the case of
series involving non-integer powers and we give as self-contained as
possible a resurgent treatment of the Airy equation.
\\[-1.5ex]

\noindent -- In Section~\ref{resurgentfreeenergy} we study the all-genus free energy of the B-model topological
string theory in Alim-Yau-Zhou's double scaling limit, and its
transseries completion,  solution to the nonlinear ODE~\eqref{eq1}.
We fully describe the summability properties and the resurgent
structure, and compute all the alien derivatives of all the components
of the two-parameter transseries, which correspond to the
singularities of the analytic continuation of their Borel transforms
and give us access to the Stokes phenomena associated with varying the
direction in which Borel-Laplace summation is performed. As an
application, real-analytic solutions can be distinguished among all
possible Borel-Laplace sums.
\\[-1.5ex]
  
\noindent -- In Section~\ref{LRL}, finally, we access the summability and
resurgence properties of the free energy in Couso-Santamar\'{i}a's
large radius limit and put \cite{[RCS]}'s statements on a solid ground
essentially by exploiting the interplay between the change of
variable~\eqref{eqdefphigs} and resurgence:
the resurgence in $z_1 = \f{1}{3}\la_s^{-2}$ of the double scaling
limit transseries automatically gives rise
to resurgence in $z_2=\f{1}{3g_s^2u^3}$ for the large radius 
transseries, which in turn can be interpreted as resurgence in
$\f{1}{g_s^2}$ for any fixed $u\in\C^*$.
%
% The last step is an elementary instance of duality between
% equational resurgence and parametric resurgence first introduced in~\cite{JEapplic}.

\smallskip

Several new objects arising from the resurgence analysis, like the
double scaling transseries~$\ti G(z,\si_1,\si_2)$ of~\eqref{eqGzsisi}
 and the large radius transseries
 $H\uu(g_s,u,\sigma)$ of~\eqref{eqdefHuugsusi} (giving rise to
 $\HHu(g_s,u,\si_1,\si_2)$ in~\eqref{eqHuugiveHHu}),
should have enumerative meaning from the geometric
point of view, and non-perturbative implications from the topological
string theory perspective---cf.\ Remarks~\ref{rationalDSL}
and~\ref{rationalLRL}. It would be interesting to make them manifest.

\smallskip

The truly challenging problem would be the complete resurgent analysis
for the HAE to understand the resurgent structure of topological
string free energy $\FF(z, S^{zz})$ and its partition
function. Hopefully, our methods together with the whole theory of
parametric resurgence can be extended to the recursive HAE. The first
attempt would be the resurgent analysis of the conjectures proposed in
\cite{IM} to compare with the singularity structure in the current
paper. We leave it for the future.

\medskip

%%%%%%%%%%%%%%%%%%%%%%%%%%%%%%%%%%%%%%%%%%%%%%%%
%%%%%%%%%%%%%%%%%%%%%%%%%%%%%%%%%%%%%%%%%%%%%%%%

\noindent {\bf Acknowledgements.} 
The 1st and 3rd author are partially supported by National Key R\&D Program
of China (2020YFA0713300), NSFC (No.s 11771303, 12171327, 11911530092, 12261131498,
11871045).
This paper is partly a result of the ERC-SyG project, Recursive and
Exact New Quantum Theory (ReNewQuantum) which received funding from
the European Research Council (ERC) under the European Union's Horizon
2020 research and innovation programm under grant agreement No
810573.
This work has been partially supported by the project
CARPLO of the Agence Nationale de la recherche (ANR-20-CE40-0007).

%%%%%%%%%%%%%%%%%%%%%%%%%%%%%%%%%%%%%%%%%%%%%%%%
%%%%%%%%%%%%%%%%%%%%%%%%%%%%%%%%%%%%%%%%%%%%%%%%

\section{A brief compendium of Resurgence theory}\label{sec-linear}

%%%%%%%%%%%%%%%%%%%%%%%%%%%%%%%%%%%%%%%%%%%%%%%%
%%%%%%%%%%%%%%%%%%%%%%%%%%%%%%%%%%%%%%%%%%%%%%%%

We have briefly alluded to the definition of $\Om$-resurgent series
and alien derivations in~\S\ref{paragResultsRes}, after the statement
of Theorem~A, and to Borel-Laplace summation
in~\S\ref{paragResultsBLconnreal} before the statement
of Theorem~B.
We will now expand on this, starting with more details on Borel-Laplace summation.

%%%%%%%%%%%%%%%%%%%%%%%%%%%%%%%%%%%%%%%%%%%%%%%%
%%%%%%%%%%%%%%%%%%%%%%%%%%%%%%%%%%%%%%%%%%%%%%%%

\subsection{Borel transform, convolution and Laplace transform} \label{secBLsum}

%%%%%%%%%%%%%%%%%%%%%%%%%%%%%%%%%%%%%%%%%%%%%%%%
%%%%%%%%%%%%%%%%%%%%%%%%%%%%%%%%%%%%%%%%%%%%%%%%

For any $\nu\in\C$, we use the notation
\begin{equation}   \label{eqdefznuCzii}
z^{-\nu}\C[[z\ii]] \defeq \Big\{\Td{\vp}(z)=\sls_{n\gqw
  0}a_nz^{-n-\nu}
\mid a_0,a_1,\ldots \in \C  \Big\}.
\end{equation}
In its simplest version, Resurgence Theory deals with the formal Borel
transform
\begin{equation}
  \BB \col \Td{\vp}(z) = \sls_{n\gqw 0}c_nz^{-n-1}\in
  z^{-1}\C[[z^{-1}]] \llra
  \wh\ph(\ze)=\sls_{n\gqw 0}c_n\f{\zeta^n}{n!}\in \C[[\zeta]].
\end{equation}
Observe that
%
% if $\Td{\vp}(z)$ has positive radius of convergence, then
% $\wh{\vp}$ defines an entire function in Borel plane, and if
%
$\wh{\vp}(\ze) \in\C\{\ze\}$ (\ie $\wh{\vp}(\ze)$ has positive radius
of convergence, and thus defines a holomorphic germ at the origin) if
and only if $\Td{\vp}(z)$ is a $1$-Gevrey formal power series, \ie
there exist $A, B>0$ such that $|c_n|\lqw AB^n n!$ for all $n\gqw 0.$
% We will denote by $\C[[z^{-1} ]]_1$ the vector space of all $1$-Gevrey formal series.

The convolution product of two holomorphic germs
$\wh{\vp},\wh{\psi}\in \C \{\ze\}$, defined as
\begin{equation}   \label{eqdefast}
  \wh{\vp}\ast\wh{\psi}(\ze) \defeq
  \int_0^\ze\wh{\vp}(\xi)\wh{\psi}(\ze-\xi)d\xi,
\end{equation}
is easily seen to be a holomorphic germ itself
(this makes $\C\{\ze\}$ a commutative associative algebra without
unit),
and $(\BB\ii\wh\ph)(\BB\ii\wh\psi)=\BB\ii(\wh\ph*\wh\psi)$.
%
% (Cauchy product in $\C[[z\ii]]$);
%

If $\RE\nu>0$, then the formal Borel transform extends to
\begin{equation}   \label{eqdefBBext}
  \BB \col \Td{\vp}(z) = \sls c_\mu z^{-\mu}\in
  z^{-\nu}\C[[z^{-1}]] \llra
  \wh{\vp}=\sls c_\mu \tf{\ze^{\mu-1}}{\Ga(\mu)}\in \ze^{\nu-1}\C[[\ze]]
\end{equation}
and, if also $\RE\nu'>0$, formula~\eqref{eqdefast} naturally extends to
\begin{align}
\label{eqdefconvintegmin}
  \wh\ph\in \ze^{\nu-1}\C\{\ze\}, \; \wh\psi \in \ze^{\nu'-1}\C\{\ze\}
  &\imp
  \wh\ph*\wh\psi \in \ze^{\nu+\nu'-1}\C\{\ze\}
\\[1ex] \label{eqmotivconv}
  &\quad \imp
  (\BB\ii\wh\ph)(\BB\ii\wh\psi)=\BB\ii(\wh\ph*\wh\psi).
\end{align}
The motivation is that the Laplace transform~\eqref{eqdefLLth}
satisfies
\[
  \LL^\th\Big( \f{\ze^{\mu-1}}{\Ga(\mu)} \Big)(z) = z^{-\mu}
  \quad\text{for any~$z$ in the half-plane $\{\RE(e^{i\th} z)>0\}$}
\]
\emph{provided $\RE\mu>0$}, and
\begin{equation}   \label{eqLaplConvol}
  \LL^\th(\wh\ph*\wh\psi) = (\LL^\th\wh\ph)(\LL^\th\wh\psi)
  \end{equation}
if $\wh\ph$ and~$\wh\psi$ can be subjected to Laplace transform, \emph{which
requires them to be integrable at~$0$}.

%%%%%%%%%%%%%%%%%%%%%%%%%%%%%%%%%%%%%%%%%%%%%%%%
%%%%%%%%%%%%%%%%%%%%%%%%%%%%%%%%%%%%%%%%%%%%%%%%

\subsection{Borel-Laplace summation of formal power series} \label{secBLsum}

%%%%%%%%%%%%%%%%%%%%%%%%%%%%%%%%%%%%%%%%%%%%%%%%
%%%%%%%%%%%%%%%%%%%%%%%%%%%%%%%%%%%%%%%%%%%%%%%%

For $\LL^\th\wh\ph$ to be defined,
even if $\wh\ph(\ze)$ is a holomorphic function regular at~$0$ and along
the ray $e^{i\th}\R_{\gqw0}$, we must impose an exponential bound of
the form
  \begin{equation}   \label{eqboundwhvpth}
    |\wh{\vp}(re^{i\th})|\lqw \be(\th) e^{\al(\th )r}
    \quad \text{for all $r>0$,}
  \end{equation}
for some $\al(\th), \be(\th)\in\R$.
%
% $|\ze|\to\infty$ (as in~\eqref{eqexpboundsJ}).
%
In fact, it is convenient to work with
\begin{Definition}\label{df_ns}
  \textbf{--} Let $I\subset \R$ denote an open interval and
  $\al:I\ra\R$ a locally bounded function.
  %
   % For any locally bounded $\be:I\ra \R_{\gqw 0}$, 
   %
   We denote by $\dst\NN(I,\al)$ % $\defeq\bigcup_{\be}\NN (I,\al,\be)$.
   % $\NN(I,\al,\be)$
   %
   the set of all $\wh{\vp}(\ze) \in \C\{\ze\}$ that
   have an analytic continuation to the open sector
  $\{\arg\ze \in I\}$
  and for which,
  for every $\eps>0$,
  there exists a locally bounded function $\be:I\ra \R_{\gqw 0}$
  such that
  \beglab{eqboundwhvpthVAR} \tag{\ref{eqboundwhvpth}'}
    |\wh{\vp}(r e^{i\th})|\lqw \be(\th) e^{(\al(\th)+\eps)r}
    \quad \text{for all $r>0$ and $\th\in I$.}
    \edla
    %
  % ~\eqref{eqboundwhvpth} holds for every $\th\in I$.
  %
%  \quad\text{and}\quad\dst\NN(I)\defeq\bigcup_{\alpha}\NN(I,\al)$.

\noindent \textbf{--} We set
$\dst \ti\NN(I,\al) \defeq \BB\ii\big(\NN(I,\al)\big) \subset z\ii\C[[z\ii]]$
and $\dst \ti\NN(I) \defeq \bigcup_\al \dst \ti\NN(I,\al)$.
\end{Definition}

Since locally bounded functions are precisely those functions that are
bounded on any compact subinterval,
imposing bounds of the form~\eqref{eqboundwhvpth}
or~\eqref{eqboundwhvpthVAR} along~$I$ with some locally bounded
functions~$\al$ and~$\be$ is equivalent to
imposing uniform bounds of the form~\eqref{eqexpboundsJ} for every
compact subinterval $J\subset\joinrel\subset I$.

Given $\wh\ph\in\NN(I,\al)$,~\eqref{eqdefLLth} yields a Laplace transform $\LL^\th\wh\ph$
holomorphic in 
\begin{equation}
  \Pialth \defeq \{ z\in\C \mid \RE(z\,e^{i\th})>\al(\th) \}
  \end{equation}
  for each $\th\in I$
  (this is the half-plane bisected by $e^{-i\th}\R_{\gqw0}$ that has
  $\al(\th)e^{-i\th}$ on its boundary---cf.\ Figure~\ref{figdomDJ}).
One can check that, for any $\th,\th'\in I$ such that $|\th'-\th|<\pi$, 
the half-planes $\Pialth$ and~$\Pialthp$ have a non-empty
intersection, in which $\LL^\th\wh\ph$ and $\LL^{\th'}\wh\ph$ coincide
(by the Cauchy theorem---cf.\ \cite[p.~142]{[M.S.]}).
We can thus glue together the various functions obtained by
varying~$\th$ continuously, but with a grain of salt if $|I|>\pi$,
because $\pi<|\th'-\th|<2\pi \iimp \Pialth\cap\Pialthp\neq\emptyset$ but
nothing guarantees that $\LL^\th\wh\ph$ and
$\LL^{\th'}\wh\ph$ agree on that subset of~$\C$.
The remedy is to consider a universal cover $\DDo(I,\al)$ of
\begin{equation}
  \DDu(I,\al) \defeq \bigcup_{\th\in I} \Pialth \subset \C.
  \end{equation}
  Note that the canonical projection $\DDo(I,\al)
  \to \DDu(I,\al)$ is a homeomorphism if $|I|<\pi$, but it may be
  many-to-one if $|I|>\pi$.
  We thus pick a lift $\Pialtho \subset \DDo(I,\al)$ of $\Pialth$
  that depends continuously on~$\th$, and define
\begin{equation}
  \DD(I,\al) \defeq \bigcup_{\th\in I} \Pialtho \subset \DDo(I,\al).
  \end{equation}

  \begin{Definition}
  \textbf{--} The Laplace transform in the directions of~$I$ is the operator
$\wh\ph\in\NN(I,\al) \mapsto \LL^I\wh\ph$, where
$\LL^I\wh{\vp}$ is the holomorphic function defined by
\begin{equation}   \label{eqdefLLI}
  z \in \DD(I,\al) \mapsto
  \LL^I\wh{\vp}(z) \defeq \LL^\th\wh{\vp}(z)
  \ens \text{for any $\th\in I$ such that $z\in\Pialtho$}
\end{equation}
(any two values of~$\th$ such that $z\in\Pialtho$ result in
the same value of $\LL^\th\wh{\vp}(z)$).

  \textbf{--} The Borel-Laplace summation operator in the directions
  of~$I$ is
  \begin{equation}  \label{eqdefASI}
    \AS^I \defeq\LL^I \circ \BB
    \col \ti\NN(I,\al) \to \gO\big(\DD(I,\al)\big).
\end{equation}
\end{Definition}

  \begin{Remark}
  If $\al\gqw0$, then $0$ is in the complement of~$\DDu(I,\al)$ and 
  we can view $\DDo(I,\al)$ as a subset of the universal cover~$\ti\C$ of
  $\C\setminus\{0\}$, \ie of the Riemann surface of the logarithm
  %  viewed as a spread domain 
  %
  % \beglab{eqtiCRsurflog}
  %
  % \blue{ $\ti\C \to \C\setminus\{0\} $}
  %
  % \edla
  %
on which there is a well-defined argument function
  $\arg\col\ti\C\to\R$.
  Our choice for the lifts $\Pialtho$ is then so that 
\begin{multline} \label{eqDDIaltiC}
I=(\th_1,\th_2) \imp
  \DD(I,\al) \defeq \{\, z\in\ti\C \mid \arg
  z\in(-\th_2-\tf\pi2,-\th_1+\tf\pi2),  \\
  \exists\,\th\in I \;\text{such that}\; \RE(z e^{i\th})>\al(\th) \,\} 
  \end{multline}
  in harmony with the convention indicated in
  footnote~\ref{ftnshiftsheet}. % (compare with~\eqref{eqdefDDJ}).
  \end{Remark}

At this point, we have the commutative diagram
\begin{equation}   \label{cdNNtiNNgO}
  \begin{tikzcd}
    \NN(I,\al)\subset\C\{\ze\} & {} & \gO\big(\DD(I,\al)\big)
    \\{}\\ \ti\NN(I,\al)\subset z\ii\C[[z\ii]]
	\arrow["\dst\LL^I", from=1-1, to=1-3]
	\arrow["{\dst\BB}"', from=3-1, to=1-1]
	\arrow["{\dst\hbox{}\hspace{-.8em}\AS^I}"', from=3-1, to=1-3]
      \end{tikzcd}
    \end{equation}
    
  It is easy to check that $\NN(I,\al)$ is stable under convolution,
thus $\ti\NN(I,\al)$ is stable under Cauchy product and~\eqref{eqLaplConvol} yields
\begin{alignat}{3}
  \wh\ph,\wh\psi &\in \NN(I,\al) &&\Imp &
  \LL^I(\wh\ph*\wh\psi) &= (\LL^I\wh\ph)(\LL^I\wh\psi) \\[1ex]
\label{eqASItiphtipsi}
  \ti\ph,\ti\psi &\in \ti\NN(I,\al) &&\Imp &
  \AS^I(\ti\ph \,\ti\psi) &= (\AS^I\ti\ph)(\AS^I\ti\psi).
\end{alignat}
It follows that $\C\oplus\ti\NN(I,\al)$ is a subalgebra of
$\C[[z\ii]]$ and we can extend~$\AS^I$ into an algebra homomorphism
\begin{equation}   \label{eqdefASIslightext}
  \AS^I \col \C\oplus\ti\NN(I,\al) \to \gO\big(\DD(I,\al)\big)
\end{equation}
by setting $\AS^I(1) \defeq 1$ (this is equivalent to~\eqref{eqfirstdefASJ}).
Correspondingly, setting $\BB 1 =\de$ and $\LL^I\de=1$, we can embed
$\NN(I,\al)$ into the convolution algebra $\C\de\oplus\NN(I,\al)$
(which amounts to adjunction of unit) and upgrade~\eqref{cdNNtiNNgO}
to a commutative diagram of unital algebras
\begin{equation}   \label{cdNNtiNNgOVAR}  \tag{\ref{cdNNtiNNgO}'}
  \begin{tikzcd}
    \C\de\oplus\NN(I,\al) & {} & \gO\big(\DD(I,\al)\big)
    \\{}\\ \C\oplus\ti\NN(I,\al)
	\arrow["\dst\LL^I", from=1-1, to=1-3]
	\arrow["{\dst\BB}"', from=3-1, to=1-1]
	\arrow["{\dst\hbox{}\hspace{-.8em}\AS^I}"', from=3-1, to=1-3]
      \end{tikzcd}
    \end{equation}

% One can extend the convolution law of $\C[[\ze]]$ to $\C\de\oplus \C[[\ze]]$ by the formula
% \begin{equation}
% (a\de+\wh{\vp})\ast(b\de+\wh{\psi})\defeq ab\de+a\wh{\psi}+b\wh{\vp}+\wh{\vp}\ast\wh{\psi}.
% \end{equation}
% This way $(\C\de\oplus \C[[\ze]],\ast)$ becomes a unital algebra with unit $\de$. By setting
% \[\BB 1\defeq\de,\]
% we extend $\BB $ as an algebra isomorphism between $\C[[z^{-1}]]$ and $\C\de\oplus \C[[\ze]].$

    \begin{Remark}   \label{remCV}
      Convergent series are $1$-summable in all directions: given an
      arbitrary interval~$I$, if a formal series $\ti\ph(z)$ is
      convergent for $|z\ii|<\rho$, then
      $\ti\ph\in\C\oplus\ti\NN(I,\al)$ for any $\al \gqw \rho\ii$ and
      $\AS^I\ti\ph(z)$ coincides with the usual sum of~$\ti\ph(z)$ for
      $z\in\DD(I,\al)$
      (because the Borel transform is then an entire function of
      bounded exponential type in all directions).
    \end{Remark}
        
%%%%%%%%%%%%%%%%%%%%%%%%%%%%%%%%%%%%%%%%%%%%%%%%
%%%%%%%%%%%%%%%%%%%%%%%%%%%%%%%%%%%%%%%%%%%%%%%%

\subsection{Extension to non-integer powers} \label{secBLsumext}

%%%%%%%%%%%%%%%%%%%%%%%%%%%%%%%%%%%%%%%%%%%%%%%%
%%%%%%%%%%%%%%%%%%%%%%%%%%%%%%%%%%%%%%%%%%%%%%%%

Since $\C\oplus\ti\NN(I,\al)\subset\C[[z\ii]]$,
so far we've been dealing with formal series involving only
non-positive integer powers,
but sometimes one needs formal power series involving positive integer
powers or even complex non-integer powers, typically
finite sums
\begin{equation}  \label{eqtiphsumnujCzii}
  \ti\ph = \ti\ph_1+\cdots+\ti\ph_N
  \quad \text{with}\ens\ti\ph_j\in
  z^{-\mu_j}\C[[z\ii]], \quad \mu_j\in\C.
\end{equation}
%
% where some of the $\mu_j$'s may not be in~$\Z$.
%
We use the notation $\sum\limits_{\mu\in\C}z^{-\mu} \C[[z\ii]]$
  for the vector space of all such expressions~$\ti\ph$
  (meant as a sum of vector spaces that is not a direct sum, due to
  the natural inclusions $z^{-\mu-\De} \C[[z\ii]]\subset z^{-\mu}
  \C[[z\ii]]$ for any $\mu\in\C$ and $\De\in\Z_{\gqw0}$---see~\eqref{eqexplqinnondir}).

\begin{Definition}\label{defbs}
  Given an open interval $I\subset\R$ and a locally bounded function $\al\col I\to\R$, we define
  $\tNNe(I,\al)$ as the set of all~$\ti\ph$ of the
  form~\eqref{eqtiphsumnujCzii} where
\begin{equation}   \label{eqcondznujtiphj}
  z^{\mu_j}\ti\ph_j \in \ti\NN(I,\al) \quad\text{for $j=1,\ldots,N$,}
  \end{equation}
  i.e.\ $\tNNe(I,\al) \defeq \sum\limits_{\mu\in\C}z^{-\mu}\ti\NN(I,\al)$.
The set of all \emph{formal series $1$-summable in the
  directions of~$I$} is defined to be
$\dst\tNNe(I) \defeq \bigcup_{\al}\tNNe(I,\al)$.
\end{Definition}

Suppose $\al\gqw0$.
We define the Borel-Laplace sum of any $\ti\ph\in\tNNe(I,\al)$ in the directions
of~$I$ as the holomorphic function
\begin{equation}    \label{eqdefextASItiph}
  \AS^I\ti\ph \defeq z^{-\mu_1}\AS^I(z^{\mu_1}\ti\ph_1) + \cdots + z^{-\mu_N}\AS^I(z^{\mu_N}\ti\ph_N)
  \in \gO\big(\DD(I,\al)\big)
\end{equation}
for any decomposition of~$\ti\ph$
satisfying~\eqref{eqtiphsumnujCzii}--\eqref{eqcondznujtiphj}---the \rhs\ of~\eqref{eqdefextASItiph}
does not depend of the choice of that decomposition because
\begin{equation}   \label{eqidentityDe}
  \ti\psi \in \ti\NN(I,\al) \Imp
  \AS^I(z^{-\De}\ti\psi) =   z^{-\De}\AS^I\ti\psi
\end{equation}
for any $\De\in\Z_{>0}$. See Appendix~\ref{appRemNext} for more details.

One can check that
 $\tNNe(I,\al)
%
% = \sum\limits_{\mu\in\C}z^{-\mu}\ti\NN(I,\al)
%
\subset \sum\limits_{\nu\in\C}z^{-\nu}\C[[z\ii]]$
inherits from the Cauchy product in~$\C[[z\ii]]$ a product law that
makes it a commutative associative algebra, and the Borel-Laplace
summation operator
\begin{equation}   \label{eqdefASIext}
  \AS^I \col \tNNe(I,\al) \to \gO\big(\DD(I,\al)\big)
\end{equation}
is an algebra homomorphism.
Moreover, $\tNNe(I,\al)$ is stable under $\frac{d\,}{dz}$ and
\beglab{eqASIdtiphdz}
  \AS^I \Big(\frac{d\ti\ph}{dz}\Big) = \frac{d\,}{dz} \big( \AS^I\ti\ph \big)
\edla
ultimately because
\beglab{eqBoreldifferential}
\ti\psi\in\C[[z\ii]] \imp \BB\Big(\frac{d\ti\psi}{dz}\Big) =
-\ze\BB\ti\psi(\ze).
\edla

Finally, one can check that $z\ti\NN(I) = \C \oplus \ti\NN(I)$ and,
in restriction to
\begin{equation}
  \tNNe^-(I,\al) \defeq
  \sum_{\RE\mu>-1} z^{-\mu}\ti\NN(I) = 
  \sum_{\RE\nu>0} z^{-\nu}\big(\C\oplus\ti\NN(I)\big),
\end{equation}
our extended Borel-Laplace summation operator~$\AS^I$ satisfies
\begin{equation}
  \AS^I = \LL^I \circ \BB
\end{equation}
with~$\BB$ as in~\eqref{eqdefBBext}, and with a convention for~$\LL^I$ naturally
deduced from~\eqref{eqdefLLth} and~\eqref{eqdefLLI}
(just because~\eqref{eqidentityDe} holds for any $\De\in\C$ with $\RE\De>0$).
However, one cannot define the Borel transform of an element of
$\tNNe(I)$ as a proper function when it does not belong to $\tNNe^-(I)$;
the Borel transform of~$1$ was defined above as~$\de$, which is a
symbol that can be identified with the Dirac mass at~$0$, and in
general one must resort to the formalism of \emph{majors}
\cite{[J.E.],DS}.

% \begin{Definition}\label{defbs}
% Given an open interval $I\subset \R$, we say that a formal series $\Td{\vp}(z)\in \C[[ z^{-1} ]]$ is $1$-summable in the direction of $I$ if $\BB \Td{\vp}\in \C\de\oplus \NN (I).$ The Borel-Laplace summation operator is defined as the composition
% \begin{equation}
% \AS^I \defeq\LL^I \circ \BB
% \end{equation}
% acting on all such formal series, which produces functions holomorphic in sectorial neighborhoods of $\infty$ of the form $\DD(I,\al)$, with locally bounded functions $\al:I \ra\R .$
% \end{Definition}

%%%%%%%%%%%%%%%%%%%%%%%%%%%%%%%%%%%%%%%%%%%%%%%%
%%%%%%%%%%%%%%%%%%%%%%%%%%%%%%%%%%%%%%%%%%%%%%%%

\subsection{Asymptotic expansion property, compatibility with
  composition}

%%%%%%%%%%%%%%%%%%%%%%%%%%%%%%%%%%%%%%%%%%%%%%%%
%%%%%%%%%%%%%%%%%%%%%%%%%%%%%%%%%%%%%%%%%%%%%%%%

Let~$I$ denote an open interval of~$\R$.
Any $1$-summable formal series $\ti\ph\in\tNNe(I)$ appears as the asymptotic
expansion at infinity of its Borel-Laplace sum~$\AS^I\ti\ph$, with
$1$-Gevrey qualification:
\[
  \AS^I\Td{\vp}(z)  \sim_1\Td{\vp }(z) \ens\text{in}\; \DD(I,\al),
  \quad\text{for some $\al\col I\to \R_{>0}$}.
  \]
When $\ti\ph(z) = \sum\limits_{n\gqw0} a_n z^{-n} \in \C\oplus\ti\NN(I)$,
this means that there exists a locally bounded function $\al\col I\to
\R_{>0}$, such that, for every $J\subset\joinrel\subset I$, there are
constants $L,M>0$ such that
\[
  | \AS^I\ti\ph(z) -a_0 -a_1z\ii - \cdots - a_{N-1} z^{-(N-1)} |
  \lqw L M^N N! |z|^{-N}
\]
for every $z\in\DD(J,\al|_J)$ and $N\in\Z_{\gqw0}$.
In the general case, the formulation of the asymptotic expansion
property must be adjusted to take into account the exponents $-\mu_j-n-1$
that stem from~\eqref{eqdefextASItiph}.

% \begin{Lemma}(\cite{[M.S.]} P.132)
% Suppose $\Td{\vp}\in z^{-1}\C[[z^{-1}]] $ and $\wh{\vp}=\BB \Td{\vp}\in \C[[\ze]] $ its Borel transform, then
% \begin{itemize}
%   \item $\pt_z\Td{\vp}\defeq\f{\pt}{\pt z}\Td{\vp}\in z^{-2}\C[[z^{-1}]]$ and $\BB (\pt_z\Td{\vp} )=-\ze \wh{\vp}(\ze)$,
%   \item for any $c\in \C$, $T_c\Td{\vp}(z)\defeq\Td{\vp}(z+c)\in z^{-1}\C[[z^{-1}]]$ and $\BB (T_c\Td{\vp}(z))=e^{-c\ze}\wh{\vp}(\ze).$
% \end{itemize}
% \end{Lemma}

Another property of the space of $1$-summable formal series that we
will use is its stability under nonlinear operations, as expressed in

\begin{Theorem}[{\cite[Theorem~5.55]{[M.S.]}}]  \label{thmStab}
  %
  % \noindent (The proof can be found in \cite[p.~159]{[M.S.]}).
%
Suppose $H(t) = \sum\limits_{n=0}^\infty H_n t^n \in \C\{t\}$, 
$\ti\ph_*\in\NN(I)$, and $\ti\ph,\ti\psi \in \C\oplus\NN(I)$.
Then the formal series
\begin{equation}
  H\circ \ti\ph_* \defeq \sls_{n=0}^\infty H_n\ti\ph_*^n
  \quad \text{and} \quad
  \Td{\psi}\circ (id+\ti\ph)\defeq\sls_{n\gqw 0}\frac{1}{n!}\ti\ph^n\pt^n\Td{\psi}
\end{equation}
are 1-summable in the directions of I, with
\begin{equation}
  \AS^I( H\circ \ti\ph_*)=H\circ(\AS^I \ti\ph_*)
  \quad \text{and} \quad
\AS^I(\Td{\psi}\circ (id +\Td\ph))=(\AS^I\Td{\psi})\circ(id
+\AS^I\Td\ph).
\end{equation}
\end{Theorem}

% \red{ Define~$\AS^I$ as soon as possible ($\AS^\th$ is just an
% intermediary step for the purpose of this paper).
% %
% We will also need the compatbility of~$\AS^I$ with composition (especially in the
% case of composition with an analytic change of variable).}

%%%%%%%%%%%%%%%%%%%%%%%%%%%%%%%%%%%%%%%%%%%%%%%%
%%%%%%%%%%%%%%%%%%%%%%%%%%%%%%%%%%%%%%%%%%%%%%%%

\subsection{Example: Asymptotics of the solutions to the Airy equation}   \label{exa1}

%%%%%%%%%%%%%%%%%%%%%%%%%%%%%%%%%%%%%%%%%%%%%%%%
%%%%%%%%%%%%%%%%%%%%%%%%%%%%%%%%%%%%%%%%%%%%%%%%

Since, according to \cite{[AYZ]}, Alim-Yau-Zhou's double scaling limit
partition function $\FZtops=\exp \FFs$ solves the Airy
equation~\eqref{eqairy} (up to an elementary factor),
we recall here the first steps of the resurgent treatment of the Airy
equation following \cite[Sec.~6.14]{[M.S.]}.
We will use the same formal series $\Td{\psi}(z)$ and
$\Td{\vp}(z)\defeq\Td{\psi}(-z)$ as in \cite{[M.S.]},
%
% (\S 6.14) as illustrative examples to explain the resurgent structure
%
which will play a key role in Sections~\ref{resurgentfreeenergy} and~\ref{LRL}.

\parag
As a preliminary step, the change of variable and unknown
\beglab{eqchgevarunk}
z=\tfrac{2}{3} w^{\f32}, \qquad y(w)=w\, e^z A(z)
\edla
is seen to bring the Airy equation $\f{d^2y}{dw^2}=wy$ to the form
\beglab{eqlinODEA}
A'' + 2 A' + \tf53 z\ii(A'+A) = 0.
\edla
For arbitrary $\nu\in\C$, we look for a solution of the form
$z^{-\nu}\big(1+O(z\ii)\big)$ to the linear ODE~\eqref{eqlinODEA} in
the space of formal series $z^{-\nu}\C[[z\ii]]$.
Since the dominant part of the \lhs\ is $2A'+\f53 z\ii A$, we must
impose $-2\nu+\f53=0$:
the only possibility is $\nu=\f56$, and one easily finds that there is a unique
formal solution~$\ti A(z)$, whose coefficients can be determined inductively.

Let us consider the Borel transform of $\ti
A(z)=z^{-\f56}\big(1+O(z\ii)\big) \in z^{-\f56}\C[[z\ii]]$:
\beglab{eqdefwhA}
\wh A(\ze) \defeq \BB \ti A =
\tfrac{\ze^{-\f16}}{\Ga(\f56)}\big( 1 + O(\ze) \big)
\in \ze^{-\f16}\C[[\ze]].
\edla
Since $\BB(\ti A')=-\ze\wh A(\ze)$ and $\BB z\ii=1$, the
formal series $\wh A(\ze)$ must be the unique solution of the
form~\eqref{eqdefwhA} to the Borel transformed equation
\begla
\ze^2 \wh A(\ze) - 2\ze \wh A(\ze) +
\tf53 1 * \big( \! -\ze \wh A(\ze) + \wh A(\ze) \big) = 0,
\edla
which is equivalent (upon differentiation \wrt~$\ze$) to
\begla
\f{d\,}{d\ze}\big( (\ze^2-2\ze)\wh A\big) + \tf53 (1-\ze)\wh A=0,
\edla
thus $\wh A(\ze)$ must be proportional to $(\ze-\ze^2/2)^{-\f16}$.
Therefore
\beglab{eqformulwhA}
\wh A(\ze) = \tf{\ze^{-\f16}}{\Ga(\f56)}  ( 1 - \tf{\ze}{2})^{-\f16}
= \sum_{n\gqw0} \frac{\Ga(n+\f16)}{2^n n! \Ga(\f16)\Ga(\f56)} \ze^{n-\f16}.
\edla
We see that $\wh A(\ze) \in \ze^{-\f16}\C\{\ze\}$ defines a holomorphic
germ on the Riemann surface of the logarithm for~$|\ze|$ small enough,
that has an analytic continuation to
$\{ \arg\ze \notin 2\pi\Z \} \subset\ti\C$.
In fact,
\begla
\wh A \in \ze^{-\f16}\NN(I_0,0)
\quad \text{with}\ens
I_0\defeq (-2\pi,0)
\edla
because $\th\in I_0 \mapsto \be(\th) \defeq \sup\limits_{\arg\ze=\th} | (1 - \f{\ze}{2})^{-\f16}|$
defines a locally bounded function,
and the formal solution~$\ti A(z)$ to~\eqref{eqlinODEA} is divergent.

\parag   \label{paragtispiB}
Since $\f{\ze^{-\f56}}{\Ga(\f16)}*\f{\ze^{n-\f16}}{\Ga(n+\f56)} =
\BB(z^{-\f16} z^{-n-\f56}) = \ze^n/n!$ by the last part of~\eqref{eqmotivconv},
we obtain integer powers by considering
\beglab{eqformulwhB}
\wh B(\ze) \defeq \f{\ze^{-\f56}}{\Ga(\f16)} * \wh A
= \sum_{n\gqw0} c_n \f{\ze^n}{n!},
\quad\text{with}\ens
c_n \defeq \frac{\Ga(n+\f16)\Ga(n+\f56)}{2^n n! \Ga(\f16)\Ga(\f56)}.
%
% \f{(\f{5}{6})_n(\f{1}{6})_n}{2^n n!}
%
\edla
The first part of~\eqref{eqmotivconv} shows that $\wh B(\ze)
\in\C\{\ze\}$ and, for $\arg \ze = \th\in I_0$,
the inequality $| \wh A(\ze) | \lqw \be(\th) |\ze|^{-\f16}/\Ga(\f56)$
entails
$| \wh B(\ze) | \lqw
\int_0^1 \f{ |t\ze|^{-\f56} }{\Ga(\f56)}
| \wh A\big( (1-t)\ze \big) \ze | dt
\lqw \be(\th)$,
whence
\begla
\wh B \in \NN(I_0,0),
\qquad
\ti B(z) \defeq \BB\ii \wh B = z^{-\f16}\ti A(z)
= \sum_{n\gqw0} c_n z^{-n-1}
\in \ti\NN(I_0,0)
\edla
(the function $\wh B(\ze)=\f{\ze^{-\f56}}{\Gamma(\f16)}*
\f{1}{\Gamma(\f56)}(\ze-\f{\ze^2}{2})^{-\f16}$
is denoted by $\wh\chi(-\ze)$ in \cite[Sec.~6.14]{[M.S.]}).

Finally, we set
\begla
\ti\psi(z) \defeq z\ti B(z) = z^{\f56}\ti A(z) 
= \sum_{n\gqw0} c_n z^{-n}.
\edla
The formal series~$\ti\psi(z)$ is divergent, and $\ti\psi \in \C\oplus\ti\NN(I_0,0)$ because
\begla
\wh\psi \defeq \BB\ti\psi = \de + \f{d\wh B}{d\ze}\in \C\de\oplus\NN(I_0,0)
\edla
(the Cauchy inequalities yield a uniform bound for
$|\f{d\wh B}{d\ze}(\ze)|$ for $\arg\ze$ restricted to any
compact subinterval $J\subset\joinrel\subset I_0$).

\parag
From $\ti\psi(z)= z^{\f56}\ti A(z)$ and~\eqref{eqlinODEA}, we deduce
that $\ti\psi(z)$ is the unique solution in $\C[[z\ii]]$
with constant term~$1$
to the linear ODE
\beglab{eqairy1}
\psi''+2\psi'+\f{5}{36}z^{-2}\psi=0.
\edla
%
% (modified Bessel equation).
%
% The formal series
% %
% \begin{equation}
%   %
%   \Td{\psi}\defeq\sls_{n=0}^{\infty}
%   %
%   \f{(\f{5}{6})_n(\f{1}{6})_n}{2^n n!}z^{-n}
%   %
%   =1+\Td{\psi}_1\in \C[[z^{-1}]]
%   %
% \end{equation}
% with $(\al)_0=1,~(\al)_n=\al(\al+1)\cdots(\al+n-1)~(n\gqw 1)$ is $1$-summable in the direction of $I_0\defeq(-2\pi, 0)$.
% Its Borel transformation is that
% \begin{equation}
% \Hat{\psi}=\de+\sls_{n\gqw 1}^\infty\f{(\f{5}{6})_n(\f{1}{6})_n}{2^n n!}\cdot\f{\ze^{n-1}}{\Ga(n)} \in \C\de\oplus \NN(I_0,0,\be_0)
% \end{equation}
% with $\be_0:I_0\ra \R_{>0}$ a locally bounded function.
%
% \bigskip
%
% However,
% the belonging relation is not immediately apparent; it requires the
% use of an analytic function
% $\hat{\chi}=\f{\ze^{-\f56}}{\Gamma(\f16)}\ast
% \f{1}{\Gamma(\f56)}(\ze-\f{\ze^2}{2})^{-\f16}$ to establish that
% $\hat{\chi}\in \NN(I_0,0, A(\th))$, with $A$ a locally bounded
% function and subsequently derive this relationship from
% $\hat{\psi}=-\f{d\hat\chi}{d\ze}$, as detailed in \cite{[M.S.]}
% (Theorem 6.98).
% %
% \marginlabel{not same~$\hat\chi$ as in~\cite{[M.S.]}!} %  I have modified}
%
By Borel-Laplace summation, we get a function
\begla
\AS^{I_0}\Td{\psi} \ens\text{holomorphic in}\ens
\DD(I_0,0)=\big\{z\in\Td{\C}\mid -\tf{\pi}{2}<\arg z<\tf{5\pi}{2}\big\}
\edla
that is $1$-Gevrey asymptotic to~$\ti\psi(z)$ and
solves~\eqref{eqairy1}
(thanks to~\eqref{eqASItiphtipsi} and~\eqref{eqASIdtiphdz}).

Undoing the change~\eqref{eqchgevarunk}, we get a particular solution~$y(w)$
to the Airy equation~\eqref{eqairy}:
\beglab{eqywGevrey}
y(w) \defeq \tf1{2i\sqrt\pi} \,w^{-\f14} \,e^{\f23 w^{3/2}}
\AS^{I_0} \ti\psi\big(\tfrac{2}{3} w^{\f32}\big)
\sim_{\f23}
\tf1{2i\sqrt\pi} \,w^{-\f14} \,e^{\f23 w^{3/2}}
(1 + \tf32 c_1 w^{-\f32} + \cdots),
\edla
where the $\f23$-Gevrey asymptotic expansion property (see
\cite[\S~6.14.2]{[M.S.]}) holds in the sector
$-\f\pi3<\arg w<\f{5\pi}3$
(in particular this $y(w)$ is exponentially small at infinity for $\f\pi3<\arg w<\pi$).

\parag
Similarly, still with the change of variable $z=\tfrac{2}{3}
w^{\f32}$, but with the change of unknown
\begla
y(w)=w\, e^{-z} A_+(z),
\edla
we get the linear ODE
\beglab{eqlinODEAp}
A_+'' - 2 A_+' + \tf53 z\ii(A_+'-A_+) = 0,
\edla
leading to the divergent formal solution
\beglab{eqformulAp}
\ti A_+(z) \defeq \BB\ii\Big[
\tf{\ze^{-\f16}}{\Ga(\f56)}  ( 1 + \tf{\ze}{2})^{-\f16} \Big]
\in z^{\f16} \ti\NN(I_\pi,0) \subset z^{-\f56} \C[[z\ii]]
\ens\text{with}\; I_\pi\defeq(-\pi,\pi).
\edla
We get
\begin{align}
\label{eqwhBpfromconvol}
\wh B_+(\ze) & \defeq
               \f{\ze^{-\f56}}{\Gamma(\f16)} *
               \f{1}{\Gamma(\f56)} \Big(\ze+\f{\ze^2}{2}\Big)^{-\f16}
               = \wh B(-\ze) \in \NN(I_\pi,0) \\[1ex]
\ti B_+(\ze) & =  z^{-1/6} \ti A_+(z) = - \ti B(-z) \in \ti\NN(I_\pi,0).
\end{align}
We thus arrive at
\begla
\ti\ph(z) \defeq z \ti B_+(z) = \ti\psi(-z) % z^{\f56} \ti A_+(z)
= \sum_{n\gqw0} (-1)^n c_n z^{-n}
\in \C\oplus \ti\NN(I_\pi,0)
\edla
divergent formal solution to
\begla % b{eqairy1}
\ph''-2\ph'+\f{5}{36}z^{-2}\ph=0
\edla
and giving rise to the analytic solution $\AS^{I_\pi}\ti\ph$,
holomorphic in 
$\DD(I_\pi,0)=\{z\in\Td{\C}\mid -\f{3\pi}{2}<\arg z<\f{3\pi}{2}\}$.
The corresponding solution to the Airy equation~\eqref{eqairy} is
\begla
y_+(w) \defeq \tf1{2\sqrt\pi} \,w^{-\f14} \,e^{-\f23 w^{3/2}}
\AS^{I_\pi} \ti\ph\big(\tfrac{2}{3} w^{\f32}\big)
\sim_{\f23}
\tf1{2\sqrt\pi} \,w^{-\f14} \,e^{-\f23 w^{3/2}}
(1 - \tf32 c_1 w^{-\f32} + \cdots),
\edla
which is nothing but the Airy function $\Ai(w)$.
It has $\f23$-Gevrey asymptotic behaviour similar to~\eqref{eqywGevrey},
but in the sector
$-\pi<\arg w<\pi$
(in particular, it is exponentially small at infinity for $-\f\pi3<\arg w<\f\pi3$).

%%%%%%%%%%%%%%%%%%%%%%%%%%%%%%%%%%%%%%%%%%%%%%%%

\begin{Remark}
  There is a relation with the hypergeometric function:
  \begla
  _2\text{\bf{F}}_1\big(\tf56,\tf16;1;-\xi\big) = \wh B_+(2\xi)
  = \tf{\xi^{-\f56}}{\Ga(\f16)} *
  \big[\tf{\xi^{-\f16}}{\Ga(\f56)}  (1 + \xi)^{-\f16}\big].
  \edla
  More generally, for any $a,b,c\in\C$,
  \begla
  \RE c > \RE a > 0 \Imp
  \xi^{1-c}\cdot {}_2\text{\bf{F}}_1(a,b;c;-\xi) =
  \tf{\xi^{c-a-1}}{\Ga(c-a)} *
  \big[\tf{\xi^{a-1}}{\Ga(a)}  (1 + \xi)^{-b}\big].
  \edla
  \end{Remark}

%%%%%%%%%%%%%%%%%%%%%%%%%%%%%%%%%%%%%%%%%%%%%%%%
%%%%%%%%%%%%%%%%%%%%%%%%%%%%%%%%%%%%%%%%%%%%%%%%

\subsection{Alien calculus for simple \texorpdfstring{$\Om$}{Ω}-resurgent series}
\label{secAlCalsimp}

%%%%%%%%%%%%%%%%%%%%%%%%%%%%%%%%%%%%%%%%%%%%%%%%
%%%%%%%%%%%%%%%%%%%%%%%%%%%%%%%%%%%%%%%%%%%%%%%%

We now give ourselves a lattice~$\Om$ of~$\C$, of rank~$1$ for the sake
of simplicity.
Thus $\Om =\om_1 \cdot \Z$, where $\om_1\in\C^\ast$ is one of the two
generators of~$\Om$.

We will give details about the alien operators labelled by the points of~$\Om$ in the case of
simple resurgent series,
as well as some indications for the more general framework
(for which the reader should consult \cite{[J.E.]}, \cite{DS}, \cite{EncyclDS}).

For $R>0$ and $\ze_0\in\C$ we use the notations
$D(\ze_0,R) \defeq \{\, \ze\in\C \mid |\ze-\ze_0|< R \,\}$,
\[
  D^*(\ze_0,R) \defeq D(\ze_0,R) \setminus\{\ze_0\}, \quad
  \D_R \defeq D(0,R), \quad
\D^*_R \defeq D^*(0,R), \quad
\Om^* \defeq \Om\setminus\{0\}.
\]

%%%%%%%%%%%%%%%%%%%%%%%%%%%%%%%%%%%%%%%%%%%%%%%%

\parag
According to Section~\ref{paragResultsRes}, \emph{the space of
  $\Om$-resurgent formal series} may be defined as
\begla
\ti\RT_\Om \defeq \BB\ii(\C\de\oplus\wh\RT_\Om) \subset \C[[z\ii]],
\edla
where the space $\wh\RT_\Om\subset\C\{\ze\}$ of
\emph{$\Om$-continuable holomorphic germs} is defined by the analytic
continuation property~\eqref{eqdefOmcont}.
Equivalently, $\wh\RT_\Om$ can be identified with the space of holomorphic
functions on a connected, simply connected Riemann surface~$\SOM$:
\begla
  \wh\RT_\Om = \gO(\SOM), \qquad
  \SOM \defeq \gP_\Om / \! \sim,
\edla
where~$\gP_\Om$ is the set of all paths $\ga \col [0,1] \to \C$ such that 
either $\ga\big([0,1]\big) = \{0\}$
or $\ga(0)=0$ and $\ga\big( (0,1] \big) \subset \C\setminus\Om$,
and the equivalence relation~$\sim$ is homotopy within~$\gP_\Om$: % with fixed endpoints
$\ga \sim \ga'$ if and only if % \quad\Longleftrightarrow\quad
\beglab{eqdefsimOm}
  \exists (\ga_s)_{s\in [0,1]} \;\text{such that}\enspace
\left| \begin{aligned}
&\text{for each $s\in [0,1]$, $\ga_s\in\gP_\Om$ and $\ga_s(1)=\ga(1)$}\\[.5ex]
& \text{$(s,t)\in [0,1]\times[0,1] \mapsto \ga_s(t)\in\C$ is
  continuous,}\\[.5ex]
& \ga_0=\ga, \; \ga_1 = \ga'.
\end{aligned} \right.
\edla
The map $\ga\in\gP_\Om \mapsto \ga(1)\in \C\setminus\Om^*$ passes to the
quotient and defines a ``projection'' 
$\pi_\Om \col \SOM \to \C\setminus\Om^*$,
which allows us to view $(\SOM,\pi_\Om)$ as a spread domain (or
\'etal\'e domain) over~$\C$,
\ie $\SOM$ is equipped with the unique structure of Riemann surface which turns~$\pi_\Om$
into a local biholomorphism.

%%%%%%%%%%%%%%%%%%%%%%%%%%%%%%

Given a path $\ga\col[0,1]\to\C$ and a holomorphic germ~$\wh\ph$
at~$\ga(0)$ that admits analytic continuation along~$\ga$, we use the
notation $\cont_\ga\wh\ph$ to denote the holomorphic germ at~$\ga(1)$
thus obtained.
An element~$\wh\ph$ of $\wh\RT_\Om$ is thus identified with the
function of $\gO(\SOM)$ whose value at the equivalence class of any
$\ga\in\gP_\Om$ is $(\cont_\ga\wh\ph)\big(\ga(1)\big)$.

%%%%%%%%%%%%%%%%%%%%%%%%%%%%%%

There is a special point~$\OOM$ in~$\SOM$:
the equivalence class of the trivial path $\ga(t)\equiv0$,
and $\pi_\Om\ii(0)=\{\OOM\}$.
That point belongs to the \emph{principal sheet of~$\SOM$}, defined as the set of
all $\ze\in\SOM$ which can be represented by a line segment (\ie such that the path
$t\in[0,1] \mapsto t\, \pi_\Om(\ze)$ belongs to $\gP_\Om$ and
represents~$\ze$).
Observe that~$\pi_\Om$ induces a biholomorphism from the principal
sheet of~$\SOM$ to % the cut plane $\dst U_\Om \defeq$
the cut plane $\C \setminus \big( \om_1[1,+\infty) \cup
(-\om_1)[1,+\infty) \big)$.

Each $\wh\ph\in\wh\RT_\Om$ has a principal branch holomorphic in the
principal sheet of~$\SOM$.
This is in contrast with the universal cover of $\C\setminus\Om$, which may be
defined as
\beglab{eqdefSOMst}
\SOM^* \defeq \gP_\Om^* / \! \sim
  \edla
where~$\gP_\Om^*$ is the set of all paths $\ga \col [0,1] \to \C\setminus\Om$ such that 
 $\ga(0)=\f14\om_1$
and the equivalence relation~$\sim$ is defined by the analogue of~\eqref{eqdefsimOm}.

For example in the case of $\Om=2\Z$, in view of~\eqref{eqformulwhA},
\begla
\wh A\in\gO(\cS_{2\Z}^*)\ens \text{but}\; \wh A\notin\gO(\cS_{2\Z}).
\edla
On the other hand, formula~\eqref{eqformulwhB} defines $\wh
B\in\C\{\ze\}$ and we will see later that $\wh B\in \gO(\cS_{2\Z}) = \wh\RT_{2\Z}$.

  The space $\wh\RT_\Om$ (clearly a linear space) happens to be
  stable under convolution (ultimately because~$\Om$ is stable under addition---cf.\ \cite[\S~6.4]{[M.S.]}),
  hence $\C\de\oplus \wh\RT_\Om$ is a convolution algebra and,
  via the isomorphism~$\BB$, we obtain that
  $\ti\RT_\Om$ is a subalgebra of $\C[[z^{-1}]]$.
  The algebra $\ti\RT_\Om$ is trivially stable under~$\f{d\,}{dz}$,
  because of~\eqref{eqBoreldifferential}, and
  contains the algebra of convergent germs at infinity, $\C\{z\ii\}$,
  since by Borel transform they yield entire functions, which are
  trivally $\Om$-continuable.

%%%%%%%%%%%%%%%%%%%%%%%%%%%%%%%%%%%%%%%%%%%%%%%%

\parag
A function~$\wh\ph(\ze)$ holomorphic on~$\SOM$ or~$\SOM^*$
can have singularities only ``above'' the points of~$\Om$ (\ie at ``boundary
points'' of~$\SOM$ or~$\SOM^*$, which project onto~$\Om$, with the
exception of~$\OOM$ in the case of a $\wh\ph\in\gO(\SOM)$).
A priori, these singularities can be of any kind. We will be
particularly interested in ``simple singularities'' in the sense of

\begin{figure}[ht]
  \caption{Analytic continuation along~$\ga$, for~$\ze$ near~$\om$.}
  \label{figpathga}
  \begin{center} 
    \includegraphics[width=.62\textwidth]{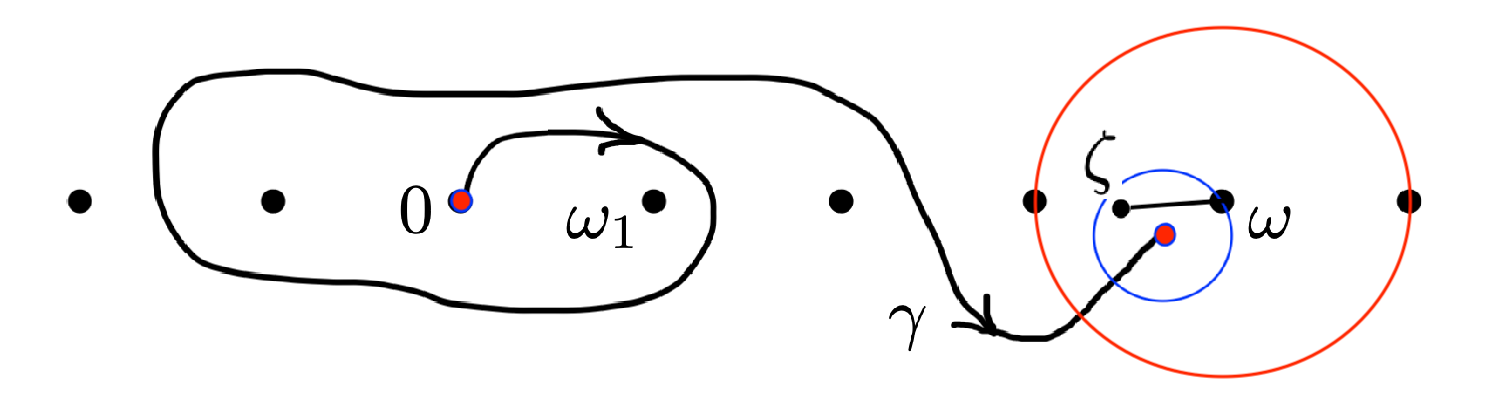}
  \end{center}
  \end{figure}

\begin{Definition}   \label{defsimpsing}
(i)  Let~$\ga$ be a non-constant path of~$\gP_\Om$ such that
  $|\ga(1)-\om|<\f12|\om_1|$ for some $\om\in\Om$;
  thus~$\om$ is uniquely determined and the analytic continuation $\cont_\ga\wh\ph$ of
  any $\wh\ph \in \wh\RT_\Om$ % $= \gO(\SOM)$ % along~$\ga$
  is holomorphic in the disc
  $D\big( \ga(1), |\ga(1)-\om| \big) \subset \C\setminus\Om$.
  We say that $\cont_\ga\wh\ph$ \emph{has a simple singularity at~$\om$}
  if one can write % , for $\ze \in D\big( \ga(1), |\ga(1)-\om| \big)$,
\begin{equation}\label{eq_ancon}
  \cont_\ga\wh{\vp}(\ze)=\f{b}{2\pi i(\ze-\om)}+\wh{\psi}(\ze-\om)\f{\log(\ze-\om)}{2\pi i}+R(\ze-\om),
\end{equation}
where~$b$ is a complex number, both~$\wh{\psi}$ and~$R$ are
holomorphic germs at~$0$, and $\log$ is any branch of the
logarithm---see Figure~\ref{figpathga}.
\smallskip

\noindent (ii) We call \emph{simple $\Om$-continuable germ} any
$\wh\ph \in \wh\RT_\Om$ all of whose branches only have simple
singularities,
and denote by $\wh\RT_\Om^\simp$ the space such germs make up.
\smallskip

\noindent (iii) We call \emph{simple $\Om$-resurgent series} any
$\Td{\vp}\in\ti\RT_\Om$ such that $\BB \Td{\vp}=a\de+\wh{\vp}$ where $a\in\C$
and $\wh\ph \in \wh\RT_\Om^\simp$.
We use the notation
\begla
\ti\RT_\Om^\simp = \BB\ii(\C\de\oplus \wh\RT_\Om^\simp)
\edla
for the space of all simple $\Om$-resurgent series.
\end{Definition}

  The space $\wh\RT_\Om^\simp$ (clearly a linear subspace of $\wh\RT_\Om$) happens to be
  stable under convolution \cite[\S~6.13]{[M.S.]},
  hence $\C\de\oplus \wh\RT_\Om^\simp$ 
  is a convolution subalgebra of $\C\de\oplus \wh\RT_\Om$ and,
  via the isomorphism~$\BB$, we obtain that
  $\ti\RT_\Om^\simp$ is a subalgebra of $\ti\RT_\Om\subset\C[[z^{-1}]]$
  (trivially stable under~$\f{d\,}{dz}$ and containing~$\C\{z\ii\}$).

 %%%%%%%%%%%%%%%%%%%%%%%%%%%%%%%%%%%%%%%%%

\parag
In the situation described in Definition~\ref{defsimpsing}(i), the
number~$b$ and the holomorphic germ~$\wh\psi$ are uniquely determined
and they depend linearly on~$\wh\ph$; indeed,
\begla
\ch\psi(\xi) \defeq \cont_\ga\wh{\vp}(\om+\xi)
\edla
can be viewed as function holomorphic in the universal cover of $\D_{|\om_1|}^*$ and
\begla
  b =  \lim_{\xi\to0} 2\pi i \xi \,\ch\psi(\xi),
  \quad
  \wh\psi(\xi) = \ch\psi(\xi) - \ch\psi(e^{-2\pi i}\xi).
  \edla
Moreover, being the difference of two branches of the analytic
continuation of a simple $\Om$-continuable germ shifted by an element
of~$\Om$, $\wh\psi$ is itself a simple $\Om$-continuable germ.
We can thus define an operator
\begla
\ZA^\ga_\om\col \C\de\oplus\wh\RT_\Om^\simp \to \C\de\oplus\wh\RT_\Om^\simp
\ens\text{such that}\ens
\ZA^\ga_\om(a \de+\wh\ph) = b \de + \wh\psi
\edla
for $a\in\C$ and $\wh\ph \in \wh\RT_\Om^\simp$, with~$b$
and~$\wh\psi$ determined by~\eqref{eq_ancon}.
This operator, which annihilates $\C\de$ and encodes the singularity at~$\om$
obtained by analytic continuation along the path\footnote{Note
that~$\ZA^\ga_\om$ is not altered if we replace~$\ga$ by any
non-constant $\ga'\in\gP_\Om$ that is homotopic to~$\ga$
or that has the property $\ga'=\ga$ on $[0,t_*]$ and
$\ga'([t_*,1])\subset D^*(\om,|\om_1|)$,
where $t_*\in(0,1)$ is such that $\ga([t_*,1])\subset
D^*(\om,|\om_1|)$.
  }~$\ga$, is called an
\emph{alien operator}.
Its counterpart $\BB\ii\circ\ZA^\ga_\om\circ \BB$ in the space of
simple $\Om$-resurgent series is denoted by the same symbol
\begla
  \ZA^\ga_\om \col \ti\RT_\Om^\simp \to \ti\RT_\Om^\simp.
\edla
From the definition and~\eqref{eqBoreldifferential},
it is easy to compute the commutator
\beglab{eqcommutZA}
\Big[\f{d\,}{dz},\ZA^\ga_\om\Big] = \om\ZA^\ga_\om.
\edla

\begin{figure}[ht]
  \centering
 \includegraphics[scale=0.065]{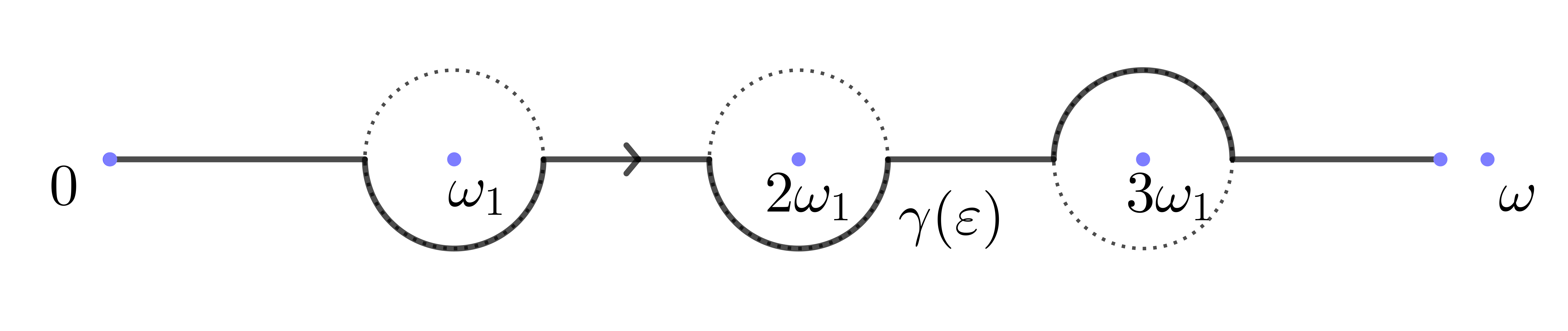}
%  \vspace{2cm}
  % 
\caption{An example of a path $\ga(\ve)$, with $r=4$ and $\ve=(+,+,-).$}
\label{fig_exa}
\end{figure}

Two families of alien operators are particularly interesting:
%
%%%%%%%%%%%%%%%%%%%%%%%%%%%%%%%%%%%%%%%%%
%
\begin{Definition}   \label{DefDeompDeom}
  Let $\om\in \Om^*$. We define
  \begla
    \De^+_\om, \; \De_\om \col
    \Td{\RT}^\simp_{\Om}\to\Td{\RT}^\simp_{\Om}
  \edla
by the formulas
\begin{equation}   \label{eqdefDeom}
  \De^+_\om\defeq\ZA_\om^{\ga(+,\cdots,+)},
  \quad
  \De_\om\defeq\sls_{\ve\in\{+,-\}^{r-1}}\f{p(\ve)!q(\ve)!}{r!}\ZA_\om^{\ga(\ve)},
\end{equation}
with notations as follows:
\begin{enumerate}[--]
\item
among the two generators of~$\Om$ we have chosen $\om_1$ so that $\om =
r\om_1$ with $r\in\Z_{\gqw1}$,
\item
  for any $\ve=(\ve_1,\cdots, \ve_{r-1})\in \{+,-\}^{r-1}$ we have
  denoted by~$p(\ve)$ and $q(\ve)=r-1-p(\ve)$ numbers of symbols~`$+$' and~`$-$',
  %in the tuple $\ve$,
  and by $\ga(\ve)$ a path that follows the line-segment $[0,(r-\f14)\om]$ except
  that it circumvents~$j\om$ to
  the right if $\ve_j=+$ and to the left if $\ve_j=-$ for any
  $j\in\{1,\ldots,r-1\}$
  (see Figure~\ref{fig_exa}).
\end{enumerate}
\end{Definition}
%
% \begin{itemize}
%   \item The operator $\ZA^{\ga(\ve)}_\om$ is a well-defined $\C$-linear operator because $\Om$ is an additive group and
% $\wh{\psi}_1$ being the difference of two branches of $\wh{\vp}$ translated by $\om$, must belong to $\wh{\RT}^\simp_\Om$(i.e. $\Td{\psi}=\BB^{-1}(b\de+\wh{\psi}_1)\in \Td{\RT}^\simp_\Om$).
% %
%   \item When {$\ve=\varnothing$ empty set}, we take $\ga(\varnothing)$ as the line-segment $[a, b]\subset]0,\om_1[$ with $a$ and $b$ sufficiently close to $0$ and $\om_1$ respectively.
% \end{itemize}
%
%
%%%%%%%%%%%%%%%%%%%%%%%%%%%%%%%%%%

One can prove that
\begla
  \De^+_\om(\ti\ph_1 \ti\ph_2) =
  (\De^+_\om\ti\ph_1 )\ti\ph_2 +
  \sls_{\substack{\om=\om_1+\om_2\\\om_1,\om_2\in \Om\,\cap\, ]0,\om[} }
  (\De^+_{\om_1}\ti\ph_1 )(\De^+_{\om_2}\ti\ph_2)
  + \ti\ph_1(\De^+_\om\ti\ph_2 ).
\edla
The reason why J.~\'Ecalle introduced the slightly more complicated
definition of~$\De_\om$ was the desire to have a family of derivations
of the algebra~$\ti\RT_\Om^\simp$:
% \ie the Leibniz rule holds.
%
\begin{Theorem}[{\cite[Theorems~6.88 and~6.91]{[M.S.]}}] \label{thmad}
  Let $\om\in\Om^*$.
  For any $\ti\ph_1$ and $\ti\ph_2\in \Td{\RT}^\simp_{\Om}$, we have
  the Leibniz rule
\begin{equation}
\De_\om(\ti\ph_1\ti\ph_2)=(\De_\om\ti\ph_1)\ti\ph_2+\ti\ph_1(\De_\om\ti\ph_2).
\end{equation}

\medskip

\noindent Furthermore, if
$\ti\ph,\ti\psi,\Td{\chi}\in \Td{\RT}^\simp_\Om$, $\ti\chi$ has no
constant term and $H(t)\in \C\{t\}$, then
\begin{align}
  \label{nolinearoperator}
  &\Td{\psi}\circ(id+\Td{\vp})\in\Td{\RT}_\Om^\simp, \quad H\circ
    \Td{\chi}\in \Td{\RT}_\Om^\simp\\[1ex] 
  \De_\om(\Td{\psi}\circ(id+\Td{\vp}))&=(\pt\Td{\psi})\circ(id+\Td{\vp})\cdot
                                        \De_\om\Td{\vp}+e^{-\om\Td{\vp}}\cdot
                                        (\De_\om\Td{\psi})\circ(id+\Td{\vp})\label{eq_nlad1}\\[1ex]  
    \De_\om(H\circ\Td{\chi} )&=(\f{dH }{dt}\circ \Td{\chi})\cdot
                               \De_\om\Td{\chi}
\end{align}
(alien chain rule).
\end{Theorem}

Apart from the commutation rule with the natural derivation
\beglab{eqcommutDeom}
\Big[\f{d\,}{dz},\De_\om\Big] = \om\De_\om,
\edla
there are no other relations between these operators and
$\f{d\,}{dz}$ or among themselves; they are called \'Ecalle's
\emph{alien derivations}.

%%%%%%%%%%%%%%%%%%%%%%%%%%%%%%%%%%%%%%%%%%%%%%%%
%%%%%%%%%%%%%%%%%%%%%%%%%%%%%%%%%%%%%%%%%%%%%%%%

\subsection{Extension to more general singularities}   \label{secgensing}

%%%%%%%%%%%%%%%%%%%%%%%%%%%%%%%%%%%%%%%%%%%%%%%%
%%%%%%%%%%%%%%%%%%%%%%%%%%%%%%%%%%%%%%%%%%%%%%%%

Returning to the general $\Om$-continuable germs of
$\wh\RT_\Om=\gO(\SOM)$, if one wants to deal with arbitrary kinds of
singularities and not just simple singularities, one may fix once for
all a generator $\om_1\in\C\setminus\R_{\lqw0}$ and consider the quotient
space
\[
 \trR_\Om \defeq \gO(\SOM^*) / \gO(\SOM)
\]
where, with reference to~\eqref{eqdefSOMst},
we identify a function $\chb\ph\in\gO(\SOM^*)$ with a holomorphic germ
at $\f14\om_1$ that has analytic continuation along any path
of~$\gP_\Om^*$,
and we view~$\gO(\SOM)$ as the subspace of those germs that have
analytic continuation in~$\D_{|\om_1|}$
(whereas the other elements of~$\gO(\SOM^*)$ are singular at~$0$).
Using the notation
\[
  \chb\ph\in\gO(\SOM^*) \mapsto
  \trb\ph = \sing_0(\chb\ph) \in \trR_\Om
\]
for the canonical projection,
we call~$\trb\ph$ an \emph{$\Om$-continuable singularity}
and~$\chb\ph$ a \emph{major} of~$\trb\ph$.
The \emph{minor} of~$\trb\ph$ is defined as the function
\[
  \wh\ph(\ze)= \minor\trb\ph(\ze) = \chb\ph(\ze) - \chb\ph(e^{-2\pi i}\ze)
  \in \gO(\SOM^*)
  \quad\text{for any major $\chb\ph$ of~$\trb\ph$.}
\]
One should think of the elements of~$\trR_\Om$ as of singularities at
the origin, among which simple singularities at~$0$ are obtained from the embedding
\beglab{eqembedCdewhRT}
\Phi\col
  a\de + \wh\ph \in \C\de \oplus \wh\RT_\Om \mapsto
  \trb\ph = \sing_0\Big( \f{a}{2\pi i\ze} + \wh\ph(\ze)\f{\log\ze}{2\pi i} \Big) \in \trR_\Om.
\edla
The formalism of $\Om$-continuable singularities will appear as
an extension of $\Om$-continuable germs.
It turns out that \emph{there exists a commutative convolution law
in~$\trR_\Om$ for which~$\Phi$ is an algebra homomorphism.}

An elementary example of $\Om$-continuable singularity at~$0$ is given by
\begla
\trn I_\nu \defeq \sing_0(\chn I_\nu), \quad
\chn I_\nu(\ze) \defeq \f{e^{i\pi\nu}\Ga(1-\nu)}{2\pi i}\ze^{\nu-1}
\quad\text{for $\nu\in\C\setminus\Z_{\gqw1}$.}
\edla
 Its minor is $\wh I_\nu(\ze) \defeq \ze^{\nu-1}/\Ga(\nu)$ (which
 is~$0$ if $\nu\in\Z_{\lqw0}$).
The singularity~$\trn I_\nu$ is not a simple singularity at~$0$ unless
$\nu=0$, in which case we find $\trn I_0 = \Phi(\de)$.
 For $\nu\notin\Z$ we use the principal branch of~$\ze^{\nu-1}$, which we
 view as element of~$\gO(\SOM^*)$ by declaring that $\arg(\f14\om_1)\in(-\pi,\pi)$.
If we extend the family $\big(\trn I_\nu\big)$ to all $\nu\in\C$ by
\[
  \chn I_n(\ze)\defeq \f{\ze^{n-1}}{\Ga(n)}\f{\log\ze}{2\pi i}
\quad\text{for $n\in\Z_{\gqw1}$,}
\]
 then the aforementioned convolution law of~$\trR_\Om$ satisfies
 $\trn I_{\nu_1}*\trn I_{\nu_2} = \trn I_{\nu_1+\nu_2}$.
More generally,
for any $\nu\in\C$ such that $\RE\nu>0$, there is an embedding
\[
  \wh\ph\in \ze^{\nu-1}\C\{\ze\} \cap \gO(\SOM^*) \mapsto
  \bem\wh\ph \in \trR_\Om
\]
such that $\minor\big(\bem\wh\ph\big) = \wh\ph$ and
$\bem\wh\ph_1 * \bem\wh\ph_2 = \bem\big(\wh\ph_1*\wh\ph_2\big)$,
where $\wh\ph_1*\wh\ph_2$ is the convolution of integrable minors
defined by~\eqref{eqdefconvintegmin},
namely
\[
  \nu\notin\Z \imp \bem\wh\ph = \sing_0\Big( \f{\wh\ph(\ze)}{1-e^{-2\pi i\nu}} \Big),
\quad
  \nu\in\Z_{\gqw1} \imp \bem\wh\ph = \sing_0\Big(
  \wh\ph(\ze)\f{\log\ze}{2\pi i}  \Big)
\]
(note that~\eqref{eqembedCdewhRT}
can be rewritten $\Phi(a\de + \wh\ph) = a \trn I_0 + \bem\wh\ph$).

\begin{figure}[ht]
  \centering
 \includegraphics[scale=0.123]{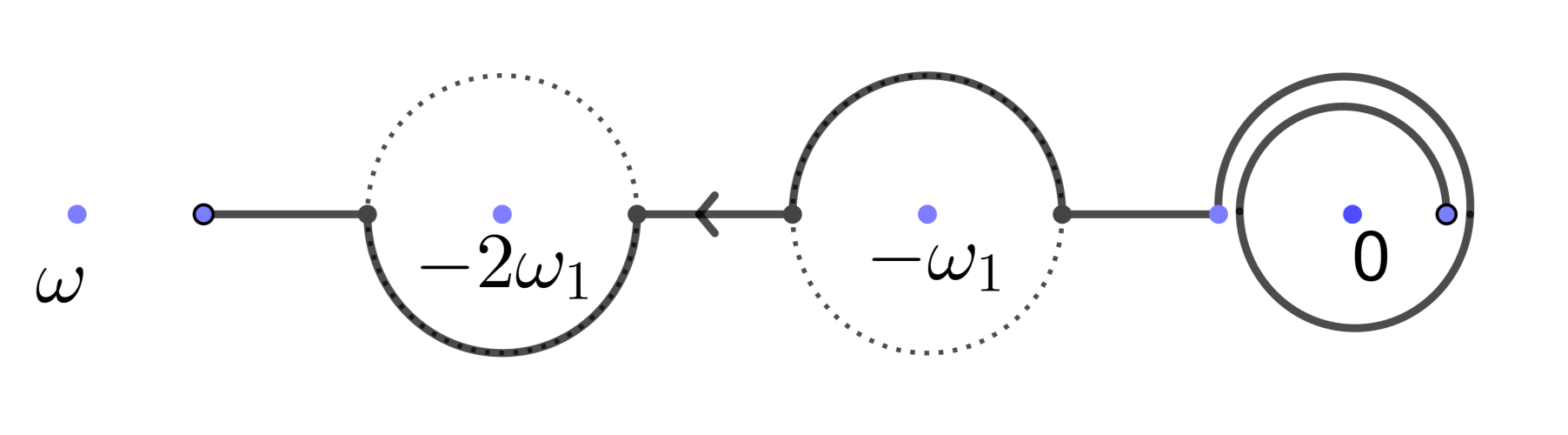} \hspace{19.5em} 
%\hspace{1.5em}

 \vspace{-8.8ex}
\hspace{16.5em} \includegraphics[scale=0.0475]{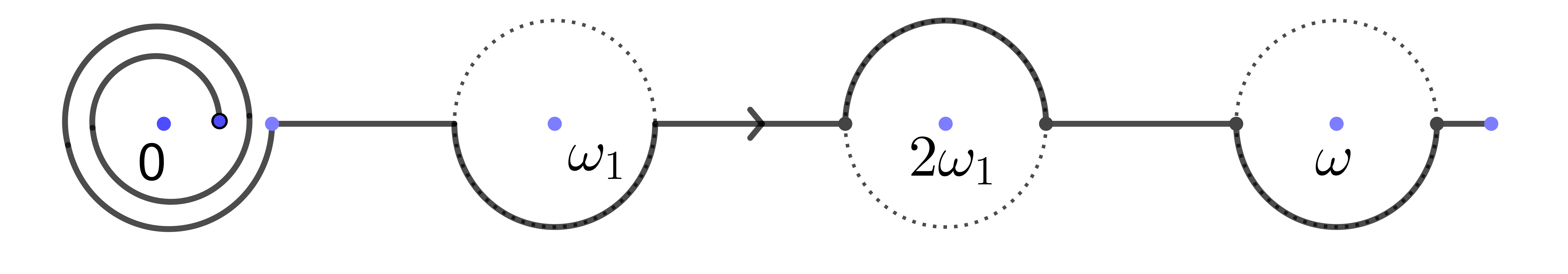}
%  \vspace{2cm}

\caption{Examples of path $\ti\ga(\ve)$, with $r=3$,
  $\ve=(+,-)$. Left: $N=3$. Right: $N=4$. In all cases
  $\ga(0)=\f14\om_1$ and $\ga(1)=\om+\f14\om_1$}
\label{fig_new_exa}
\end{figure}

We can now extend Definition~\ref{DefDeompDeom} and define operators
of~$\trR_\Om$ that measure singularities at certain ``boundary
points'' of~$\SOM$.
These new operators~$\De^+_\tom$ and~$\De_\tom$ will be indexed by all $\tom\in\TOM^*$,
%
% $\De_\tom^+, \De_\tom \col \trR_\Om\to\trR_\Om$ 
%
where $\TOM^*$ is the lift~$\pi\ii(\Om^*)$ of~$\Om^*$ to the Riemann
surface of the logarithm
$\pi\col\ti\C\to\C^*$,
and they will boil down to the previous~$\De^+_\om$ and~$\De_\om$ in
the case of simple singularities in the sense that
\[
  \De_\tom^+ \circ \Phi = \Phi\circ \De_\om^+, \quad
  \De_\tom \circ \Phi = \Phi\circ \De_\om.
\]
%
% so that if $a\de+\wh\ph\in\C\de\oplus\wh\RT_\Om^\simp$ is identified
% with~$\trb\ph$ as in~\eqref{eqembedCdewhRT}, then
% %
% $\De_{\pi(\tom)}(a\de+\wh\ph)$ gets identified with $\De_\tom\trb\ph$.
%
To proceed, we write $\tom = e^{i\pi N} r \om_1$ with
$r\in\Z_{\gqw1}$ and $N\in\Z$ and let
\begin{gather*}
  \De_\tom^+ \trb\ph \defeq \sing_0\!\Big( (
  \cont_{\ti\ga(+,\cdots,+)}\minor\trb\ph)(\tom+\ze) \Big)
  \\[1ex]
  \De_\tom \trb\ph \defeq \sls_{\ve\in\{+,-\}^{r-1}}\f{p(\ve)!q(\ve)!}{r!}\sing_0\!\Big( (
  \cont_{\ti\ga(\ve)}\minor\trb\ph)(\tom+\ze) \Big)
\end{gather*}
%
% using $\ga_+(\ve)$ if~$N$ is even and $\ga_-(\ve)$ if~$N$ is odd,
%
where~$\ti\ga(\ve)$ goes from~$\f14\om_1$ to $\f14 e^{i\pi N}\om_1$
turning around the origin ($N$ half-turns) and follows
the line-segment $[\f14 e^{i\pi N}\om_1,(r-\f14)e^{i\pi N}\om_1]$ except
  that it circumvents~$j\om$ to
  the right if $\ve_j=+$ and to the left if $\ve_j=-$ for any
  $j\in\{1,\ldots,r-1\}$, and we add a half-turn around~$\om$ from
  $(r-\f14)e^{i\pi N}\om_1$ to $(r+\f14)e^{i\pi N}\om_1$ if~$N$ is
  even
  (so that in all cases~$\ti\ga(\ve)$ ends at
  $\om+\f14\om_1$, where~$\om$ is the projection of~$\tom$ in~$\C^*$,
  and there is a natural way of viewing
  $(\cont_{\ti\ga(\ve)}\minor\trb\ph)(\tom+\ze)$ as element of
  $\gO(\SOM^*)$).
  See Figure~\ref{fig_new_exa}.

One can view $\C\de\oplus\wh\RT_\Om^\simp$ as the largest
subspace of $\C\de\oplus\wh\RT_\Om$ whose image by~$\Phi$ is stable
under all operators~$\De_\tom$.

%%%%%%%%%%%%%%%%%%%%%%%%%%%%%%%%%%%%%%%%%%%%%%%%
%%%%%%%%%%%%%%%%%%%%%%%%%%%%%%%%%%%%%%%%%%%%%%%%

\subsection{Resurgence in the Airy equation}

%%%%%%%%%%%%%%%%%%%%%%%%%%%%%%%%%%%%%%%%%%%%%%%%
%%%%%%%%%%%%%%%%%%%%%%%%%%%%%%%%%%%%%%%%%%%%%%%%

    \label{exa3}
%
%  According to \cite[p.~259]{[M.S.]}...
%
In Section~\ref{exa1} we have introduced various formal series in
relation with the Airy equation;
in view of~\eqref{eqformulwhA} and~\eqref{eqformulAp},
\[
\wh A(\ze) = \tf{\ze^{-\f16}}{\Ga(\f56)}  ( 1 - \tf{\ze}{2})^{-\f16}
\quad\text{and}\quad
\wh A_+(\ze) = \tf{\ze^{-\f16}}{\Ga(\f56)}  ( 1 + \tf{\ze}{2})^{-\f16}
\]
may be considered as integrable $2\Z$-continuable minors (that are not
regular at the origin),
and~\eqref{eqformulwhB} and~\eqref{eqwhBpfromconvol} can be rewritten
\begin{multline*}
  \wh B = \wh I_{1/6} *\wh A, \ens
  \wh B_+ = \wh I_{1/6} *\wh A_+, \\[1ex]
\text{whence} \quad
  \bem\wh B = \trn I_{1/6} *\bem\!\wh A, \ens
  \bem\wh B_+ = \trn I_{1/6} *\bem\!\wh A_+ \in \trR_{2\Z}.
\end{multline*}
Applying the above recipe and taking care of identifying the right
branches of the analytic continuation, we get
\begin{multline*}
  \De_{2e^{i0}} \bem\!\wh A = \sing_0\bigg(
  \f{(2+\ze)^{-\f16}}{\Ga(\f56)}\Big(\f{e^{i\pi}\ze}{2}\Big)^{-\f16}
  \bigg)
  = e^{-i\pi/6} \sing_0\big( \wh A_+(\ze) \big) \\[1ex]
  = e^{-i\pi/6} (1-e^{2\pi i/6}) \,\bem\!\wh A_+ = -i\,\bem\!\wh A_+
\end{multline*}
and, similarly,
\[
 \De_{2e^{i\pi}} \bem\!\wh A_+ = \sing_0\bigg(
  \f{\big((2e^{i\pi})(1-\f\ze2)\big)^{-\f16}}{\Ga(\f56)}\Big(\f\ze2\Big)^{-\f16}
  \bigg)
  = e^{-i\pi/6} \sing_0\big( \wh A(\ze) \big) \\[1ex]
  =  -i\,\bem\!\wh A.
\]
Since~$\De_{2e^{i0}}$ and~$\De_{2e^{i\pi}}$ are derivations that
annihilate~$\trn I_{1/6}$, it follows that
\[
  \De_{2e^{i0}} \bem\wh B=  -i\,\bem\wh B_+,
  \quad
  \De_{2e^{i\pi}} \bem\wh B_+=  -i\,\bem\wh B
  \quad \text{in the algebra~$\trR_{2\Z}$,}
\]
which boils down to 
\[
  \De_{2} \wh B=  -i\,\wh B_+,
  \quad
  \De_{-2} \wh B_+=  -i\,\wh B
  \quad \text{in}\ens \C\de\oplus\RT_{2\Z}^\simp.
\]
Finally, since $\ti B = z\ii\ti\psi$ and $\ti B_+ = z\ii\ti\ph$, we
can rewrite this as
\begin{equation}   \label{eqDepsiph}
\ti\psi, \ti\ph \in \ti\RT_{2\Z},  \quad
  \De^+_2\Td{\psi} = \De_2\Td{\psi} =-i\Td{\vp},
  \quad
  \De^+_{-2}\Td{\vp}=\De_{-2}\Td{\vp}=-i\Td{\psi}
\end{equation}
(using the fact that $\De^+_\om= \De_\om$ when~$\om$ is a generator
of~$\Om$).
On the other hand,
\beglab{eqotherderivz}
\om\in 2\Z^*\setminus\{2\} \imp \De_\om\ti\psi = 0,
\qquad
\om\in 2\Z^*\setminus\{-2\} \imp \De_\om\ti\ph = 0.
\edla

%%%%%%%%%%%%%%%%%%%%%%%%%%%%%%%%%%

%%%%%%%%%%%%%%%%%%%%%%%%%%%%%%%%%%%%%%%%%%%%%%%%
%%%%%%%%%%%%%%%%%%%%%%%%%%%%%%%%%%%%%%%%%%%%%%%%

\subsection{Simple \texorpdfstring{$\Om$}{Ω}-resurgent transseries}
\label{secResTransseries}

%%%%%%%%%%%%%%%%%%%%%%%%%%%%%%%%%%%%%%%%%%%%%%%%
%%%%%%%%%%%%%%%%%%%%%%%%%%%%%%%%%%%%%%%%%%%%%%%%

We now explain the interplay between Borel-Laplace summation and alien
calculus in the case of simple $\Om$-resurgent series (but much of what
follows can be adapted to the case of more general singularities).

\parag
We first fix $\om_1\in\C^*$ and $\Om = \Z\, \om_1$ as in
Sections~\ref{secAlCalsimp}--\ref{secgensing}, and set
\begin{equation}\label{eq_o}
  d \defeq \R_{\gqw0} \,\om_1, \quad
  \Om^+ \defeq d \cap \Om^* = \{m\,\om_1\mid m\in\Z_{\gqw1}\}.
\end{equation}

\begin{Definition}\label{omts}
We call \emph{simple $\Om$-resurgent transseries} any expression of
the form
\[
  \Td{\Psi} = \sls_{m\gqw0} e^{-m\, \om_1 z} \Td{\psi}_m(z)
\]
where $(\Td{\psi}_m)_{m\gqw 0}$ is a sequence in~$\ti\RT_\Om^\simp$.
\end{Definition}

The space of all simple $\Om$-resurgent transseries
can be viewed as a completed graded algebra
\[
  \Td{\RT}_\Om^\simp[[e^{-\om_1z}]] = \bigoplus_{m\gqw 0}^{\wedge}e^{-m\,\om_1 z}\Td{\RT}_\Om^\simp,
\]
\ie we can manipulate infinite sums thanks to the notion of formal
convergence induced by the $m$-grading.

\begin{Theorem}[{\cite[p.~226]{[M.S.]}}]
  \label{thm1}
  Consider the two operators of $\ti\RT_\Om^\simp[[e^{-\om_1z}]]$
  defined by
  \beglab{eqDDdDDdp}
    \DDE_d \defeq \sls_{\om\in \Om^+} e^{-\om z}\De_\om,
    \quad
    \DDE_d^+ \defeq \ID+\sls_{\om\in \Om^+}e^{-\om
      z}\De_\om^+
  \edla
  with the convention $\De_\om(\sum e^{-m\, \om_1 z} \ti\psi_m) \defeq \sum
  e^{-m\, \om_1 z} \De_\om\ti\psi_m$ and similarly for $\De_\om^+$.
  Then
\smallskip

\emph{(i)}
%
% the first one,
$\DDE_d$ is a derivation that commutes with the
  natural derivation~$\f{d\,}{dz}$;
\smallskip

\emph{(ii)} 
% the second one,
$\DDE_d^+$ is an algebra automorphism that commutes with~$\f{d\,}{dz}$;
\smallskip

\emph{(iii)}
  moreover, 
\begin{equation}\label{relationexp}
  \DDE_d^+=\exp(\DDE_d)=\sls_{s\gqw0}\f{1}{s!}\left(\DDE_d\right)^s.
\end{equation}
\end{Theorem}

The operator $\DDE_d$ is called the \emph{symbolic Stokes
  infinitesimal generator for the direction~$d$} and the operator
$\DDE_d^+$ is called the \emph{symbolic Stokes automorphism for the
  direction~$d$}.
Note that the \rhs\ of~\eqref{relationexp} makes sense because
$\DDE_d$ increases the $m$-grading.
This relation implies that the family of operators
$(\De_{m\om_1}^+)_{m\gqw1}$ can be expressed in terms of the family
$(\De_{m\om_1})_{m\gqw1}$,
\begin{multline*}
  \De_{\om_1}^+ = \De_{\om_1}, \quad
  \De_{2\om_1}^+ = \De_{2\om_1} + \tf1{2!} \De_{\om_1}\circ \De_{\om_1}, \\[1ex]
  \De_{3\om_1}^+ = \De_{3\om_1} + \tf1{2!} (\De_{2\om_1}\circ
  \De_{\om_1} + \De_{\om_1}\circ \De_{2\om_1})
  + \tf1{3!}\De_{\om_1}\circ \De_{\om_1}\circ \De_{\om_1},\quad \text{etc.}
\end{multline*}
and vice versa.
The commutation rules~\eqref{eqcommutZA} and~\eqref{eqcommutDeom} show
that each ``homogeneous operator'' $e^{-\om z}\De_\om^+$ or $e^{-\om z}\De_\om$ commutes with~$\f{d\,}{dz}$.

%%%%%%%%%%%%%%%%%%%%%%%%%%%%%%%%%%%%%%

\parag
Let us write $\om_1 = |\om_1| e^{i\th^*}$ with $\th^*\in\R$ and
consider an interval $I=(\th^*-\de,\th^*+\de)$ of length $\lqw\pi$.
We will be interested in formal power series that are $1$-summable in
the directions of $I\setminus\{\th^*\}$, \ie in the directions of
\beglab{eqdefILIR}
  I_R \defeq (\th^*-\de,\th^*)
  \ens \text{and} \ens
  I_L \defeq (\th^*,\th^*+\de)
\edla
but not necessarily in the direction~$\th^*$; supposing them to be
$\Om$-resurgent our aim is to compare the action of the summation
operators~$\AS^{I_R}$ and~$\AS^{I_L}$ defined
by~\eqref{eqdefASIslightext}. %  on them.

We thus give ourselves a locally bounded function $\al\col I_R\cup I_L
\to\R_{\gqw0}$ and consider the space
\[
  \ti\RT_\Om^\simp(I,\al) \defeq  \ti\RT_\Om^\simp \cap
  \big( \C \oplus \ti\NN(I_R,\al) \big) \cap
  \big( \C \oplus \ti\NN(I_L,\al) \big),
\]
which is a subalgebra of $\C[[z\ii]]$.
Note that
\[
  \DD^* \defeq \DD(I_R,\al) \cap \DD(I_L,\al)
  %
  % \ens \text{is contained in the half-plane $\{\RE(z e^{i\th_*})>0\}$}
  %
\]
contains sectors bisected by $e^{-i\th^*}\R_{\gqw0}$ of any opening
$<\pi$,
and is contained in the half-plane $\{\RE(z e^{i\th_*})>0\}$,
whence $e^{-\om_1z}$ is exponentially small at infinity in that domain.

For any $\ti\ph \in \ti\RT_\Om^\simp(I,\al)$, we want to compare the 
functions
\[
  \AS^{I_R}\ti\ph \in \gO(\DD(I_R,\al))
  \quad \text{and} \quad \AS^{I_L}\ti\ph \in \gO(\DD(I_L,\al)),
  \]
  whose difference is exponentially small on~$\DD^*$.
  This is possible if we assume that $\De_\om^+\ti\ph \in \C
  \oplus \ti\NN(I_L,\al)$ for each $\om\in\Om^+$.
  The result is then
\beglab{eqStkTruncatedmst}
z \in \DD^* \imp
  \AS^{I_R}\ti\ph(z) = \AS^{I_L}\ti\ph(z) +
  \sum_{m=1}^{m_*} e^{-m\,\om_1 z} \AS^{I_L} \De_{m\om_1}^+\ti\ph(z)
  + O( |e^{-\mu\,\om_1z}|) % O( e^{-\mu \RE(\om_1 z)})
\edla
for any integer $m_*\gqw1$ and any real $\mu\in(m_*,m_*+1)$.

% Here is a possible statement for simple $\Om$-resurgent transseries
% depending on a parameter~$\si$:

% \begin{Theorem} % (\cite{[M.S.]} P.230)
%   %
%   \label{th_ssa}
%   %
% The space $\ti\RT_\Om^\simp(I,\al)\{\si e^{-\om_1 z}\}$ consisting
% of all
% %
% \[
%   %
%   \Td{\Psi} = \sls_{m\gqw0} \si^m e^{-m\, \om_1 z} \Td{\psi}_m(z)
%   %
%   \quad\text{with each}\ens \ti\psi_m \in \ti\RT_\Om^\simp(I,\al)
%   %
% \]
% %
% such that, for every $\eps>0$, there exist
% $B,C>0$ such that the constant term of each~$\ti\psi_m$ satisfies
% %
% $|\ti\psi_{m,0}|\lqw B C^m$
% %
% and
% %
% \[
%   %
%   | \BB(\ti\psi_m - \ti\psi_{m,0})(\ze) | \lqw B C^m e^{(\al(\arg\ze)+\eps)|\ze|}
%   %
%   \quad \text{for all $\ze\in S(I_L)\cup S(I_R)$}
%   %
% \]
% %
% is an algebra, on which we have two algebra homomorphisms~$\AS^{I_L}$ and~$\AS^{I_R}$
% %
% defined by the formula
% %
% \[
%   %
%   \AS^J\ti\Psi(z,\si) \defeq \sum_{m\gqw0} \si^m e^{-m\, \om_1 z}
%   \AS^J\ti\psi_m(z),
%   %
% \]
% %
%  \\[1ex]
%   %
% holomorphic function of $(z,\si)$ for $z\in\DD(J,\al)$ and
% $|\si|e^{-\RE(\om_1 z)}<\f1C$.
% %
% % with $J=I_L$ or~$I_R$.
% \smallskip

% The space $\{ \ti\Psi\in\ti\RT_\Om^\simp(I,\al)\{\si e^{-\om_1 z}\}
% \mid \DDE_d^+\ti\Psi\in\ti\RT_\Om^\simp(I,\al)\{\si e^{-\om_1 z}\} \}$
% %
% is a subalgebra, on which
% % 
%   \begin{equation}   \label{eqPassageRelation}
%     %
%     \AS^{I_R} = \AS^{I_L} \circ \DDE_d^+
%     %
%     \quad \text{in restriction to $z\in\DD^*$.}
%     %
%   \end{equation}
% %
% \end{Theorem}

\begin{figure}[ht]
  \caption{Decomposition of integration contour~$\Gadiff$ for the computation of
  $\AS^{I_R}\ti\ph - \AS^{I_L}\ti\ph$.}
  \label{figHankelcontours}
  \begin{center} 
    \includegraphics[width=.64\textwidth]{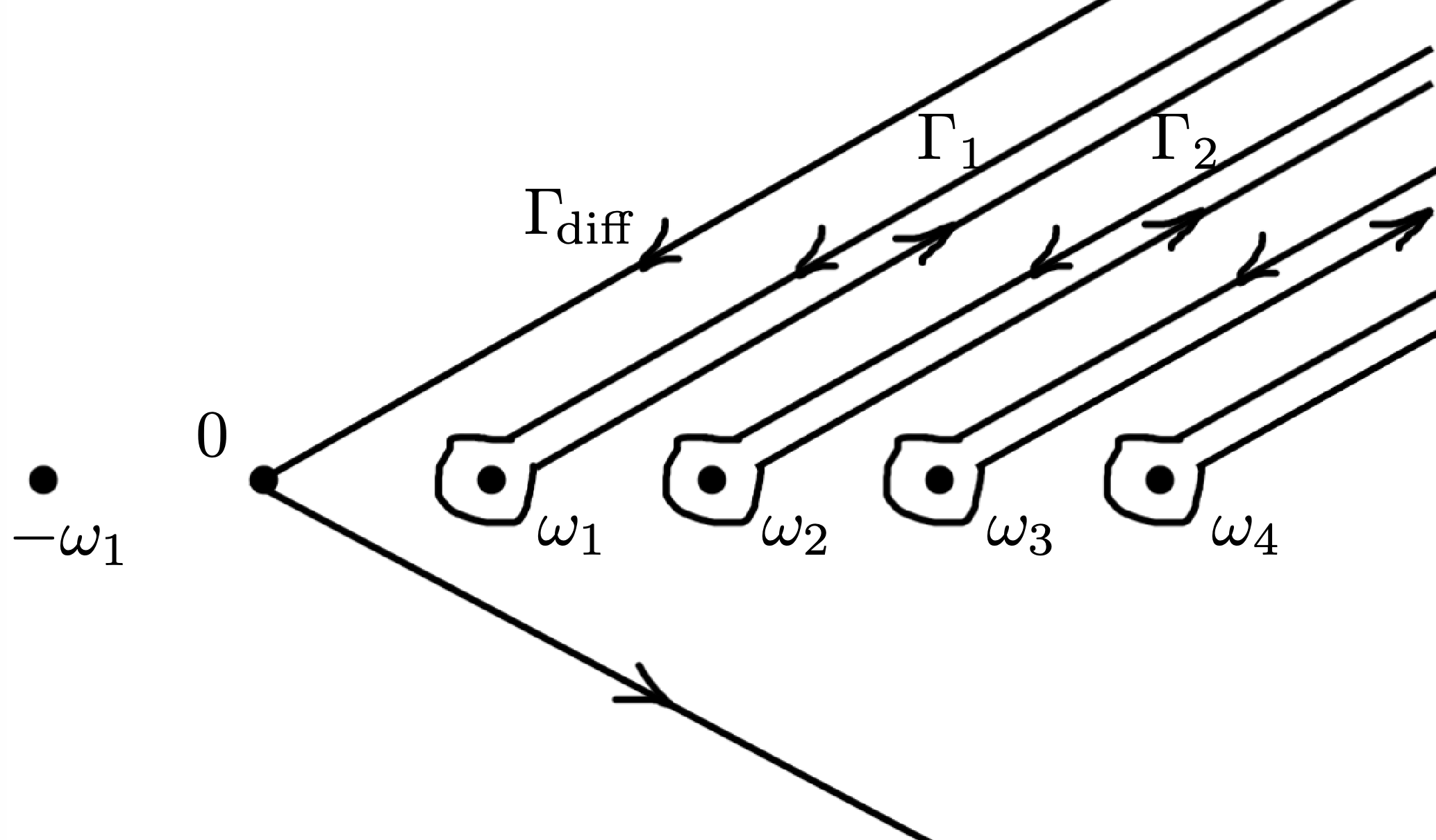}
  \end{center}
  \vspace{-4ex}
  \end{figure}

The idea of the proof of~\eqref{eqStkTruncatedmst} is to write
  $\AS^{I_R}\ti\ph(z) - \AS^{I_L}\ti\ph(z)$ as a
Laplace-like integral on a contour~$\Gadiff$ that can be decomposed
in a sum of Hankel contours $\Ga_1,\Ga_2,\ldots$ as illustrated on
Figure~\ref{figHankelcontours}.

  It is sometimes possible to let~$m_*$ tend to
infinity and get % , for the resurgent summable series at hand
  \begin{equation}   \label{eqPassageRelation}
     \AS^{I_R} \ti\ph(z) = \AS^{I_L} \circ \DDE_d^+ \ti\ph(z) 
    \quad \text{in restriction to $z\in\DD^*$,}
  \end{equation}
  where the \rhs\ involves the action of the summation
    operator~$\AS^{I_L}$ on a simple $\Om$-resurgent
    transseries,\footnote{defined termwise:
      $\AS^J(\sum e^{-m\, \om_1 z} \Td{\psi}_m) \defeq \sum e^{-m\, \om_1 z} \AS^J\Td{\psi}_m$}
but this usually requires some justification. We will see two examples
where~\eqref{eqPassageRelation} holds in
Section~\ref{resurgentfreeenergy}
(in one case because there are only finitely many values of~$m$ for
which~$\De_{m\om_1}^+$ does not annihilate~$\ti\ph$, % in~\eqref{eqStkTruncatedmst},
in the other case because both sides of~\eqref{eqPassageRelation} are
solutions to the same ODE and it is thus easier to prove them equal). 

\begin{Remark}
  One can also make sense of~\eqref{eqPassageRelation} in the algebra
  $\ti\RT_\Om^\simp$ without summability assumption (\ie without any
  exponential bound~\eqref{eqboundwhvpth} or~(\ref{eqboundwhvpth}')),
  at the price of replacing holomorphic functions on~$\DD^*$ by
  \emph{exponential evanescence classes} \cite{CNP}.
\end{Remark}

% \begin{Remark}
%   %
%   We introduce $I_R$ and $I_L$ to emphasize the “local” nature of
%   the Stokes “jumps” along the direction $\th^*$ used to measure the
%   discrepancy of the Laplace transforms on the left and the right of
%   the direction $\th^*$. For readers’ convenience we list here
%   several related intervals mentioned before:
%   $$I_0=(-2\pi, 0),\,\, I_\pi=(-\pi,\pi);$$ and
% $$I^+=(-\pi, 0)=I_\pi\cap I_0,\,\, I^-=(0,\pi)=I_\pi \cap (2\pi+I_0).$$ 
% \end{Remark}

\begin{Example}\label{exa4}
  For the formal series related to the Airy equation, we recall from
  Section~\ref{exa1} that
  \[
  \ti\psi \in \C\oplus\ti\NN(I_0,0), \quad I_0=(-2\pi, 0),
  \qquad
  \ti\ph \in \C\oplus \ti\NN(I_\pi,0), \quad I_\pi=(-\pi, \pi)
  \]
  and \eqref{eqDepsiph}--\eqref{eqotherderivz} yield
%
%  Combining Section~\ref{exa1} and Example~\ref{exa3} with Theorem \ref{thm1}, we can calculate
%
  \begin{equation}\label{eqp}
    \DDE_{\R_{\gqw0}}^+\Td{\psi}=\Td{\psi}-ie^{-2z} \Td{\vp},
    \qquad
    \DDE_{\R_{\lqw 0}}^+\Td{\vp}=\Td{\vp}-ie^{2z} \Td{\psi}.
\end{equation}
Let us use $\th^*=0$ or~$-\pi$. For $z\in\ti\C$ (Riemann surface
of the logarithm), we get
%
% Further by using Theorem \ref{th_ssa} we obtain the following Stokes phenomena: for any $z\in\Td{\C}$,
%
\begin{align}
  \label{eqfirstiden}
  \arg z\in \big(\hspace{-.3em}-\tf{\pi}{2},\tf{\pi}{2}\big) 
& \imp \AS^{I_0}\Td{\psi}(z) =
      \AS^{I_0}\Td{\psi}(e^{2\pi i}z)-ie^{-2z}\AS^{I_\pi}\Td{\vp}(z) \\[1ex]
  \label{eqsecondiden}
  \arg z\in \big(\tf{\pi}{2},\tf{3\pi}{2}\big)
  &\imp  \AS^{I_\pi}\Td{\vp}(e^{-2\pi i}z)=
    \AS^{I_\pi}\Td{\vp}(z)-ie^{2z}\AS^{I_0}\Td{\psi}(z) 
\end{align}
Indeed, \eqref{eqfirstiden} is obtained with $\th^*=0$,
$I_R=(-\de,0)\subset I_0$ and
$I_L=(0,\de) \subset (2\pi + I_0)\cap I_\pi$,
using $\AS^{2\pi+I_0}\Td{\psi}(z) = \AS^{I_0}\Td{\psi}(e^{2\pi i}z)$
(cf.\ footnote~\ref{ftnshiftsheet}).
For \eqref{eqsecondiden}: $\th^*=-\pi$, $I_R\subset -2\pi+I_\pi$,
$I_L \subset I_\pi \cap I_0$. 
\end{Example}

%\section{Resurgent Analysis of the Universal Structures in B-model Topological String Theory}
%
%%%%%%%%%%%%%%%%%%%%%%%%%%%%%%%%%%%%%%%%%%%%%%%
%%%%%%%%%%%%%%%%%%%%%%%%%%%%%%%%%%%%%%%%%%%%%%%
\section{Resurgent analysis of the free energy in the double
  scaling limit}
\label{resurgentfreeenergy}
%%%%%%%%%%%%%%%%%%%%%%%%%%%%%%%%%%%%%%%%%%%%%%%
%%%%%%%%%%%%%%%%%%%%%%%%%%%%%%%%%%%%%%%%%%%%%%%

In this section we apply resurgence theory to the study of the free
energy as obtained by Alim-Yau-Zhou's double scaling
limit~(\cite{[AYZ]}).  
% which is supposed to be the universal structures in the B-model topological string theory.
%
Following background and motivations from \S\ref{paragonetwo}, we consider the
second-order ordinary differential equation~\eqref{eq1}
%
% \begin{equation}\label{eq_ageq}
% \th^2_{\la_s}\FF+(\th_{\la_s}\FF)^2+2\left(1-\f{2}{3\la_s^2}\right)\th_{\la_s}\FF+\f{5}{9}=0,~~\th_{\la_s}\defeq\la_s\f{\pt}{\pt \la_s}
% \end{equation}
%
derived from the holomorphic anomaly equations % (HAE)
in their polynomial form.  As shown in \cite{[AYZ]}, the all-genus
free energy $$\FFs(\la_s)=\sum_{g=2}^\infty a_g \la_s^{2(g-1)}$$
(cf.~\eqref{eqdefFFs})
is the only solution to equation~\eqref{eq1} in $\la_s^2\C[[\la_s^2]]$.
More precisely, we have

\begin{Lemma}   \label{lemformsolns}
The formal solutions to Equation~\eqref{eq1} in $\C[[\la^2_s]]$
are the formal series
\[\si+\FFs(\la_s), \qquad \si\in \C.\]
\end{Lemma}

To apply the resurgence theory to study $\FFs$, we first change the
variable to $z=\f{1}{3\la_s^2}$.
From now on, we will systematically use the variable~$z$ rather than~$\la_s$.
Then~$\FFs$ becomes $\Td{g}(z)=\Td{g}(\f{1}{3\la_s^2})=\FFs(\la_s)$, i.e.\
\beglab{eqreltigFFs}
\Td{g}(z) \defeq \FFs( (3z)^{-1/2})
= \sls_{g=2 }^\infty   3^{-(g-1)} a_g z^{-(g-1)}
=\sls_{n=1}^\infty  b_nz^{-n},
\qquad b_n\defeq3^{-n}a_{n+1}.
\edla
Our change of variable changes the ODE~\eqref{eq1} into
%
% Now $\Td{g}$ satisfies the following differential equation
%
\begin{equation}\label{eq_ageq2}
  g''+(g')^2+2g'+\f{5}{36}z^{-2}=0,
  %
  % z^2(g''+(g')^2+2g')+\f{5}{36}=0,
  %
\end{equation}
the formal solutions of which are thus $\si+\ti g(z)$, $\si\in\C$.

The proof of Theorem~A will result from a succession of
propositions to be found in Sections~\ref{DSL}--\ref{secsymbolSt}.
Theorem~B will be proved in Sections~\ref{secsummaSt}--\ref{secNLSt},
and Theorem~C in Section~\ref{real}.
% \blue{MAYBE MORE SUBSECTIONS... TO BE SEEN}

%%%%%%%%%%%%%%%%%%%%%%%%%%%%%%%%%%%%%%%%%%%%

\subsection{Link with the series~$\ti\psi(z)$ of
\S~\ref{exa1} and first summability result}\label{DSL}

%%%%%%%%%%%%%%%%%%%%%%%%%%%%%%%%%%%%%%%%%%%%

The formal series~$\ti\psi(z)$ and~$\ti\ph(z)$ introduced in
Section~\ref{exa1} can be written
\beglab{eqtipsitiphone}
\ti\psi(z) = 1+ \ti\psi_1(z), \ens
\ti\ph(z) = 1+ \ti\ph_1(z) \ens
\text{with}\ens
\ti\psi_1(z) \defeq \sum_{n\gqw1} c_n z^{-n},
\ens \ti\ph_1(z) \defeq \ti\psi_1(-z).
\edla
It is observed in \cite{[AYZ]} that the change of unknown $g=\log\psi$
changes~\eqref{eq_ageq2} into the linear ODE~\eqref{eqairy1}, of
which~$\ti\psi$ is the unique formal solution with constant term~$1$.
Thus,
%
% the formal power series $\Td{g}$ can be written as
% $$\Td{g}=\log(\Td{\psi})=\log(1+\Td{\psi}_1),$$
% the logarithm of a formal power series $\Td{\psi}=1+\Td{\psi}_1$ in
% Example \ref{exa1}. Here the logarithm is understood as the following
% formally convergent series of formal series
%
\beglab{eqtiglogtipsi}
\Td{g}=\log\ti\psi=\sum_{n= 1}^\infty \f{(-1)^n}{n}(\ti\psi_1)^n.
\edla
We will also consider
\beglab{eqdeftif}
\Td{f}(z) \defeq \log\Td{\vp} = \Td{g}(-z) = \sum_{n\gqw1} (-1)^n b_n z^{-n}.
\edla
%
% with $\Td{\vp}=1+\Td{\vp}_1$ at the same time.
%
% Since the methods to study $\Td{g}$ and $\Td{f}$ are similar, we
% will only focus on $\Td{g}$ and are content to give c% orresponding statements for $\Td{f}$.

\begin{Proposition}\label{propseries}
  The formal power series~$\Td{g} $ is $1$-summable in the directions
  of $I_0=(-2\pi,0)$,
  with Borel transform $\wh g=\BB\Td{g}\in\NN(I_0,\be_0)$
  % $\BB\Td{g}\in\C\de\oplus\NN(I_0,\be_0)$
  for some locally bounded function $\be_0\col I_0\ra\R_{\gqw 0}$,
  and thus has a Borel sum $\AS^{I_0}\Td{g}$ holomorphic in the domain
  $\DD(I_0,\be_0)$ defined as in~\eqref{eqDDIaltiC}.
  Moreover,
  \begin{multline*}
  z\in \DD(I_0,\be_0) \imp |\AS^{I_0}\ti\psi_1(z)|<1
  \ens\text{and}\\[1ex]
  \AS^{I_0}\Td{g}(z) = \log \big( 1 + \AS^{I_0}\ti\psi_1(z) \big)
  \ens\text{(principal branch).}
  \end{multline*}

%=\{z\in \Td{\C}|-\f{\pi}{2}<arg z<\f{5\pi}{2},~\Re
%e(ze^{i\th})>\be_g(\th),\th\in I_g\}$, where $\Td{\C} is the %Riemann
%surface of $\log z$.\\
  
Similarly,~$\Td{f} $ is $1$-summable in the
directions of $I_\pi=(-\pi,\pi)$,
with Borel transform
$\wh f = \BB\Td{f}\in\NN(I_\pi,\be_\pi)$
% $\BB\Td{f}\in\C\de\oplus\NN(I_\pi,\be_\pi)$
%
for some locally bounded function $\be_\pi\col I_\pi\ra\R_{\gqw 0}$,
and has a Borel sum
$\AS^{I_\pi}\Td{f}(z) = \log \big( 1 + \AS^{I_\pi}\ti\ph_1(z) \big)$
holomorphic in $\DD(I_\pi,\be_\pi)$.
%=\{z\in \Td{\C}|-\f{3\pi}{2}<arg z<\f{3\pi}{2},\Re e(ze^{i\th})>\be_f(\th),\th\in I_f\}.
%

 One can choose~$\be_0$ and~$\be_\pi$ so that their $2\pi$-periodic
extensions 
\begla
 \be_0\col \R\setminus 2\pi\Z \to \R_{\gqw0},
\quad
\be_\pi\col \R\setminus (\pi+2\pi\Z) \to \R_{\gqw0} 
\edla
 are even.
\end{Proposition}

\begin{proof}
  In fact, this is a particular case of Theorem~\ref{thmStab}, but for
  the sake of completeness we
   give details for~$\ti g(z)$.
%
%From the relationship between formal series $\Td{g}$ and $\Td{\psi}$, we have
% By taking the Borel transform of $\Td{g}$
% \begin{equation}\label{eq_gs}
% \Td{g}=\log(1+\Td{\psi}_1)=\sls_{n=1}^{\infty}\f{(-1)^{n-1}}{n}\Td{\psi}_1^n\in z^{-1}\C[[z^{-1}]],
% \end{equation}
%For formula (\ref{eq_gs}), we take his Borel transformation as
From~\eqref{eqtiglogtipsi} we deduce
\begin{equation}\label{eq_shf}
  \wh{g}=\sls_{n=1}^{\infty}\f{(-1)^{n-1}}{n}\wh{\psi}_1^{\ast n}
  \ens\text{with}\ens
  \wh{\psi}_1^{\ast n}=\underbrace{\wh{\psi}_1\ast\wh{\psi}_1\ast\cdots \ast \wh{\psi}_1}_{n~factors}\in \ze^{n-1}\C\{\ze\}.
\end{equation}

According to \S~\ref{paragtispiB}, $\wh\psi_1 = \f{d\wh B}{d\ze}$
where $\wh B \in \gO\big(\D_2\cup \{\arg\ze\in I_0\} \big)$ and there is
a locally bounded function $\be\col I_0\to\R_{>0}$ such that
$|\wh B(\ze) | \lqw \be(\arg\ze)$ for $\arg\ze\in I_0$.
For arbitrary $R\in(0,2)$, we can thus find $M>0$ such that
\beglab{ineqwhpsioneDR}
|\wh\psi_1| \lqw M \ens\text{on $\D_R$.}
\edla
We can also, by the Cauchy inequalities, find a locally bounded
function $\be_0\col I_0\to\R_{>0}$ such that
\beglab{ineqwhpsioneIzero}
|\wh\psi_1(\ze)| \lqw \be_0(\arg\ze) \ens\text{for $\arg\ze\in I_0$}
\edla
 and its $2\pi$-periodic extension is even
(replacing~$\be_0(\th)$ by $\be_0(\th)+\be_0(2\pi-\th)$ if necessary).
Inequality~\eqref{ineqwhpsioneIzero} implies that, if $\th\in I_0$,
\beglab{ineqASIztipsiun}
|\LL^{I_0}\wh\psi_1(z)| \lqw \f{\be_0(\th)}{\RE(z e^{-i\th})}
\ens\text{in the half-plane $\{ \RE(z e^{-i\th}) > 0 \}$.}
\edla

% $\Td{\psi}_1(z)$ is $1$-summable in the directions of $I_0=(-2\pi,0)$. 
% Therefore, there is a locally bounded function $\be_0(\th) (\th\in I_0$),
% which is the function $\be(\th)$ mentioned in Example \ref{exa1}, 
% such that
% $\wh{\psi}_1\in \NN (I_0,0,\be_0(\th))$, \ie there exists $R>0$
% smaller than the radius of convergence of $\wh{\psi}_1$ such that
% %
% \begin{align*}
% |\wh{\psi}_1(\ze)|&\lqw M, & \ze&\in D(0,R),\\
% |\wh{\psi}_1(\ze)|&\lqw \be_0(\th),&\ze&\in \Si=\{re^{i\th}|r>0,\th\in I_0\}
% \end{align*}
% with $M$ some positive real number.

Since the domain $\D_R\cup \{\arg\ze\in I_0\} \big)$ is star-shaped \wrt\ the
origin, one can easily check that
% Because $\wh{\psi}_1$ is a holomorphic function in $D(0,R)\cup \Si$,
each $\wh{\psi}_1^{\ast n}$ is also holomorphic in that domain, with
% $D(0,R)\cup \Si$ {by the convolution property (\cite{[J.E.]}, \cite{[M.S.]})}.
%
%So the \rhs\ of~\eqref{eq_shf} is a series of holomorphic functions.
%
%Next, we need to check that it is {uniformly convergent} in each compact subset of
%$D(0,R)\cup \Si$ and gives an appropriate bound.
%By calculations, we have the following inequalities:
%
\begin{align}
  \left|\wh{\psi}^{\ast n}_1(\ze)\right| &\lqw
M^n\f{|\ze|^{n-1}}{(n-1)!} & & \hspace{-6em}\text{for} \,\ze\in \D_R,\label{eq_cc0}\\
  \left|\wh{\psi}^{\ast n}_1(\ze)\right| &\lqw
\be_0(\th)^n\f{|\ze|^{n-1}}{(n-1)!} & & \hspace{-6em}\text{for} \, \arg\ze\in I_0.\label{eq_cc}
\end{align}
This shows that the series of holomorphic functions
$\sum \f{(-1)^{n-1}}{n}\wh{\psi}_1^{\ast n}$ is
uniformly convergent in every compact subset of $\D_R\cup \{\arg\ze\in
I_0\}$.
Further, inequality~\eqref{eq_cc0} guarantees that~$\wh g$ is the
Taylor expansions at~$0$ of the resulting holomorphic function,
thus $\wh{g}\in\C\{\ze\}$ and $\wh{g}$ extends
analytically to $\D_R\cup \{\arg\ze\in
I_0\}$, inequality~\eqref{eq_cc} yielding
\begin{equation}\label{neq1}
|\wh{g}(\ze)|\lqw \sls_{n=1}^\infty \be_0(\th)^n\f{|\ze|^{n-1}}{n!}\lqw \sls_{n=1}^\infty \be_0(\th)^n\f{|\ze|^{n-1}}{(n-1)!}=\be_0(\th)e^{\be_0(\th)|\ze|}
\end{equation}
for $\th=\arg\ze\in I_0$.
%
% Hence $\wh{g}\in \NN (I_0,\be_0(\th),\be_0(\th))\subset \NN(I_0)$ with $I_0=(-2\pi,0)$.
%
Moreover, for every $z\in\DD(I_0,\be_0)$, \eqref{ineqASIztipsiun}
shows that $|\LL^{I_0} \wh\psi_1(z)|<1$ and we see that
$\AS^{I_0}\Td{g}= \LL^{I_0}\wh g$ coincides with
$\log(1+\LL^{I_0}\wh{\psi}_1 )=\log(1+\AS^{I_0}\Td{\psi}_1 )$ in
$\DD(I_0,\be_0)$,
where the logarithm series gives rise to the principal branch.
%
% is holomorphic in $\DD(I_0,\be_0(\th))$ and
% \begin{equation}\label{eq_bsas}
%   \AS^{I_0}\Td{g}(z)~\sim_1~\Td{g}(z),~for~z\in \DD(I_0,\be_0(\th)).
% \end{equation}

In the case of $\ti f(z) = \log\ti\ph(z)$, we have
\beglab{eqrelwhphwhpsi}
\wh\ph_1(\ze) = - \wh\psi_1(-\ze), \quad \wh f(\ze) = - \wh g(-\ze)
\edla
due to~\eqref{eqtipsitiphone} and~\eqref{eqdeftif}, whence
\beglab{ineqwhphIpi}
 |\wh\ph_1(\ze)| \lqw \be_\pi(\arg\ze) \ens\text{for $\arg\ze\in
    I_\pi$} 
\edla
with $\be_\pi(\th) \defeq \be_0(\th-\pi)$ and the conclusion follows.
\end{proof}

%For the formal power series $\Td{f}$ we have the same proof process because $\Td{f}(z)=\Td{g}(-z)$ can ensure that we have the same inequality as (\ref{eq_cc0}) and (\ref{eq_cc}) after completing Borel transformation for this series.\qqed

%%%%%%%%%%%%%%%%%%%%%%%%%%%%%%%%%%%%%%%%%%%%%%%%%% 

\subsection{Resurgent structure, formal integral and Bridge Equation}
\label{bridgetransseries}

%%%%%%%%%%%%%%%%%%%%%%%%%%%%%%%%%%%%%%%%%%%%%%%%%% 

\begin{Proposition}\label{th_omrs}
  The formal power series $\Td{g}$ and $\Td{f}$ are simple
$2\Z$-resurgent series. Their alien derivatives are

\begin{equation}\label{eq_adv}
\De_{2m}\Td{g}= \begin{cases*}
  -i\,e^{\Td{f}-\Td{g}} & for $m=1$ \\
  \hspace{1em} 0  & for $m\in\Z^*\setminus\{1\}$
\end{cases*}
\qquad
\De_{-2m}\Td{f}= \begin{cases*}
  -i\,e^{\Td{g}-\Td{f}} & for $m=1$ \\
  \hspace{1em} 0  & for $m\in\Z^*\setminus\{1\}$
\end{cases*}
\end{equation}
\end{Proposition}

\begin{proof}
According to \S~\ref{exa3}, $\ti\psi_1$ is a simple $2\Z$-resurgent series,
thus Theorem~\ref{thmad} with $H(t)=\log(1+t)$ implies that~$\ti g$ is a simple $2\Z$-resurgent simple
  series and
  \[
    \De_{\om}\ti g = \De_\om\log(1+\Td{\psi}_1)
    = \f{\De_\om\ti\psi_1}{1+\ti\psi_1}
    \quad \text{for any $\om\in 2\Z^*$.}
    \]
  Using~\eqref{eqDepsiph}--\eqref{eqotherderivz}, we get
% to derive the calculation formula for the derivative, we can conclude that:
%
  \[
    \De_2\ti g = \f{-i\ti\ph}{\ti\psi}=-ie^{\Td{f}-\Td{g}},
    \qquad \om\in2\Z^*\setminus\{2\} \imp \De_\om\ti g = 0.
  \]
  
  The case of~$\ti f$ is similar.
\end{proof}

Note that Proposition~\ref{th_omrs} % \ref{propseries}
amounts to Point~(i) and a little part of Point~(iii) of Theorem~A.
We now prove a result that contains Point~(ii) of Theorem~A:

%%%%%%%%%%%%%%%%%%%%%%%%%%%%%%

\begin{Proposition}\label{propkey0}
On~$\ti g$, the actions of the algebra automorphisms $\exp(\si
e^{-2z}\De_2)$ and
  $(\DDE_{\R_{\gqw 0}}^+)^\si=\exp(\si \DDE_{\R_{\gqw 0}})$
  of $\ti\RT_{2\Z}[[\si,e^{-2z}]]$ coincide and define a sequence of
  simple $2\Z$-resurgent series
  $\ti G_0=\ti g, \ti G_1, \ti G_2,\ldots$ by the formula
  \beglab{eqdeftiGn}
\exp(\si e^{-2z}\De_2) \ti g =  (\DDE_{\R_{\gqw 0}}^+)^\si\,\ti g 
  = \sum_{n\gqw0} (-i\si)^n e^{-2nz} \ti G_n(z).
  \edla
For any $\si_1,\si_2\in\C$,
  \beglab{eqGzsisi}
  \ti G(z,\si_1,\si_2) \defeq
  \exp(i\si_2 e^{-2z} \De_2)(\si_1+\ti g) =
  \si_1 + \sum_{n\gqw0} \si_2^n e^{-2nz} \ti G_n(z)
  \edla
  is a $2\Z$-resurgent transseries solution to
  Equation~\eqref{eq_ageq2}.
  Moreover,
  \beglab{eqDDEpsiG}
  (\DDE_{\R_{\gqw 0}}^+)^\si \ti G(z,\si_1,\si_2)=\ti G(z,\si_1,\si_2-i\si)
  \quad\text{for any $\si\in\C$.}
\edla
\end{Proposition}

\begin{proof}
We choose $\om_1=+2$ as generator of $2\Z$ so as to put ourselves in the framework of Section~\ref{secResTransseries}.

  Proposition~\ref{th_omrs} shows that the only alien derivation with a
non-trivial action on~$\ti g$ is $\De_2$, hence
$\si \DDE_{\R_{\gqw 0}}\ti g = \si e^{-2z} \De_2\ti g$
and, since the result is proportional to $e^{-2z} e^{\ti f -\ti g}$,
alien calculus shows that $(\si \DDE_{\R_{\gqw 0}})^r\ti g = (\si e^{-2z}
\De_2)^r\ti g$ by induction on $r\gqw1$.
We thus obtain~\eqref{eqdeftiGn} with a certain sequence
$(\ti G_n)_{n\gqw0}$ of $\ti\RT_{2\Z}^\simp$ starting with $\ti
G_0=\ti g$.

The resurgent transseries~\eqref{eqGzsisi} is nothing but $\exp(i\si_2
e^{-2z} \De_2)(\si_1+\ti g)$.
It is a solution to~\eqref{eq_ageq2} because $\si_1+\ti g$ is a
solution (Lemma~\ref{lemformsolns}) and $\exp(\si e^{-2z} \De_2)$ is an algebra
automorphism that commutes with $\f{\pt}{\pt z}$ and acts trivially on
every convergent series.

Finally,
$(\DDE_{\R_{\gqw 0}}^+)^\si \ti G(z,\si_1,\si_2)=
(\DDE_{\R_{\gqw 0}}^+)^{\si+i\si_2} (\si_1+\ti g) =
(\DDE_{\R_{\gqw 0}}^+)^{i(\si_2-i\si)} (\si_1+\ti g) =
\ti G(z,\si_1,\si_2-i\si)$.
\end{proof}

%%%%%%%%%%%%%%%%%%%%

The two-parameter resurgent transseries~$\ti G$ is nothing but the
``formal integral''~$\GG$ of Theorem~A(ii) written in the variable
$z=\f{1}{3\la_s^2}$:
\beglab{eqreltiGnGGn}
\GG(\la_s,\si_1,\si_2) = \ti G\big(\tfrac{1}{3\la_s^2},
\si_1,\si_2\big), \qquad
\ti G_n(z) = \GG_n\big( (3z)^{-1/2} \big)
\quad \text{for}\ens n\in\Z_{\gqw0}.
\edla

%%%%%%%%%%%%%%%%%%%%

\begin{Proposition}
For every $\si\in\C$,
  \begin{equation}\label{stk2}
    (\DDE_{\R_{\gqw 0}}^+)^\si \, \Td{g}=\Td{g}-\sum_{m=1}^{\infty}\f{(i\si)^m}{m}e^{-2mz}e^{m(\Td{f}-\Td{g})},
  \end{equation}
  whence
  \beglab{eqidentifytiGm}
  \ti G_m = \f{(-1)^{m-1}}{m} e^{m(\ti f-\ti g)}
  \quad \text{for all $m\gqw1$.}
  \edla
Moreover, % For the operators $\De_\om^+$, we have
\begin{equation}\label{alienfg}
  \De_{2m}^+\Td{g}=-\f{i^m}{m}e^{m(\Td{f}-\Td{g})}, \quad
  \De_{-2m}^+\Td{f}=-\f{i^m}{m}e^{m(\Td{g}-\Td{f})}
  \quad \text{for all $m\in\Z_{\gqw1}$,}
\end{equation}
while $\De_{-2m}^+\Td{g}=0$
and $\De_{-2m}^+\Td{f}=0$.
\end{Proposition}

\begin{proof}
  %
%  It can be calculated directly by formula~\eqref{stk1} and $(\DDE_{\R_{\gqw 0}}^+)^\si=\exp(\si\DDE_{\R_{\gqw 0}})$.
%
Using \eqref{eqDepsiph}--\eqref{eqotherderivz} we compute
  \[
    (\DDE_{\R_{\gqw0}}^+)^\si\Td{\psi}= \sum_{r\gqw0} \f{\si^r}{r!}
    (\DDE_{\R_{\gqw0}})^r \Td{\psi}
    = \Td{\psi}-i\si e^{-2z} \Td{\vp}
   =  e^{\ti g} ( 1 - i\si e^{-2z} e^{\Td{f}-\Td{g}} ) 
  \]
  (note that the terms with $r\gqw2$ do not contribute).
Since $(\DDE_{\R_{\gqw 0}}^+)^\si$ is an algebra automorphism of
  $\ti\RT_{2\Z}[[e^{-2z}]]$, we deduce that
  \begin{equation} % \label{stk1}
    (\DDE_{\R_{\gqw 0}}^+)^\si\,\Td{g} = (\DDE_{\R_{\gqw 0}}^+)^\si(\log\ti\psi) =
    \log\big( (\DDE_{\R_{\gqw 0}}^+)^\si\ti\psi\big) =
    %
    %\log(1+\DDE_{\R_{\gqw 0}}^+\Td{\psi}_1)
   %
   \Td{g}-\sum_{m=1}^{\infty}\f{(i\si)^m}{m}e^{-2mz}e^{m(\Td{f}-\Td{g})}.
\end{equation}
 %due to the fact that operator $\DDE_{\R_{\gqw 0}}^+$ is an automorphism;
 %
When $\si=1$, the homogeneous components of the latter identity % \eqref{stk1}
yield~\eqref{alienfg}.
%
% Moreover, one has $\DDE_{\R_{\gqw 0}}^+\Td{g}=\Td{g}+\sls_{n\gqw 1}^\infty e^{-2nz}\De_{2n}^+\Td{g}$.
% % by the definition of operator $\DDE_{\R_{\gqw 0}}^+$ in Theorem \ref{thm1}.
% Putting together, we obtain the desired result
% \[\De_{2n}^+\Td{g}=-\f{i^n}{n}e^{n(\Td{f}-\Td{g})}~for~n\gqw 1.\]
%
\end{proof}

%%%%%%%%%%%%%%%
%%%%%%%%%%%%%%%
\iffalse
%%%%%%%%%%%%%%%
%%%%%%%%%%%%%%%

\begin{Corollary}\label{transalien}
The formal series $e^{m(\Td{g}-\Td{f})}$ and $e^{m(\Td{f}-\Td{g})}$ ($m\gqw 1$) are simple $2\Z$-resurgent.
When $n\gqw 1$,
\begin{itemize}
    \item the n-th alien derivative of the formal series $\Td{g}$ and $\Td{f}$ is
\begin{equation}
    \begin{aligned}
        \De_2^n\Td{g}=-i^n\Ga(n)e^{n(\Td{f}-\Td{g})},~~\De_{-2}^n\Td{f}=-i^n\Ga(n)e^{n(\Td{g}-\Td{f})};
    \end{aligned}
\end{equation}
    \item  the n-th alien derivative of the formal series $e^{m(\Td{f}-\Td{g})}$ is
\begin{equation}\label{alienn1}
\De_2^n e^{m(\Td{f}-\Td{g})}=i^n\f{\Ga(m+n)}{\Ga(m)}e^{(m+n)(\Td{f}-\Td{g})},
\end{equation}
\begin{equation}\label{alienn2}
\De_{-2}^n e^{m(\Td{f}-\Td{g})}=\left\{\begin{array}{cc}
    (-i)^n\f{\Ga(m+1)}{\Ga(m-n+1)}e^{(m-n)(\Td{f}-\Td{g})} &m\gqw n\\
    0&m<n,
\end{array}\right.
\end{equation}
and the n-th alien derivative of the formal series $e^{m(\Td{g}-\Td{f})}$ is
\begin{equation}\label{alienn3}
\De_{-2}^n e^{m(\Td{g}-\Td{f})}=i^n\f{\Ga(m+n)}{\Ga(m)}e^{(m+n)(\Td{g}-\Td{f})},
\end{equation}
\begin{equation}\label{alienn4}
\De_2^n e^{m(\Td{g}-\Td{f})}=\left\{\begin{array}{cc}
    (-i)^n\f{\Ga(m+1)}{\Ga(m-n+1)}e^{(m-n)(\Td{g}-\Td{f})} &m\gqw n\\
    0&m<n.
\end{array}\right.
\end{equation}
\end{itemize}
\end{Corollary}
\begin{proof}
It is the combination of Proposition \ref{th_omrs}  and Theorem \ref{thmad}.
\end{proof}

%%%%%%%%%%%%%%%
%%%%%%%%%%%%%%%
\fi
%%%%%%%%%%%%%%%
%%%%%%%%%%%%%%%

We are now ready to prove the ``Bridge Equation'', \ie Theorem~A(iii).

\begin{Proposition}\label{propbridge}
  %
% For the formal integral~\eqref{eq_solu1}, there holds the following bridge equations:
%
For every $\si_1\in\C$, the following identities hold in $\ti\RT_{2\Z}^\simp[[\si_2,e^{-2z}]]$:
\begin{align}
  \Delta_2 \ti G(z,\si_1,\si_2) &=-i e^{2z}\f{\pt}{\pt \si_2} \ti G(z,\si_1,\si_2)\label{eqbridge1}  \\
  \Delta_{-2}\ti G(z,\si_1,\si_2) &=-ie^{-2z}\Big(\si_2\f{\pt}{\pt \si_1}\ti G(z,\si_1,\si_2)-\si_2^2\f{\pt}{\pt\si_2}\ti G(z,\si_1,\si_2)\Big)\label{eqbridge2}
\end{align}
and $\De_\om \ti G = 0$ for $\om\in 2\Z^*\setminus\{-2,2\}$.
\end{Proposition}

\begin{proof}
We first see that
\[
\DDE_{\R_{\gqw0}}\ti G = - i \f{\pt}{\pt\si_2}\ti G,
\]
because of~\eqref{eqDDEpsiG}, since $\DDE_{\R_{\gqw0}}$ is the
  infinitesimal generator \wrt~$\si$ of the one-parameter group of automorphisms
  $\big( (\DDE_{\R_{\gqw 0}}^+)^\si \big)_{\si\in\C}$
(just evaluate the derivative in~$\si$ of~\eqref{eqDDEpsiG} at $\si=0$).
This yields the desired result for $\De_{2m}\ti G$ for all $m\gqw1$.
  
  For the case $m\lqw-1$, we write
  \[
    \ti G = \si_1 + \sum_{n\gqw0} \si_2^n e^{-2nz} \ti G_n(z)
    = \si_1 + \Td{g} +\sum_{n\gqw1} \f{(-1)^{n-1}}{n} \si_2^n e^{-2nz}e^{n(\Td{f}-\Td{g})}
  \]
  (since the partial derivative $\f{\pt}{\pt\si_1} \ti G=1$ will be
  involved, we no longer treat~$\si_1$ as a constant:
  we rather work in $\ti\RT_{2\Z}^\simp[[\si_2,e^{-2z}]][\si_1]$).
In view of Proposition~\ref{th_omrs}, $\De_{2m}$ annihilates~$\ti G$ for all $m\lqw-2$,
while $e^{2z}\De_{-2}\ti g=0$ and
  $e^{2z}\De_{-2}\ti f = -i e^{2z} e^{\ti g-\ti f}$ imply
  $e^{2z}\De_{-2}(e^{n(\Td{f}-\Td{g})}) = -i n e^{2z}
  e^{(n-1)(\ti g-\ti f)}$, hence
  \begin{multline*}
e^{2z}\De_{-2}\ti G = -i \sum_{n\gqw1} (-1)^{n-1} \si_2^n e^{-2(n-1)z}
  e^{(n-1)(\ti g-\ti f)}\\%[1ex]
  = -i \si_2 \Big( 1 + \sum_{n\gqw0} (-1)^n \si_2^n e^{-2nz} e^{n(\ti g-\ti f)}\Big)
  = -i \si_2 \big( \f{\pt}{\pt\si_1} \ti G - \si_2\f{\pt}{\pt\si_2} \ti G \big).
    \end{multline*}
\end{proof}

%%%%%%%%%%%%%%%%%%%%%%%%%%%%%%%%%%%%%%%%%%%

%%%%%%%%%%%%%%%%%%%%%%%%%%%%%%%%%%%%%%%%%%%%%%%%%%

\subsection{Another view on the formal integral}\label{secotherview}

%%%%%%%%%%%%%%%%%%%%%%%%%%%%%%%%%%%%%%%%%%%%%%%

As already mentioned, the change of unknown $g=\log \psi$ transforms
equation~\eqref{eq_ageq2} into the linear ODE~\eqref{eqairy1},
with~$\ti\psi$ as unique formal solution with constant term~$1$.
Since $e^{-2z}\Delta_2$ is a derivation that commutes with $\f{\pt}{\pt z}$ and acts trivially on
every convergent series, by applying this operator
to~$\Td{\psi}$, we get another solution, $e^{-2z}\Delta_2\ti\psi=-ie^{-2z}\Td{\vp}$.
%
% because the Wronskian of $(\Td{\psi},-ie^{-2z}\Td{\vp})$ is
% \red{non-zero} and equal to $2ie^{-2z}$. In this way,
%
In fact, we can view
\begin{equation}\label{eq_solu}
  \Psi(z,c_1,c_2) =c_1\Td{\psi} +c_2 e^{-2z}\Td{\vp}
\end{equation}
as the general transseries solution of~\eqref{eqairy1}, depending on
two free parameters, $c_1,c_2\in \C$.
We may thus consider $\log(c_1\Td{\psi}+c_2e^{-2z}\Td{\vp})$
(at least if $c_1$ and $c_2$ are not both zero)
as a general formal solution of~\eqref{eq_ageq2}.

When $c_1\neq 0$, this solution reads
\begin{align}
S_1(z,c_1,c_2)&= \log c_1 + \log\Td{\psi}+\log\left(1+\f{c_2}{c_1}e^{-2z}\f{\Td{\vp} }{\Td{\psi}}\right)\\
            &=\log c_1 + \Td{g}+\log\left(1+\f{c_2}{c_1} e^{-2z}e^{\Td{f}-\Td{g}}\right)\\
            &=\log c_1+\Td{g}+\sls_{n\gqw 1}\f{(-1)^{n-1}}{n}\left(\f{c_2}{c_1}\right)^ne^{-2nz}e^{n(\Td{f}-\Td{g})}.
\end{align}
Similarly, when $c_2\neq 0$, the formal solution is
\begin{equation}
S_2(z,c_1,c_2) = -2z+ \log c_2 + \Td{f}+\sls_{n\gqw 1}\f{(-1)^{n-1}}{n}\left(\f{c_1}{c_2}\right)^ne^{2nz}e^{n(\Td{g}-\Td{f})}.
\end{equation}
%
% Note that the logarithm of the formal series appearing in
% $S_1(z,c_1,c_2)$ (resp. $S_2(z,c_1,c_2) $) is defined in the sense of
% the formally convergent series in
% $\C[[z^{-1},e^{-2z}]]$(resp. $\C[[z^{-1},e^{2z}]]$).

The solution $S_1(z,c_1,c_2)$ gives rise to % of the equation can be simply represented as
\begin{align}
   \ti G(z,\si_1,\si_2)&=\si_1+\Td{g}+\sls_{n\gqw 1}\f{(-1)^{n-1}}{n}\si_2^ne^{-2nz}e^{n(\Td{f}-\Td{g})}\label{eq_solu1},\\
                   &=\si_1+\sls_{n=0}^\infty \si_2^n e^{-2nz} \ti G_n
                     \quad (\si_1=\log{c_1},\si_2=\tfrac{c_2}{c_1}),
\end{align}
%
%where $G_0=\Td{g}$ and $G_n=\f{(-1)^{n-1}}{n}e^{n(\Td{f}-\Td{g})},~n\gqw 1.$
%
while $S_2(z,c_1,c_2)$ gives rise to % is expressed as:
\begin{equation}
   \ti F(z,\delta_1,\delta_2)=-2z+\de_1+\Td{f}+\sls_{n\gqw 1}\f{(-1)^{n-1}}{n}\delta_2^ne^{2nz}e^{n(\Td{g}-\Td{f})}
 \end{equation}
with $\delta_1=\log{c_2}$ and $\delta_2=\f{c_1}{c_2}$. %  \ln -> \log  OK!

% Shanzhong:
Proposition~\ref{propkey0} with~\eqref{eqidentifytiGm} on the one hand
and Formula~\eqref{eq_solu1} on the other hand provide different
perspectives on the formal integral~$\ti G$ of % for the same formula~\eqref{tssss} in
Theorem~A(ii), which can be viewed as a two-parameter transseries solution
that belongs to $\ti\RT_{2\Z}[[e^{-2z}]]$.
%
% Proposition~\ref{propkey0} is the transseries completion, while the
% expression in~\eqref{eq_solu1} is a "formal integral".
%
The previous discussion shows that the transseries~$\ti G$ coexists with a
transseries solution of a different nature,~$\ti F$,
depending on two parameters too,
but belonging to $\ti\RT_{2\Z}[[e^{2z}]]$ up to the term~$-2z$.
In the forthcoming sequel \cite{LSS}, we will investigate
the Borel sums of $\ti G(z,\si_1,\si_2)$ and \( \ti F(z,\de_1,\de_2) \) 
and their analytic continuation, %  in various regions of directions.
% the directions $ I_0 $ and $ I_\pi $, which involves considering its
% corresponding analytic function.
% Roughly %speaking, this analytic function will be analytic in the
% right half of the complex plane (we will elaborate on this in
% %detail later). Additionally, the analytic function corresponding to
% \( \ti F(z,\de_1,\de_2) \) will also be analytic in the %left half of
% the complex plane. In fact, 
%
so as to connect one with the other.
%
% More interestingly these two analytic functions can be connected to
% each other through analytic continuations, thereby enlarging the
% domain of the analyticity.

%%%%%%%%%%%%%%%%%%%%%%%%%%%%%%%%%%%%%%%%%%%%%%%%%%

\subsection{Action of the symbolic Stokes automorphism}\label{secsymbolSt}

%%%%%%%%%%%%%%%%%%%%%%%%%%%%%%%%%%%%%%%%%%%%%%%%

We will now prove Point~(iv) of Theorem~A, \ie compute the action of
$\DDE_{\R_{\gqw 0}}^+ = \exp\big(\DDE_{\R_{\gqw 0}}\big)$
and $\DDE_{\R_{\lqw 0}}^+ = \exp\big(\DDE_{\R_{\lqw 0}}\big)$
on the formal integral $\ti G(z,\si_1,\si_2)$.

\parag
The case of $\DDE_{\R_{\gqw 0}}^+$ is easier because, as explained in
footnote~\ref{ftnrestr} and Section~\ref{secResTransseries}, it can be
defined as the exponential of 
$\DDE_{\R_{\gqw 0}} = \sum_{m\gqw1} e^{-2mz} \De_{2m}$
in the algebra % $\ti\RT_{2\Z}^\simp[[e^{-2z}]]$,
%
% or even in the algebra
$\ti\RT_{2\Z}^\simp[[\si_2,e^{-2z}]]$.
Specifically, given an arbitrary
\beglab{eqwritePhi}
\ti\Phi = \sum_{k\gqw0} \, \sum_{n\gqw0} \si_2^k \, e^{-2n z} \, \ti\Phi_{k,n}(z)
\in \ti\RT_{2\Z}^\simp[[\si_2,e^{-2z}]],
\edla
the action of $\DDE_{\R_{\gqw 0}}$ yields a well-defined transseries
\begin{multline}
\DDE_{\R_{\gqw 0}}\ti\Phi = \bigg( \sum_{p\gqw1} e^{-2pz} \De_{2p} \bigg)
\bigg( \sum_{k\gqw0} \, \sum_{q\gqw0} \si_2^k e^{-2q z} \ti\Phi_{k,q}
\bigg) \\
\label{eqDDEPhi}
= \sum_{k\gqw0} \, \sum_{n\gqw1} \si_2^k \, e^{-2n z} \bigg(
\sum_{p\gqw1,q\gqw0\atop q+p=n} \De_{2p} \ti\Phi_{k,q}\bigg)
\end{multline}
(notice that for each pair $(k,n)$ the sum over $(p,q)$ is a finite sum)
with an increase of the $n$-grading: if the sum in~\eqref{eqwritePhi}
involves only $n\gqw n_0$, then~\eqref{eqDDEPhi} involves only $n\gqw
n_0+1$.
Therefore $(\DDE_{\R_{\gqw 0}})^r$ increases the $n$-grading by at
least~$r$ units for every $r\in\Z_{\gqw0}$ and the exponential series
$\sum \f1{r!}(\DDE_{\R_{\gqw 0}})^r$
is a formally convergent series of operators and makes sense as an operator of $\ti\RT_{2\Z}^\simp[[\si_2,e^{-2z}]]$.

\begin{Proposition}\label{propstkp}
  We have
  \beglab{stk3}
    \DDE_{\R_{\gqw 0}}^+\ti G(z,\si_1,\si_2)=\ti G(z,\si_1,\si_2-i).
    \edla
  \end{Proposition}
  
  \begin{proof}
    This is just the particular case $\si=1$ in~\eqref{eqDDEpsiG}.
    \end{proof}

\begin{Remark}
Equation~\eqref{stk3} amounts to giving $\De_{2n}^+\ti G_k$ for all
  $n\gqw1$ and $k\gqw0$ as follows: 
  \begin{equation}
    \De_{2n}^+ \ti G_k = (-i)^n\binom{k+n}{n}\ti G_{k+n}.
  \end{equation}
\end{Remark}

In fact, since $\DDE_{\R_{\gqw 0}}$ and $\f{\pt}{\pt \si_2}$ are two operators
of $\ti\RT_{2\Z}^\simp[[\si_2,e^{-2z}]]$
that commute, one can iterate the Bridge Equation~\eqref{eqbridge1}:
\[
  (\DDE_{\R_{\gqw 0}})^r \ti G =\Big(\!-i \f{\pt}{\pt \si_2}\Big)^r \ti G
  \quad \text{for all $r\in\Z_{\gqw0}$},
\]
thus the derivation $\DDE_{\R_{\gqw 0}}$ can be seen as a vector field
whose action on~$\ti G$ coincides with that of $-i \f{\pt}{\pt \si_2}$,
and the action of the flows of these vector fields must coincide too,
as expressed by~\eqref{eqDDEpsiG}.

\parag
Things are different with
$\DDE_{\R_{\lqw 0}} = \sum_{m\gqw1} e^{2mz} \De_{-2m}$,
which is well defined in $\ti\RT_{2\Z}^\simp[[\si_2,e^{2z}]]$ but not
in $\ti\RT_{2\Z}^\simp[[\si_2,e^{-2z}]]$.
Indeed, the analogue of~\eqref{eqDDEPhi} would be
\begla
\DDE_{\R_{\lqw 0}}\ti\Phi = 
\sum_{k\gqw0} \, \sum_{n\in\Z} \si_2^k \, e^{-2n z}
\, \raisebox{1.3ex}{``} \! \bigg(
\sum_{p\gqw1,q\gqw0\atop q-p=n} \De_{-2p} \ti\Phi_{k,q}
\bigg) \!\raisebox{1.3ex}{''}
\edla
but that formula usually does not make sense, because the inner
summation over $(p,q)$ may involve infinitely many terms.
Moreover, even if that first obstacle were overcome, $n=q-p$ might sometimes
be negative and the result would not necessarily stay in our algebra $\ti\RT_{2\Z}^\simp[[\si_2,e^{-2z}]]$.

We thus need the remedy alluded to in Footnote~\ref{ftnsubalgforDDElqw}.
The details are as follows:
for any~$\ti\Phi$ as in~\eqref{eqwritePhi}
and $(k,n)\in\Z_{\gqw0} \times\Z$,
we set
\begla
M_{k,n}(\ti\Phi) \defeq \{\, (p,q)\in\Z_{\gqw1}\times\Z_{\gqw0} \mid
q-p=n \; \text{and} \; \De_{-2p}\ti\Phi_{k,q} \neq0 \,\}.
\edla
\begin{Lemma}
  The set
  \begin{multline}
    \cA_0 \defeq \{\, \ti\Phi\in \ti\RT_{2\Z}^\simp[[\si_2,e^{-2z}]] \mid
    M_{k,n}(\ti\Phi) \; \text{is empty for all $(k,n)\in\Z_{\gqw0}\times\Z_{<0}$} \\[1ex]
    \text{and finite for all $(k,n)\in\Z_{\gqw0}\times\Z_{\gqw0}$} \,\}
  \end{multline} 
  is a subalgebra of $\ti\RT_{2\Z}^\simp[[\si_2,e^{-2z}]]$, on which
  the formula
  \begla
    \DDE_{\R_{\lqw 0}}\ti\Phi = 
\sum_{k\gqw0} \, \sum_{n\gqw0} \si_2^k \, e^{-2n z}
\bigg( \sum_{(p,q)\in M_{k,n}(\ti\Phi)} \De_{-2p} \ti\Phi_{k,q} \bigg)
\edla
defines a $\C$-linear derivation $\DDE_{\R_{\lqw 0}} \col \cA_0 \to
\ti\RT_{2\Z}^\simp[[\si_2,e^{-2z}]]$.
\end{Lemma}

\begin{proof}
  The set~$\cA_0$ is clearly a linear subspace of
  $\ti\RT_{2\Z}^\simp[[\si_2,e^{-2z}]]$ containing the series~$1$.
  Let $\ti\Phi,\ti\Psi\in\cA_0$. Their product $\ti\Phi\ti\Psi$ belongs to~$\cA_0$ too
  because, for each $(k,n)\in\Z_{\gqw0} \times\Z$, the set
  $M_{k,n}(\ti\Phi\ti\Psi)$ is contained in
  \beglab{eqfiniteset}
  \{\, (0,q_2) + (p,q_1) \mid
  0\lqw q_2\lqw n, \;
  (p,q_1) \in \bigcup_{0\lqw k_1 \lqw k} \, \bigcup_{0\lqw n' \lqw n}
  M_{k_1,n'}(\ti\Phi) \cup M_{k_1,n'}(\ti\Psi) \,\},
  \edla
  which is obviously finite and empty if $n<0$.
  We check the inclusion as follows: Suppose $(p,q) \in M_{k,n}(\ti\Phi\ti\Psi)$. Then $q-p=n$ and,
  since $\De_{-2p}$ satisfies the Leibniz rule,
  \[
     \sum_{k_1,k_2\gqw0\atop k_1+k_2=k}
    \sum_{q_1,q_2\gqw0 \atop q_1+q_2=q}
    (\De_{-2p} \ti\Phi_{k_1,q_1}) \ti\Psi_{k_2,q_2}
    +
    \sum_{k_1,k_2\gqw0\atop k_1+k_2=k}
    \sum_{q_1,q_2\gqw0 \atop q_1+q_2=q}
    \ti\Phi_{k_2,q_2} (\De_{-2p} \ti\Psi_{k_1,q_1}) 
    \neq0, 
  \]
  whence there exists $q_1,q_2\in\{0,\ldots,q\}$ and
  $k_1\in\{0,\ldots,k\}$ such that
  $q=q_1+q_2$ and $\De_{-2p} \ti\Phi_{k_1,q_1} \neq0$
  or $\De_{-2p} \ti\Psi_{k_1,q_1}\neq0$.
  This means $(p,q_1) \in M_{k_1,n'}(\ti\Phi) \cup M_{k_1,n'}(\ti\Psi)$
  with $n'\defeq q_1-p = n-q_2$,
  thus necessarily $n'\gqw0$, and we see that $q_2\lqw n$ and $n'\lqw n$.
  Since $(p,q) = (0,q_2)+(p,q_1)$, this proves that
  $M_{k,n}(\ti\Phi\ti\Psi)$ is contained in the set~\eqref{eqfiniteset}.

  It is easy to check that, for each $\ti\Phi\in\cA_0$,
  $(e^{2mz}\De_{-2m}\ti\Phi)_{m\gqw1}$ is a summable family of
  $\ti\RT_{2\Z}^\simp[[\si_2,e^{-2z}]]$ (for the metrizable topology
  induced by the total order \wrt~$\si_2$ and~$e^{-2z}$),
  whose sum is $\DDE_{\R_{\lqw 0}} \ti\Phi$.
 The operator $\DDE_{\R_{\lqw 0}}$ is the sum of a formally convergent series of
 derivations of~$\cA_0$, and thus a derivation itself.
\end{proof}

The operator $\DDE_{\R_{\lqw 0}} $ is thus well-defined on~$\cA_0$, but
to iterate it we need to restrict to a smaller subspace.
We thus define inductively
\beglab{eqdefcAr}
\cA_r \defeq \{\, \ti\Phi \in \cA_{r-1} \mid
(\DDE_{\R_{\lqw 0}})^r \ti\Phi \in \si_2^r \cA_0 \,\}
\quad \text{for $r\gqw1$}
\edla
(notice that, by induction on~$r$, $(\DDE_{\R_{\lqw 0}})^r$ is
well-defined on~$\cA_{r-1}$ and~\eqref{eqdefcAr} makes sense).
Here, we denote by $\si_2^r \cA_0$ the subspace of those elements of~$\cA_0$ that are
divisible by~$\si_2^r$, \ie whose partial order in~$\si_2$ is at
least~$r$;
this condition ensures that the series of operators
$\sum \f1{r!}(\DDE_{\R_{\lqw 0}})^r$ is formally convergent on
\begla
\cA_\infty \defeq \bigcap_{r\gqw0} \cA_r.
\edla

\begin{Lemma}   \label{lemcAsubalgforDDElqw}
The set $\cA_\infty$ is a subalgebra of
$\Td\RT_{2\Z}^\simp[[\si_2,e^{-2z}]]$,
the operator $\DDE_{\R_{\lqw 0}}$ induces a derivation
of~$\cA_\infty$,
with a well-defined exponential
\begla
\DDE_{\R_{\lqw 0}}^+ \defeq \sum_{r\gqw0} \f1{r!}(\DDE_{\R_{\lqw 0}})^r \col \cA_\infty\to\cA_\infty
\edla
that is an algebra automorphism.
\smallskip

Moreover, $\ti G(z,\si_1,\si_2) \in \cA_\infty[\si_1]$ with affine
dependence in~$\si_1$ and
\beglab{eqiterDDEmG}
  (\DDE_{\R_{\lqw 0}})^r \ti G = D^r \ti G \quad\text{for all $r\gqw0$},
  \quad \text{where}\ens D \defeq -i \si_2
  \Big(\f{\pt}{\pt
    \si_1}-\si_2\f{\pt}{\pt\si_2}\Big).  
\edla
\end{Lemma}

\begin{proof}
We first check by induction on~$r$ that each~$\cA_r$ is a subalgebra
of~$\cA_0$:
suppose that $\ti\Phi,\ti\Psi\in\cA_r$, then their product is in~$\cA_{r-1}$
by the induction hypothesis and, since $\DDE_{\R_{\lqw 0}}$ is a
derivation on~$\cA_{r-1}$,
\[
  (\DDE_{\R_{\lqw 0}})^r(\Td\Phi\Td\Psi) = \sum_{r=r_1+r_2}
  \binom{r}{r_1} \big( (\DDE_{\R_{\lqw 0}})^{r_1}\Td\Phi\big)
  \big((\DDE_{\R_{\lqw 0}})^{r_2}\Td\Psi\big),
\]
which is in $\si_2^r\cA_0$ because~$\cA_0$ is stable under
multiplication and $\ti\Phi,\ti\Psi\in\cA_r$.
Therefore $\ti\Phi\ti\Psi\in\cA_r$.

Since $(\cA_r)_{r\gqw0}$ is a decreasing sequence of subalgebras, so
is their intersection~$\cA_\infty$.
By restriction, we have a derivation $\DDE_{\R_{\lqw 0}} \col
\cA_\infty \to \cA_\infty$,
and since its exponential is a convergent series of operators, it is
an algebra automorphism (general property of the exponential series).

% {
% At the same time, 
% notice that as the $\si_2$-valuation $r$ increases, the degree of $\si_2$ in $(\DDE_{\R_{\lqw 0}})^r(\Td\vp\Td\psi)$ also grows which means that the sum $\sls_{n\gqw 0}^\infty\f{1}{n!}(\DDE_{\R_{\lqw 0}})^n(\Td\vp\Td\psi)$
% is formally convergent series within $\Td\RT_{2\Z}^\simp[[\si_2,e^{-2z}]]$.
% }

We now verify that $\ti G(z,\si_1,\si_2)\in\cA_\infty[\si_1]$ and prove~\eqref{eqiterDDEmG}. % , with affine dependence in~$\si_1$.

From Proposition~\ref{propbridge}, we easily get $\ti G\in\cA_0[\si_1]$
(with $M_{k,n}(\ti G)\neq\emptyset$ if and only if $n\gqw0$ and $k=n+1$,
and $M_{n+1,n}(\ti G)=\{(1,n+1)\}$),
and we find $\DDE_{\R_{\lqw 0}} \ti G = D \ti G$ with~$D$ as in~\eqref{eqiterDDEmG}.
Since the operators~$\DDE_{\R_{\lqw 0}}$ and~$D$ commute, and
since~$D$ maps $\si_2^{r-1} \cA_0[\si_1]$  in $\si_2^r \cA_0[\si_1]$  for each $r\gqw1$, we find that
\[
  \ti G \in \cA_{r-1}[\si_1], \qquad
  (\DDE_{\R_{\lqw 0}})^r \ti G = D^r \ti G \in \si_2^r \cA_0[\si_1]
\]
by induction on $r\gqw1$.
This shows that $\ti G\in\cA_\infty[\si_1]$ and proves~\eqref{eqiterDDEmG}.
\end{proof}

%
%%%%%%%%%%%%%%%%%%%%%%%%%%%%%%%%%%%%%%%%%%%

\begin{Proposition}\label{propstk}
  We have
\beglab{stk4}
\DDE_{\R_{\lqw 0}}^+\ti G(z,\si_1,\si_2)=
\ti G\big(z,\si_1+\log(1-i\si_2),\f{\si_2}{1-i\si_2}\big)
\edla
\end{Proposition}

\begin{proof}
  According to~\eqref{eqiterDDEmG}, the derivation $\DDE_{\R_{\lqw 0}}$ can be seen as a vector field
whose action on~$\ti G$ coincides with that of~$D$,
thus the action of the flows of these vector fields must coincide too.

We can easily compute the flow of~$D$, by solving the Cauchy problem % system of ODEs with a given initial value
\[
  \left\{\begin{aligned}
      \dot{x}_1&=-ix_2,\\[.5ex]
      \dot{x}_2&=ix_2^2,\\[.5ex]
      x_1(0)&=\si_1,\ens x_2(0)=\si_2.
    \end{aligned} \right.
\]
The solution is
\[
  \left\{\begin{aligned}
      x_1(t) &=\si_1+\log(1-it\si_2),\\[.5ex]
x_2(t)&=\f{\si_2}{1-it\si_2}.
\end{aligned}\right.
\]
We conclude that
\[
  \exp(t\DDE_{\R_{\lqw 0}})\ti G = \exp(tD)\ti G =  \ti G\big(z,
    \si_1+\log(1-it\si_2),\f{\si_2}{1-it\si_2}\big) 
\]
and get~\eqref{stk4} by making $t=1$.
\end{proof}

This completes the proof of Theorem~A.

%%%%%%%%%%%%%%%%%

\begin{Remark}
   Equation~\eqref{stk4} amounts to giving $\De_{-2n}^+\ti G_k$ for all
  $n\gqw1$ and $k\gqw0$ as follows: 
  \beglab{eqvanishDDnpGk}
   n > k \imp \De_{-2n}^+ \ti G_k = 0 
  \edla
   (in particular $\De_{-2n}^+ \ti G_0 = 0$ for all $n\gqw1$),
    and 
  \beglab{eqDDnpGkcnk}
     1\lqw n=k \imp \De_{-2k}^+ \ti G_k = -\f{i^k}k,
  \quad
  1\lqw n< k \imp \De_{-2n}^+ \ti G_k = % c_{n,k}\,
  i^n \binom{k-1}{n}
  \ti G_{k-n} .
  \edla
  %
  % \blue{ with coefficients $(c_{n,k})$ defined by the generating
  %   series }
  %
  % \begla
  %   %
  %   \blue{ r \gqw 1 \imp
  %   %
  %   \Big( \sum_{k\gqw1} i^{k-1} \si^k \Big)^r =
  %   %
  %   \si^r + \sum_{k>r} c_{k-r,k} \si^k. }
  %   %
  % \edla
  %  
\end{Remark}

%%%%%%%%%%%%%%%%%

% In general, we have 
% %
% \begin{Proposition}\label{propkey}
% When $\si\in\C$, for the transseries $\ti G(z,\si_1,\si_2)$ we have
% \begin{equation}
% (\DDE_{\R_{\gqw 0}}^+)^\si \ti G(z,\si_1,\si_2)=\ti G(z,\si_1,\si_2-i\si).
% \end{equation}
% \end{Proposition}
% \begin{proof}
%   The proof follows immediately from Propositions \ref{propkey0} and Proposition \ref{propstk}.
% \end{proof}
% %
% \blue{  This proposition should perhaps appear earlier...  }

% It means that formulas \eqref{stk1} are just particular cases with $\si_1=\si_2=0~\text{and}~\si=1.$ 

%%%%%%%%%%%%%%%%%%%%%%%%%%%%%%%%%%%%%%%%%%%%%%%%%%
%\newpage

\subsection{Summability of the formal integral} \label{secsummaSt}

% Now we turn to the Stokes phenomenon of the formal integral $G(z,\si_1,\si_2).$

We now prove Theorem~B(i).
In view of~\eqref{eqreltigFFs}, \eqref{eqreltiGnGGn} and~\eqref{eqidentifytiGm}, we have
\begla
\GG_0(\la_s) = \FFs(\la_s) = \ti g(z), \qquad
\GG_n(\la_s) = \ti G_n(z) = \f{(-1)^{n-1}}n (\ti G_1)^n \quad\text{for $n\gqw1$}
\edla
with $\ti G_1 = e^{\ti f - \ti g}$.
We have already seen in Proposition~\ref{propseries} that
$\ti g \in \ti\NN(I_0,\be_0)$, with a locally bounded function
$\be_0 \col I_0=(-2\pi,0)\to\R_{\gqw0}$
 whose $2\pi$-periodic extension (still denoted by~$\be_0$) is even.
Since the Borel transform~$\wh g$ is regular at $\ze=0$, we can as
well say that  $\ti g \in \ti\NN(2k\pi+I_0,\be_0)$ for any $k\in\Z$.
%
% extending~$\be_0$ by $2\pi$-periodicity.

Recall that an even locally bounded function $\be_\pi \col
I_\pi=(-\pi,\pi)\to\R_{\gqw0}$ was also introduced in Proposition~\ref{propseries}.
%
% Theorem~B(i) is a direct consequence of
%
\begin{Proposition} \label{propabs}
Each~$\ti G_n$, $n\gqw1$, is $1$-summable in the directions of
\beglab{eqdefIpmagain}
I^+=(-\pi,0)=I_\pi\cap I_0
\quad \text{and} \quad
I^-=(0,\pi)=I_\pi\cap (2\pi+I_0).
\edla
For each choice of sign, `$+$' or~`$-$', the Borel-Laplace sums
$\AS^{I^\pm}\ti G_n$ is analytic in $\DD(I^\pm,\be_0)$,
and we have $\ti G_n\in\C\oplus\ti\NN(I^\pm, \al)$
with $\al\defeq 2\be_0+\be_\pi$ and 
% the Borel sum $\AS^{I^\pm}\ti G_n$ is thus holomorphic in $\DD(I^\pm,2\be_0+\be_\pi)$.
%
% Moreover,
%
\beglab{ineqASGn}
|\AS^{I^\pm}\ti G_n(z)| \lqw \f{2^n}{n}
\quad \text{for any $z\in\DD(I^\pm,\al)$ and $n\gqw1$.}
\edla
 The series of holomorphic functions
\begin{equation}\label{eq_abs}
 G^\pm(z,\si_1,\si_2)  \defeq % \AS^{I^\pm}\ti G(z,\si_1 ,\si_2) \defeq
  \si_1 + \sum_{n\gqw0} \si_2^n e^{-2nz} \AS^{I^\pm}\ti G_n(z)
  %
% \AS^{I^\pm} \Td{g}+\si_1+\sls_{n\gqw 1}\f{(-1)^{n-1}}{n}\si_2^n e^{-2nz} e^{n(\AS^{I^\pm} \Td{f}-\AS^{I^\pm} \Td{g}) 
%
\end{equation}
is convergent and holomorphic in the domain
\[
  \big\{ (z,\si_1,\si_2)\in\ti\C\times\C\times\C\mid
  z\in\DD(I^\pm,\al),\; \RE(z)>\tf{1}{2}\ln |2\si_2| \big\}
  \]
and defines a two-parameter family of analytic solutions to~\eqref{eq_ageq2}
(recall that~$\ti\C$ denotes the Riemann surface of the logarithm and
$\DD(I^\pm,\al)$ is defined by~\eqref{eqDDIaltiC}).
\end{Proposition}

%%%%%%%%%%%%%%%%%%

\begin{proof}
  Using \eqref{eqtipsitiphone}--\eqref{eqdeftif}, we can write
\begla
\ti G_1 = 1 + \ti h, \qquad \ti h \defeq \f{\ti\ph_1-\ti\psi_1}{1+\ti\psi_1}.
\edla
%
% \begin{Lemma}\label{lem_1}
%
% For any $n\in\N^\ast$, the formal power series $e^{n(\Td{f}-\Td{g})} $
% is $1$-summable in the directions $I\subset\R\setminus (\pi\Z)$ with
% Borel transform
% $\BB(e^{n(\Td{f}-\Td{g})})\in \C\de\oplus \NN(I,\al)$ and
% Borel sum $\AS^I(e^{n(\Td{f}-\Td{g})})=e^{n(\AS^I\Td{f}-\AS^I\Td{g})}$
% holomorphic in $\DD(I,\al)$ where $\be_0$ and $\be_\pi$ are
% functions given in the Proposition \ref{propseries}. In fact, these
% two functions are periodic of period $2\pi.$
%
% \end{Lemma}
%
%
% Without loss of generality, we may take $I=(0,\pi)$.
%
% The formal power series $e^{n(\Td{f}-\Td{g})}$ can be written as
% %
% \begin{equation}\label{eq_bie}
% e^{n(\Td{f}-\Td{g})}=(1+\Td{h})^n,~\Td{h}=\f{\Td{\vp}_1-\Td{\psi}_1  }{1+\Td{\psi}_1}.
% \end{equation}
%
The Borel transform of $\Td{h}$ is
$\wh{h}=(\wh{\vp}_1-\wh{\psi}_1)\ast
(\de-\wh{\psi}_1+\wh{\psi}_1^{\ast 2}-\wh{\psi}_1^{\ast 2}+\cdots)\in
\C[[\ze]]$
(formal convergence ensured by $\wh\psi_1^{*n}\in\ze^{n-1}\in\C[[\ze]]$).
Here $\wh{\vp}_1,\wh{\psi}_1$ and $\wh{\psi}_1^{\ast k}$ are
holomorphic in $D(0,R)$ for any $0<R<2$, implying that $\wh{h}$ is
holomorphic in $D(0,R)$ and can be analytically continued to
$D(0,R)\cup \Si$, where
\begla \Si \defeq \{ \arg \ze \in I^-\cup I^+ \}, \edla
thanks to~\eqref{eq_cc}. By \eqref{eqrelwhphwhpsi}--\eqref{ineqwhphIpi},
we get
\begin{equation}\label{eq_cc1}
  |\wh{h}|\lqw \left(\be_\pi(\th)+\be_0(\th)\right)\cdot\Big(1+\sls_{k\gqw 1}\be_0(\th)^k\f{|\ze|^k}{k!} \Big)=(\be_\pi(\th)+\be_0(\th))\cdot e^{\be_0(\th) |\ze|}
\end{equation}
for all $\ze\in\Si$.
This proves that $\AS^{I^\pm} \ti h$ is holomorphic in $\DD(I^\pm,
\be_0)$,
as well as 
$\AS^{I^\pm} \ti G_n = \f{(-1)^{n-1}}n (1+\AS^{I^\pm} \ti h)^n$.
Moreover,~\eqref{eq_cc1} yields
\begla
| \AS^{I^\pm} \ti h(z) | \lqw \f{ \be_0(\th)+\be_\pi(\th)}{\RE(z
  e^{-i\th}) - \be_0(\th)}
\ens \text{in}\;
\{ \RE(z e^{-i\th}) > \be_0(\th) \}
\ens \text{for any $\th\in I^\pm$.}
\edla
In particular, $z \in \DD(I^\pm,\al) \imp
| \AS^{I^\pm} \ti h(z) | \lqw 1
 \imp | \AS^{I^\pm} \ti G_1(z) | \lqw 2$.
 This implies~\eqref{ineqASGn} and the condition
\beglab{cdt0}
  |\si_2 e^{-2z}| < \tf12
  \quad\Longleftrightarrow\quad
  \RE(z)>\tf{1}{2}\ln |2\si_2|
\edla
 ensures the convergence of
 the series of functions~\eqref{eq_abs},
 which gives rise to analytic solutions of the ODE~\eqref{eq_ageq2} by
 virtue of the algebra homomorphism property of~$\AS^{I^\pm}$ and~\eqref{eqASIdtiphdz}.
 
 % Furthermore, we write the binomial expansion of~\eqref{eq_bie} as
 %
One can check that each $\ti G_n \in\C\oplus\ti\NN(I^\pm,
 \al)$ as follows:
\[
(-1)^{n-1} n \, \ti G_n = 
  (\ti G_1)^n=1+\binom{n
  }{1}\Td{h}+\binom{n}{2}\Td{h}^2+\cdots+\binom{n}{n}\Td{h}^n=1+\Td{K}_n
\]
  and the Borel transform $\wh{K}_n$  of $\Td{K}_n$ is holomorphic in
  $D(0,R)$ with analytic continuation to $D(0,R)\cup \Si$.
  Finally, by~\eqref{eq_cc1}, we obtain that, for all $\ze\in\Si$,
\begin{align*}
  |\wh{K}_n| &\lqw 2^n\cdot \big(|\wh{h}|+|\wh{h}^{\ast 2}|+\cdots+|\wh{h}^{\ast n}|\big)  \\
             &\lqw 2^n\cdot(\be_\pi+\be_0)e^{\be_0|\ze|}\Big(1+(\be_\pi+\be_0)|\ze|+\cdots +(\be_\pi+\be_0)^{n-1}\f{|\ze|^{n-1}}{(n-1)!} \Big)\\
             &\lqw 2^n\cdot (\be_\pi(\th)+\be_0(\th))e^{(2\be_0(\th) +\be_\pi(\th))|\ze|}.\label{eq_ebd}
\end{align*}
%
%
% $\BB (e^{n(\Td{f}-\Td{g})})\in \C\de\oplus \NN(I,\al)$ and
% $\AS^I (e^{n(\Td{f}-\Td{g})})=e^{n(\AS^I\Td{f}-\AS^I\Td{g})}$ is
% holomorphic in $\DD(I,\al).$
%
\end{proof}

% \begin{proof}
% %
% ... we see that $\BB(e^{n(\Td{f}-\Td{g})})\in\C\de\oplus \NN(I^\pm,\al)$ for all $n\in \N^\ast$, so $\AS^{I^\pm} G_n$ is holomorphic in $\DD(I^\pm,\al).$ On the other hand, inequality~\eqref{eq_ebd} implies that we need to require
% \begin{equation}\label{cdt0}
% |\si_2e^{-2z}|<\f{1}{2},~\big(\ \text{or~equivalently}\ \Re e(z)>\f{1}{2}\ln |2\si_2|\big)
% \end{equation}
% to ensure that the Borel sum $\AS^{I^\pm}G(z,\si_1 ,\si_2)$ is analytic in
% $\DD(I^\pm,\al,\si_2)$.
% %
% \end{proof}

%%%%%%%%%%%%%%%%%%

Theorem~B(i) is a direct consequence of the proposition we just
proved:
we can take $\DD_{I^\pm} \defeq \DD(I^\pm,\al)$, i.e.
%
% Indeed, the sectorial neighbourhoods of infinity mentioned in the
% statement of Theorem~B(i) can thus be taken to be
%
\beglab{eqDDIpm}
\begin{aligned}
\DD_{I^+} &= \bigcup_{\th\in (-\pi,0)} \{\, z\in\ti\C \mid
\arg z \in J^+,\; \RE(z e^{i\th})>\al(\th) \,\}
\ens &\text{with}\ens J^+ &\defeq \big(\!-\tf\pi2,\tf{3\pi}2\big) \\[1ex]
\DD_{I^-} &= \bigcup_{\th\in (0,\pi)} \{\, z\in\ti\C \mid
\arg z \in J^-,\; \RE(z e^{i\th})>\al(\th) \,\}
\ens &\text{with}\ens J^- &\defeq \big(\!-\tf{3\pi}2,\tf\pi2\big)
\end{aligned}
\edla
Using the notation % $\DD_{I^\pm}$ as in~\eqref{eqDDIpm}
$\DD^\pm(\si_2)$ as in~\eqref{eqdefDDpmsi2},
we thus have analytic solutions
\beglab{eqGpmsol}
z\in \DD^\pm(\si_2) \mapsto G^\pm(z,\si_1,\si_2) % \defeq \AS^{I^\pm}\ti G(z,\si_1,\si_2)
\edla
to the ODE~\eqref{eq_ageq2}.
Notice that for every $\th\in J^\pm$,
  the intersection $\DD_{I^\pm}\cap e^{i\th}\R_{>0}$ is a half-line of
  the form $e^{i\th}\big( \al'(\th),\infty)\subset\ti\C$, for some $\al'(\th)\gqw0$.
  For every $\th\in(-\f\pi2,\f\pi2)$ and $\si_2\in\C$, % the intersection
  $\DD^\pm(\si_2)\cap e^{i\th}\R_{>0}$ is a half-line of the same
  form, along which $e^{-2z}$ is % is
  exponentially decaying at infinity. 

 One may view the parameters~$\si_1$ and~$\si_2$ as
  \emph{boundary conditions at infinity relative to
    $\AS^{I^\pm}\ti g$} in the following sense: 
\begin{Proposition}   \label{PropcharacterizGpm}
   \emph{For each $(\si_1,\si_2)\in\C^2$ and $\th\in(-\f\pi2,\f\pi2)$,
      the function~\eqref{eqGpmsol} is the unique solution
      to~\eqref{eq_ageq2} such that} 
  \beglab{eqcharacterizGpm}
   G^\pm(z,\si_1,\si_2) \xrightarrow[z\to\infty]{} \si_1, \quad
  e^{2z} \big( G^\pm(z,\si_1,\si_2) -\si_1 - \AS^{I^\pm}\ti
  g(z)\big) \xrightarrow[z\to\infty]{} \si_2
  \edla
   \emph{where the limits are taken along the half-line
    $\DD^\pm(\si_2)\cap e^{i\th}\R_{>0}$.} 
\end{Proposition}

\begin{proof}
   The solution~$G^\pm$ obviously
  satisfies~\eqref{eqcharacterizGpm}. The uniqueness can be obtained
  as follows.
  Recast the ODE~\eqref{eq_ageq2} as a non-autonomous vector field, 
  \beglab{eqODEasvf}
     y_2 \f{\pa\,}{\pa y_1} - (y_1^2+2y_1+\tf5{36}z^{-2})\f{\pa\,}{\pa
      y_2} 
  \edla
  (by setting $y_1=g$ and $y_2=g'$).
The transformation 
  \beglab{eqtrsfoyY}
     y_1 = Y_1 +  \sum_{n\gqw0} Y_2^n \AS^{I^\pm}\ti G_n(z),
    \quad
    y_2 = \sum_{n\gqw0} Y_2^n (-2n+\f{d\,}{dz})\AS^{I^\pm}\ti G_n(z), 
  \edla
induces a biholomorphism between two neighbourhoods of
$\C\times\{0\}\times\{\infty\}$ in $\C\times\C\times\big(\{\arg
z\in(-\de,\de)\}\cap \DD^\pm(\si_2)\big)$
and conjugates~\eqref{eqODEasvf} to the normal form
  $-2Y_2\f{\pa\,}{\pa Y_2}$,
whose solutions are the curves $z\mapsto (\si_1,\si_2 e^{-2z})$.
Now take an arbitrary solution~$G(z)$ to~\eqref{eq_ageq2} analytic along the
half-line $\DD^\pm(\si_2)\cap e^{i\th}\R_{>0}$. 
The image of $(G(z),G'(z))$ by the inverse of~\eqref{eqtrsfoyY} is one
of the solutions of the normal form, thus there exists
$(\si_1,\si_2)$ such that~$G(z)$ is of the form
$\si_1 + \AS^{I^\pm}\ti G_0(z) + \si_2 e^{-2z}\AS^{I^\pm}\ti G_1(z) +
O(e^{-4z})
= \si_1 + \AS^{I^\pm}\ti g(z) + \si_2 e^{-2z} \big(1+O(z\ii) \big)+
O(e^{-4z})$, which implies~\eqref{eqcharacterizGpm}.
\end{proof}
%

%%%%%%%%%%%%%%%%%%%%%%%%%%%%%%%%%%%%%%%%%%%%%%%%%%%%%%%%%

\subsection{Nonlinear Stokes phenomenon} \label{secNLSt}

%%%%%%%%%%%%%%%%%%%%%%%%%%%%%%%%%%%%%%%%%%%%%%%%%%%%

We recall that, according to~\eqref{eqdefIpm} or~\eqref{eqdefIpmagain},
$I^+ = (-\pi,0)$ and $I^- = (0,\pi)$
and Proposition~\ref{propabs} has introduced an even locally bounded
function $\al\col I^+\cup I^-\to\R_{\gqw0}$.
The next proposition gives the proof of Theorem B(ii).

\begin{figure}[ht]
  \centering
\includegraphics[scale=0.245]{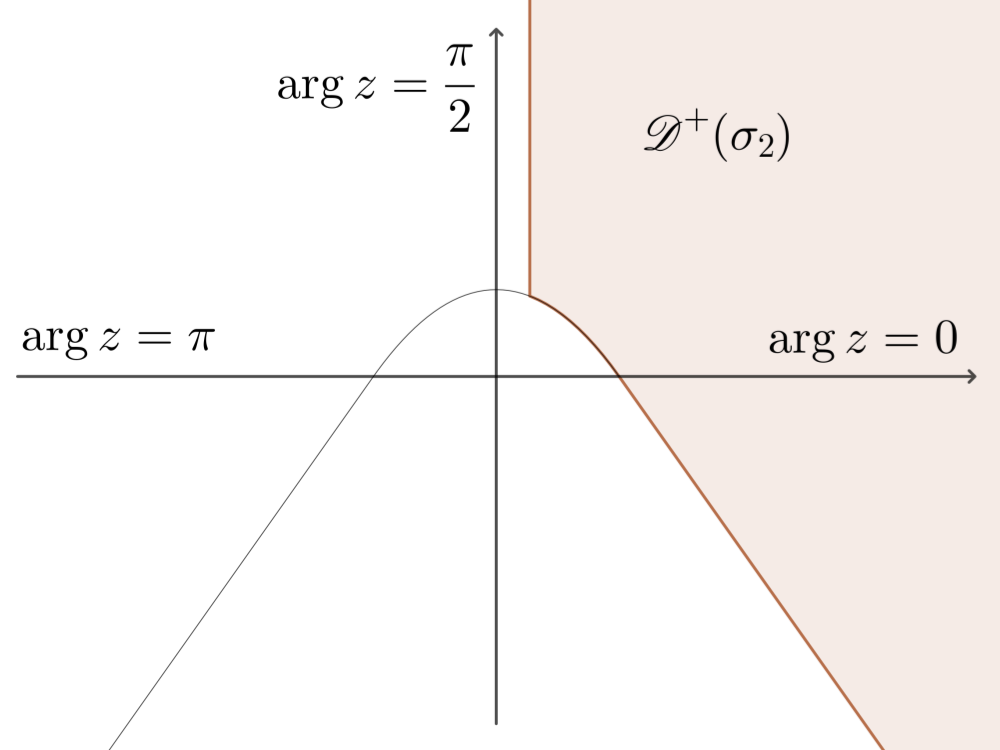}
\hspace{.5em}
\includegraphics[scale=0.245]{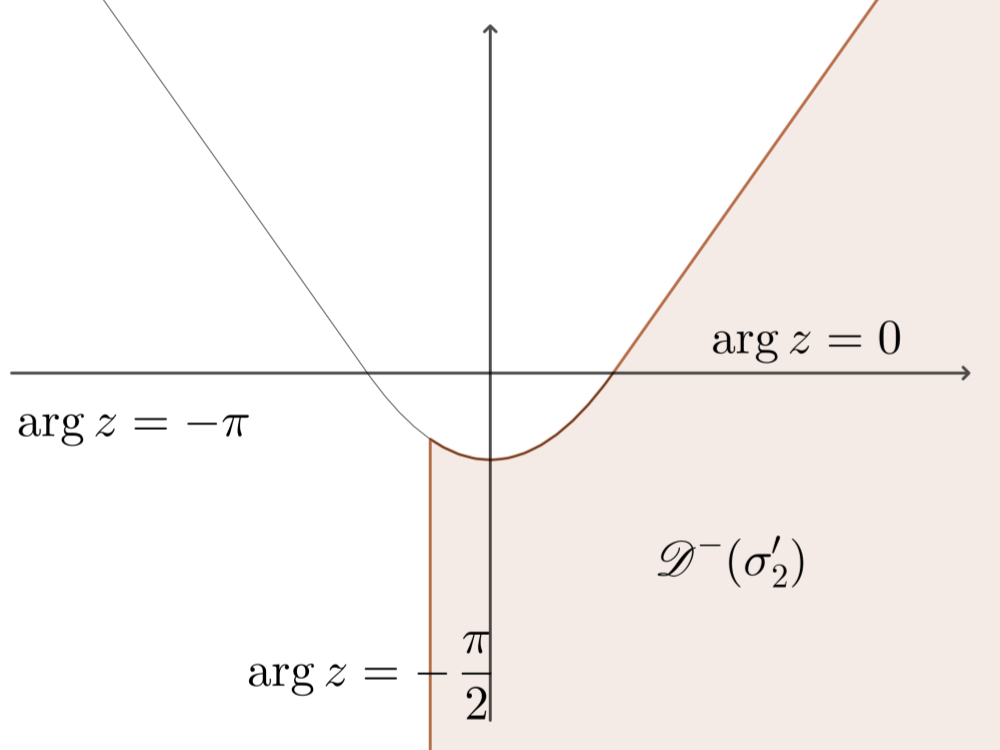}

\vspace{4ex}

\includegraphics[scale=0.245]{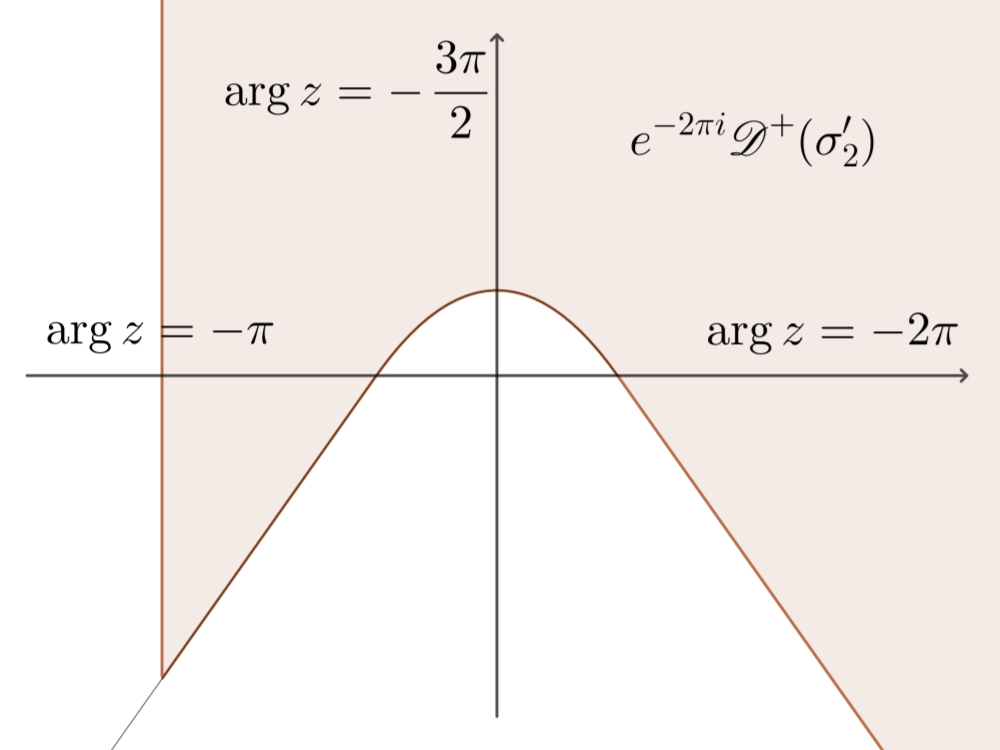}
\hspace{.5em}
\includegraphics[scale=0.245]{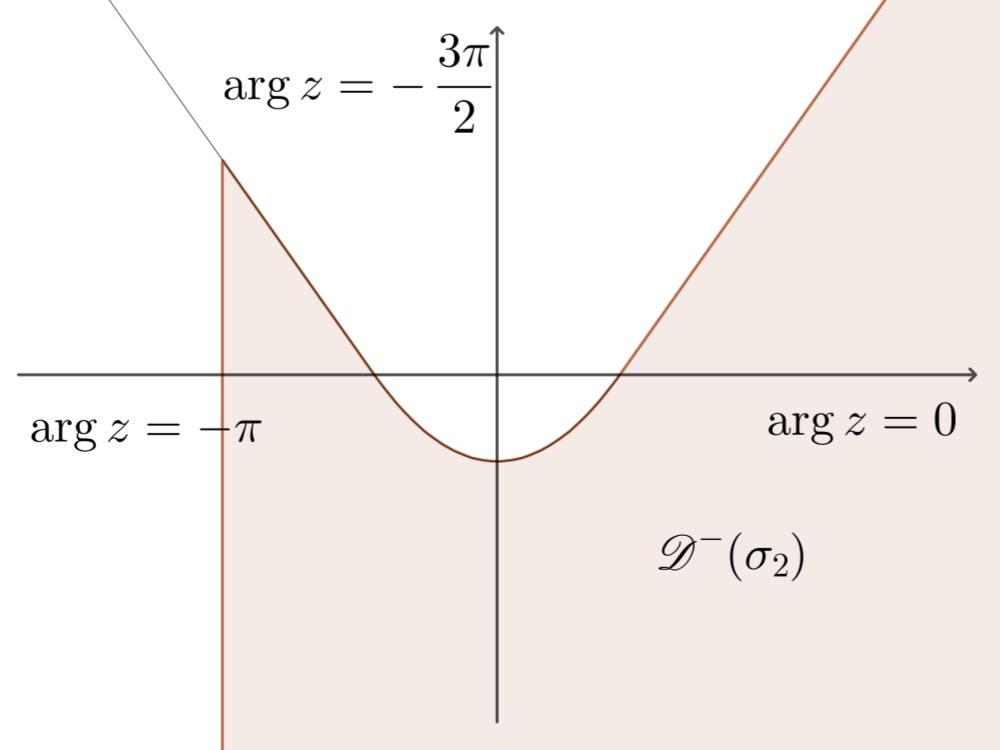}
\caption{Top: Domains of analyticity near $\arg z = 0$ with
    arbitrary~$\si_2$ and~$\si_2'$.
  Bottom: Domains of analyticity near $\arg z = -\pi$ for~$|\si_2|$
  and~
  $|\si_2'|$
  small enough so as to yield non-empty intersection with $e^{-i\pi}\R_{>0}$.}
%
% The first two pictures are the analytic regions corresponding
%  to direction $\arg z=0$, and the last two pictures are the analytic
%  regions corresponding to direction $\arg z=-\pi$.}
%
\label{figdomainLEFT}
\end{figure}

\begin{figure}[ht]
  \centering
\includegraphics[scale=0.245]{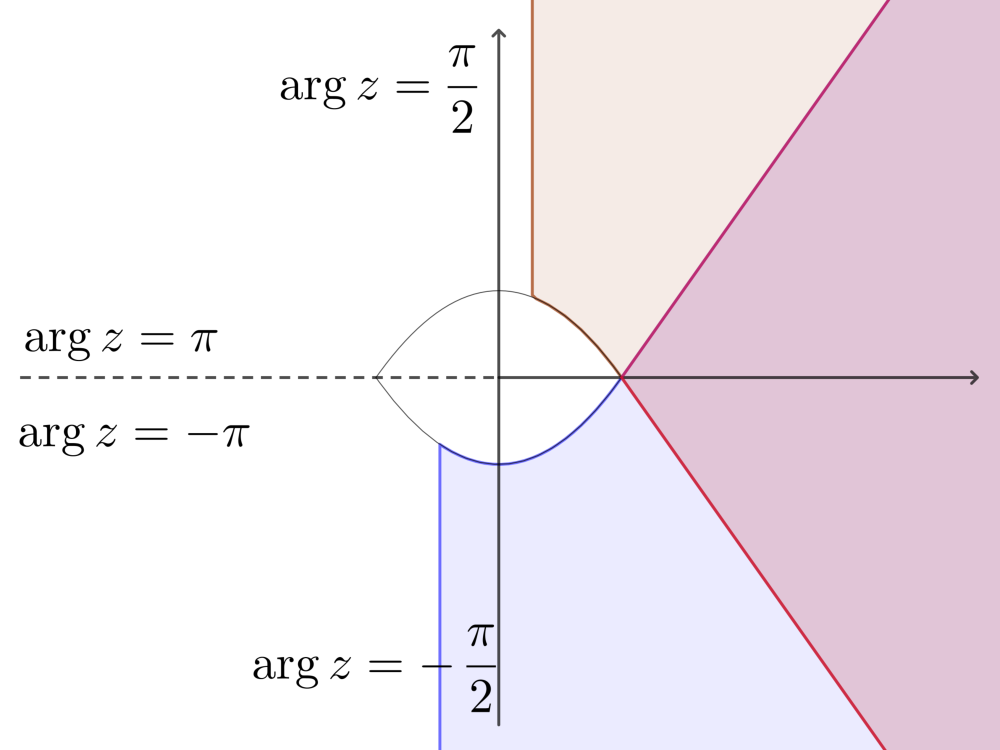}
\hspace{1em}
\includegraphics[scale=0.245]{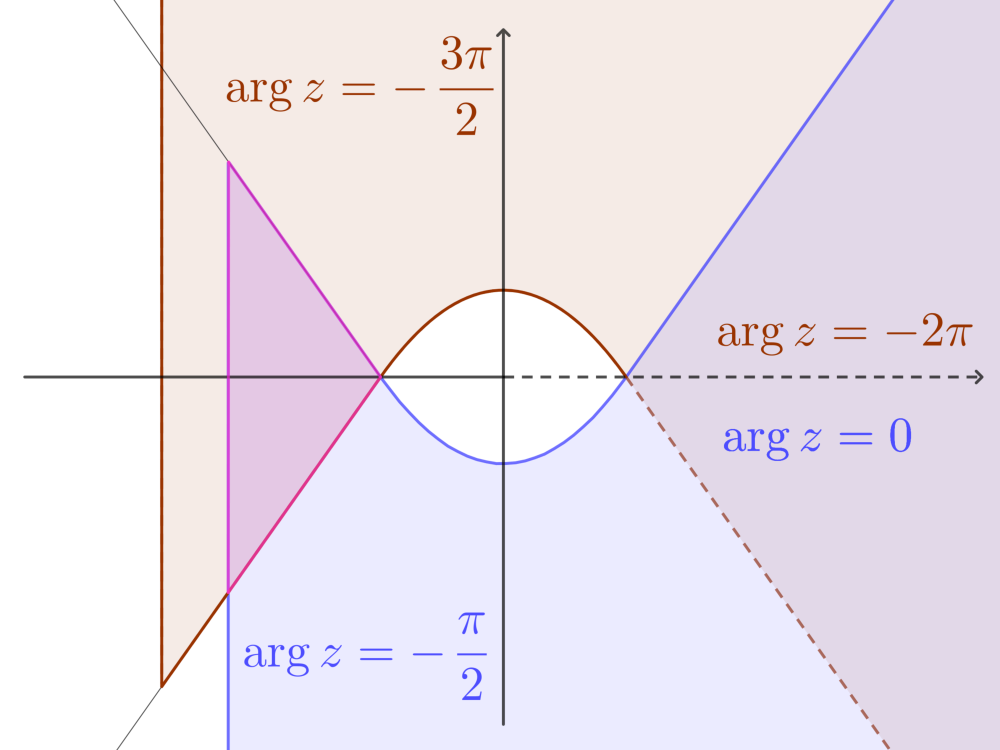}
\caption{Left: Overlap of
  $\DD^+(\si_2)$ and $\DD^-(\si_2-i)$.
  Right: The overlap of $\DD^-(\si_2)$ and $e^{-2\pi i} \DD^+(\f{\si_2}{1-i\si_2})$
   is only the bounded domain on the left.
  \label{figdomainINTERSEC}}
\end{figure}

\begin{Proposition}[Connection formula around the direction $\arg z=0$]\label{sph2}
%
  %
  % Let $I^+$ and $I^-$ be the same as in Proposition \ref{propabs}.
%
Let 
\begla
 \al_0 \defeq
  \inf \bigg\{ \f{\al(\th)}{\cos\th} \mid
\th\in \big(\!-\tf\pi2,0\big) \bigg\}
= \inf \bigg\{ \f{\al(\th)}{\cos\th} \mid
\th\in \big(0,\tf\pi2\big) \bigg\}. 
%
% \inf \bigg\{ \f{\al(\th)}{\cos\th} \mid
% %
% \th\in \big(\!-\tf\pi2,\tf\pi2\big)\cap I^\pm \bigg\}
%
\edla
%
% with~$\al$ as in Proposition~\ref{propabs}.
For any $\si_2,\si_2'\in\C$, $\DD^+(\si_2) \cap \DD^-(\si_2')$
contains the half-line $(x_0,+\infty) \subset e^{i0} \R_{>0}$, where
$x_0\defeq \max\big\{ \f12\ln|2\si_2|, \f12\ln|2\si_2'|, \al_0
% \al_0^+,\al_0^-
\big\}$,
and
%
% Near the direction $\arg z=0$, the two families of solutions are connected by
%
\begin{equation}\label{eq_right}
  G^+(z,\si_1,\si_2)=G^-(z,\si_1,\si_2-i)
  \quad \text{for} \ens
  z\in \DD^+(\si_2)\cap \DD^-(\si_2-i)
\end{equation}
(see top of Figure~\ref{figdomainLEFT} and left of Figure~\ref{figdomainINTERSEC}).
\end{Proposition}

\begin{proof}
Since $\DD_{I^+}$ contains the half-plane $\{ \RE(z
e^{-i\th})>\al(\th) \}$ for any $\th\in I^+$, it contains all the
half-lines
$(\f{\al(\th)}{\cos\th},+\infty) \subset e^{i0}\R_{>0}$,
$\th\in\big(-\f\pi2,\f\pi2\big)\cap I^+$,
 and thus their union $(\al_0,+\infty)$.
% $(\al_0^+,\infty)$.
%
Similarly $\DD_{I^-}$ contains the half-line $(\al_0,+\infty)$.
% $(\al_0^-,+\infty)$.
%
This implies the statement about $\DD^+(\si_2)\cap \DD^-(\si_2')$.

The connection formula is obtained by applying the theory of
Section~\ref{secResTransseries} to~$\ti G$, viewing it as a simple
$2\Z$-resurgent transseries
and using $\om_1=2$ as generator of $\Om=2\Z$.
Indeed, set
\begla
 \th^* \defeq 0, \quad
I_R \defeq (-\tf\pi2,0) \subset I^+, \quad
I_L \defeq (0,\tf\pi2) \subset I^-. 
\edla
 Then, in view of~\eqref{stk3}, \eqref{eqStkTruncatedmst} with
$m_*=1$ shows that $G^+(z,\si_1,\si_2)$ cannot differ from
$G^-(z,\si_1,\si_2-i)$ by more than $O( e^{-2\mu \RE z})$,
for any $\mu\in(1,2)$, 
and Proposition~\ref{PropcharacterizGpm} thus yields the conclusion.
\end{proof}

%%%%%%%%%%%%%%%%%%%%%%%%%%

To obtain Theorem~B(iii), we now use the two-parameter
family of solutions
\begla
z \in e^{-2\pi i}\DD^+(\si_2) \mapsto G^+(e^{2\pi
  i}z,\si_1,\si_2) = \AS^{2\pi+I^+}\ti G(z,\si_1,\si_2)
\edla
(cf.\ footnote~\ref{ftnshiftsheet}).

\begin{Proposition}[Connection formula around the direction $\arg
  z=-\pi$]
  \label{propleft}
Let % $x_* \defeq \max\{ \al_\pi^+, \al_\pi^- \}$, where
\begla
 \al_\pi \defeq \inf \bigg\{ \f{\al(\th)}{|\cos\th|} \mid
\th\in \big(\!-\pi,-\tf\pi2\big) \bigg\}
= \inf \bigg\{ \f{\al(\th)}{|\cos\th|} \mid
\th\in \big(\tf\pi2,\pi\big) \bigg\}. 
%
% \al_\pi^+ \defeq \inf \bigg\{ \f{\al(\th)}{|\cos\th|} \mid
% %
% \th\in \big(\!-\pi,-\tf\pi2\big) \bigg\}, \quad
% %
% \al_\pi^- \defeq \inf \bigg\{ \f{\al(\th)}{|\cos\th|} \mid
% %
% \th\in \big(\tf\pi2,\pi\big) \bigg\}.
%
\edla
%
% with~$\al$ as in Proposition~\ref{propabs}.
%
Then, for any $\si_2,\si_2'\in\C$ such that
$s\defeq \max\big\{ |\si_2|,|\si_2'| \big\} <  \f12 e^{-2 \al_\pi}$,
the intersection $\DD^-(\si_2')\cap ( e^{-2\pi i} \DD^+(\si_2) )$
contains the non-trivial line-segment
%
% \begla
%
$ e^{-i\pi}\big(\al_\pi,x(s)\big)  \subset e^{-i\pi} \R_{>0}$, where $x(s)\defeq\tf12\ln\tf1{2s}$.
%
% \edla

Moreover, for any $\si_1\in\C$, if both
$|\si_2|$ and $\left|\f{\si_2}{1-i\si_2}\right| <  \f12 e^{-2
    \al_\pi}$,
then
\begin{equation}\label{eq_left}
  G^-(z,\si_1,\si_2) =
  G^+\Big(e^{2\pi i}z,\si_1+\log(1-i\si_2),\f{\si_2}{1-i\si_2}\Big) 
\end{equation}
 in the non-empty domain
$z\in\DD^-(\si_2)\cap ( e^{-2\pi i} \DD^+(\f{\si_2}{1-i\si_2}) )$ 
(see bottom of Figure~\ref{figdomainLEFT} and right of
Figure~\ref{figdomainINTERSEC}). 
% $\Re e z> \f12\max\big\{\ln|2\si_2|,\ln\f{|2\si_2|}{|\si_2+i|}\big\}.$
%
\end{Proposition}

\begin{proof}
   One can check that $\DD_{I^+}$ contains all the half-lines
$e^{i\pi}(\f{\al(\th)}{|\cos\th|},+\infty) \subset e^{i\pi}\R_{>0}$,
$\th\in\big(-\pi,-\f\pi2)$,
 and thus their union $e^{i\pi}(\al_\pi,+\infty)$.
Similarly $\DD_{I^-}$ contains
the half-line $\newc{ e^{-i\pi}(\al_\pi,+\infty) }$. Therefore 
\begla
  e^{-i\pi} (\al_\pi,+\infty) \subset \DD_{I^-}\cap ( e^{-2\pi i}\DD_{I^+} ). 
\edla
 This implies the statement about $\DD^-(\si_2')\cap (e^{-2\pi
  i}\DD^+(\si_2))$. 

We now apply~\eqref{eqStkTruncatedmst} to~$\ti G_k$
  ($k\gqw0$) with $\om_1=-2$ and 
  \begla
 \th^* \defeq \pi, \quad % \th^* \defeq -\pi, \quad
I_R \defeq (\tf{\pi}2,\pi) \subset I^-, \quad
% I_R \defeq (-\tf{3\pi}2,-\pi) \subset -2\pi+I^-, \quad
%
I_L \defeq (\pi,\tf{3\pi}2) \subset 2\pi+I^+. 
% I_L \defeq (-\pi,-\tf\pi2) \subset I^+,
%
\edla
 Extending~$\al$ by $2\pi$-periodicity, we thus take~$z$ in the domain
$\DD(I_R,\al)\cap \DD(I_L,\al) \subset
\DD_{I^-}\cap(e^{-2\pi i}\DD_{I^+})$.
% (e^{2\pi i}\DD_{I^-})\cap\DD_{I^+}$.
  %
  Note that, by footnote~\ref{ftnshiftsheet}, we have 
  \begla
   \AS^{I_L}\ti G_k(z) = \AS^{2\pi+I^+}\ti G_k( z)
  = \AS^{I^+}\ti G_k(e^{2\pi i} z). 
  \edla
 In view of~\eqref{eqvanishDDnpGk} there are only finitely many values of~$n$ for
which~$\De_{-2n}^+$ does not annihilate~$\ti G_k$, we thus get an exact
formula
\begla \begin{aligned}
    \AS^{I^-}\ti G_k(z) = \AS^{I_R}\ti G_k(z) 
  %
  % \AS^{I^-}\ti G_k(e^{-2\pi i} z) = \AS^{I_R}\ti G_k(z) }
  %
  &  = \AS^{I_L}\ti G_k(z) + \sum_{1\lqw n\lqw k} e^{2nz}
  \AS^{I_L}\De_{-2n}\ti G_k(z) \\
&   = \AS^{I^+}\ti G_k(e^{2\pi i} z) + \sum_{1\lqw n\lqw k}
  e^{2nz} \AS^{I^+}\De_{-2n}\ti G_k(e^{2\pi i} z).
  %
  % \AS^{I^+}\ti G_k(z) + \sum_{1\lqw n\lqw k} e^{2nz} \AS^{I^+}\De_{-2n}\ti G_k(z).}
%
\end{aligned} \edla
 Multiplying this by~$\si_2^k$, we can take the sum over all $k\gqw0$
  and get a convergent series if $|\si_2|$ is small enough: we just
  need both $\tf12\ln|2\si_2|$ and $\tf12\ln\big| \tf{2\si_2}{1-i\si_2}
  \big| < \RE z
  $, which ensures
  \begla
   z\in \DD^-(\si_2)\cap ( e^{-2\pi i}\DD^+\big(\tf{\si_2}{1-i\si_2}\big) ) 
  %
  % (e^{2\pi i}\DD^-(\si_2))\cap\DD^+\big(\tf{\si_2}{1-i\si_2}\big) }
  %
  \edla
 and, by virtue of~\eqref{eqvanishDDnpGk}--\eqref{eqDDnpGkcnk}, after
  adding~$\si_1$ the result is~\eqref{eq_left}.
%
% \begla
% %
% \blue{ G^-(e^{-2\pi i} z,\si_1,\si_2) =
% %
% G^+\Big(z,\si_1+\log(1-i\si_2),\f{\si_2}{1-i\si_2}\Big), }
% %
% \edla
% %
% \blue{ which is equivalent to the connection formula~\eqref{eq_left}.}
  %
\end{proof}

Inverting the map
$(\si_1,\si_2) \mapsto \big(\si_1+\log(1-i\si_2),\f{\si_2}{1-i\si_2}\big)$,
we find that~\eqref{eq_left} is equivalent to
\begin{equation}\label{eq_left_inv}
  G^+(e^{2\pi i}z,\si_1,\si_2) =
  G^-(z,\si_1+\log(1+i\si_2),\f{\si_2}{1+i\si_2}\Big) 
\end{equation}
 in the non-empty domain
$z\in\DD^-(\f{\si_2}{1+i\si_2})\cap ( e^{-2\pi i} \DD^+(\si_2) )$ 
if both
$|\si_2|$ and $\left|\f{\si_2}{1+i\si_2}\right| <  \f12 e^{-2
  \al_\pi}$,
which gives rise to the connection formula of Theorem~B(iii).

%%%%%%%%%%%%%%%
%%%%%%%%%%%%%%%
\iffalse
%%%%%%%%%%%%%%%
%%%%%%%%%%%%%%%
\begin{Corollary}
Let $I^+$ and $I^-$ be as above and $\si_1,\si_2=0$. Near the direction $\arg z=0$, the Stokes phenomenon for the formal power series $\Td{g}$ is as follows:
\begin{equation}
  \AS^{I^+}\Td{g}=\AS^{I^-}\Td{g}-\sls_{n\gqw 1}\f{i^n}{n}e^{-2nz}e^{n(\AS^{I^-}\Td{f}-\AS^{I^-}\Td{g})}
\end{equation}
%
for $z\in\DD(I^+,\al)\cap \DD(I^-,\al)$.
%
\end{Corollary}
\begin{proof}
The value of function $\be_0(\th)$ becomes very large as $\th$ tends
to 0, so $z\in\DD(I^+,\al)\cap \DD(I^-,\al)$
implies the inequality $\Re e z> \f12\ln2$.
%
\blue{I suppose that what you mean is that, without loss of
  generality, we can assume $\al\gqw \f12 \ln2+\eps$ near
  $\th=0$, and this implies $\RE z\gqw \f12 \ln2+\eps$.
  %
  But... WHAT'S THE POINT OF THAT COROLLARY? Is it needed somewhere...?}
  %
\end{proof}
%%%%%%%%%%%%%%%
%%%%%%%%%%%%%%%
\fi
%%%%%%%%%%%%%%%
%%%%%%%%%%%%%%%

%%%%%%%%%%%%%%%%%%%%%%%%%%%%%%%%
%\newpage

\subsection{Real analytic solutions and rationality of coefficients} \label{real}

%%%%%%%%%%%%%%%%%%%%%%%%%%%%%%%%

A natural question is:
For which values of~$\si_1$
and~$\si_2$ are the Borel sums $G^\pm(z,\si_1,\si_2) $ real analytic?
This plays a crucial role in perturbation theory, as
emphasized and analyzed in \cite{AS}.

The answer will be obtained as a consequence of the connection
formulas of the previous section.
We will find for which $(\si_1,\si_2)$ the function
$G^-(z,\si_1,\si_2)$ is real for~$z$ real, i.e., since we must take
$z\in\DD^-(\si_2)$, for $\arg z=0$ or $\arg z=-\pi$.

We first observe that, as noticed in \cite{[AYZ]}, the coefficients of $\FFs(\la_s) =
\ti g(z)$ are real and rational. More generally,

\begin{Lemma}   \label{lemrqtionalGk}
  For each $k\gqw0$, $\ti G_k(z) \in \Q[[z\ii]]$.
\end{Lemma}

\begin{proof}
  The coefficients~$c_n$ in~\eqref{eqformulwhB} belong to~$\Q$, thus
  the series~$\ti\ph(z)$ and~$\ti\psi(z)$ of Section~\ref{exa1} belong
  to $\Q[[z\ii]]$.
  The same is true for $\ti g = \ti G_0$ and~$\ti f$ by
  \eqref{eqtiglogtipsi}--\eqref{eqdeftif}, and thus for all~$\ti G_k$,
  $k\gqw1$, by~\eqref{eqidentifytiGm}.
\end{proof}

Now, if a summable formal series $\ti a(z) = \sum_{n\gqw0} a_n z^{-n}
\in \C\oplus\ti\NN(J,\be)$ has real coefficients, that does not imply
that $\AS^J\ti a(z)$ is real whenever~$z$ is real. % real analytic function.
%
% , unless the interval~$J$ contains~$0$.
%
What is always true is that the Borel transform $\wh a(\ze) =
\sum_{n\gqw1} a_n \f{\ze^{n-1}}{(n-1)!} \in \R\{\ze\}$ is real analytic:
$\wh a(\ze) = \ov{\wh a(\ov\ze)}$, hence the conjugate of
$\AS^J \ti a(\ov z) = a_0 + \int_0^{+\infty} e^{-\ov z\,e^{i\th}t} \, \wh
a(e^{i\th} t) e^{i\th} dt$ (with appropriate $\th\in J$) is seen to be
\begla
\ov{\AS^J \ti a(\ov z)} = \AS^{-J}\ti a(z).
\edla
In the case of the Borel sums of the formal integral~$\ti G$, since
$I^- = - I^+$ and $\al\col I^-\cup I^+\to\R_{\gqw0}$ is even, we obtain
$z\in \DD^-(\si_2) \;\Longleftrightarrow\; \ov z\in \DD^+(\ov\si_2)$
and, when these equivalent conditions are fulfilled,
\begin{equation}\label{eq_cm}
  %
  % \ov{G^+(\ov z,\si_1,\si_2)}=G^-(z,\ov {\si_1},\ov {\si_2}),
  %
  % \quad
  %
  \ov{G^-(z,\si_1,\si_2)}=G^+(\ov z,\ov {\si_1},\ov {\si_2}).
\end{equation}
%
% with $z\in \DD^\pm(\si_2) \;\Longleftrightarrow\; \ov z\in \DD^\mp(\ov\si_2)$.

%We now turn to the proof of Theorem~C. 
%We return to the original
%formal solution~$\Td{g}$ whose coefficients are rational numbers, and
%We have derived the Borel sum of the formal integral $\ti
%G(z,\si_1,\si_2) $.

%%%%%%%%%%%%%%%%%%%%%%%%%%%%%%%%
\medskip

\noindent \textbf{Proof of Theorem~C(i)}.

\medskip
%%%%%%%%%%%%%%%%%%%%%%%%%%%%%%%%

We first focus on the case $\arg z=0$ and write $z=r e^{i0}$ with $r>0$. The proof of
Proposition~\ref{sph2} shows that
$\DD_{I^-}\cap\DD_{I^+}$ contains $(\al_0,+\infty) \subset
e^{i0}\R_{>0}$
%
% with $\al_0 \defeq \max\{\al_0^-,\al_0^+\}$
%
and~\eqref{eq_cm} shows that, for any $(\si_1,\si_2)\in\C^2$,
the function $r\in \DD^-(\si_2)\cap e^{i0}\R_{>0} \mapsto
G^-(r,\si_1,\si_2)$ is real-valued if and only if
\beglab{eqcondreal}
  G^-(r,\si_1,\si_2) = G^+(r,\ov {\si_1},\ov {\si_2})
  \quad \text{for} \ens
  r > \max\{\al_0,\tf12\ln|2\si_2|\}.
  \edla
The \rhs\ of~\eqref{eqcondreal} is $G^-(r,\ov{\si_1},\ov{\si_2}-i)$ by~\eqref{eq_right}, 
  thus this is equivalent to
  \begla
  \si_1 = \ov{\si_1}, \quad \si_2 = \ov{\si_2}-i,
  \edla
  which amounts to $(\si_1,\si_2) = (a,b-\f{i}{2})$ where $a,b\in \R$.

  Since $\ln|2(b-\f{i}{2})| = \f12\ln(1+4b^2)$, we see that the
  function
  $G^-(z,a,b-\f{i}{2})$ is analytic in $\{\, z\in\DD_{I^-} \mid \RE z>\f14\ln(1+4b^2)\,\}$;
  it coincides with $G^+(z,a,b+\f{i}{2})$ and is thus also analytic in $\{\, z\in\DD_{I^+} \mid \RE z>\f14\ln(1+4b^2)\,\}$.

% , and we also need to require
% that  $|b+\f{i}{2}|<\f{1}{2}e^{\Re e(2z)}$ and
% $|b-\f{i}{2}|<\f{1}{2}e^{\Re e(2z)}$ (\ie $b^2<\f{1}{4}(e^{4\Re e(z)}-1) $). Under these conditions the Borel sum $G^+(z,a,b+\f{i}{2} )=G^-(z,a,b-\f{i}{2})$ is real analytic.

% Realness means that
% \begin{equation}\label{eq_rean}
%   \overline{G^+(z,\si_1,\si_2)}=G^+(z,\si_1,\si_2)
% \end{equation}
% in $z\in \DD(I^+,\al)\cap \DD(I^-,\al)\cap \R.$

% First, by taking the complex conjugate we have
% \begin{equation} % \label{eq_cm}
%  \overline{G^+(z,\si_1,\si_2)}=G^-(\ov{z},\ov {\si_1},\ov {\si_2}).
% \end{equation}

\begin{Remark}
This is an instance of \emph{median summation}.
Indeed, as discussed earlier, the formal integral $\ti G(z,\si_1,\si_2)$
belongs to an algebra of simple $2\Z$-resurgent transseries on which
$\AS^{I^+} = \AS^{I^-} \circ \DDE_{\R_{\gqw 0}}^+$ in restriction to
$z\in e^{i0}\R_{>0}$. %  with~$|z|$ large enough.
In that context, we can introduce the median summation operator
relative to the direction $\th = 0$:
\begla
\ASm^0 \defeq \AS^{I^-} \circ (\DDE_{\R_{\gqw 0}}^+)^{1/2} =
\AS^{I^+} \circ (\DDE_{\R_{\gqw 0}}^+)^{-1/2}.
\edla
Then,~\eqref{eqDDEpsiG} shows that the above real analytic solutions are
nothing but
\begla
\ASm^0 \ti G(z,a,b) = G^-(z,a,b-\tf i2) = G^+(z,a,b+\tf i2).
  \edla
  For $a,b\in\R$, $\ti G(z,a,b)$ is a real transseries solution to~\eqref{eq_ageq2}, and~$\ASm^0$
  belongs to a family of summation operators that preserve realness as
  well as the fact of being solution to a nonlinear ODE.\footnote{%
  \'Ecalle's
  `well-behaved real-preserving averages' offer an alternative
  approach to real summation---see \cite{EM96}
  and \cite{MeGUA}.
}
\end{Remark}

\medskip

%%%%%%%%%%%%%%%%%%%%%%%%%%%%%%%%
\noindent \textbf{Proof of Theorem~C(ii)}.
\medskip
%%%%%%%%%%%%%%%%%%%%%%%%%%%%%%%%

We now consider the case $\arg z = -\pi$ and write $z=r e^{-i\pi}$
with $r>0$.
The proof of Proposition~\ref{propleft} shows that
$\DD_{I^-}\cap(e^{-2\pi i}\DD_{I^+})$ contains $e^{-i\pi}(\al_\pi,+\infty)$.
For any $(\si_1,\si_2)\in\C^2$ with
$|\si_2| < \tf12 e^{-2\al_\pi}$ (so that
$\tf12\ln\tf{1}{|2\si_2|}>\al_\pi$),
the restriction of $G^-(\cdot\,,\si_1,\si_2)$
to $\DD^-(\si_2)\cap e^{-i\pi}\R_{>0}$ gives rise to the function
\begla
r\in (\al_\pi, \tf12\ln\tf{1}{|2\si_2|})
\mapsto
G^-(r e^{-i\pi},\si_1,\si_2).
\edla
In view of~\eqref{eq_cm}, this function is real-valued if and only if
\beglab{eqcondrealleft}
  G^-(r e^{-i\pi},\si_1,\si_2) = G^+(r e^{i\pi},\ov {\si_1},\ov {\si_2})
  \quad \text{for} \ens
  r\in (\al_\pi, \tf12\ln\tf{1}{|2\si_2|}).
  \edla
It follows from~\eqref{eq_left_inv} that, for $|\si_2|$ small enough, the \rhs\
of~\eqref{eqcondrealleft} is
$G^-(r e^{-i\pi},\ov{\si_1}+\log(1+i\ov{\si_2}),\f{\ov{\si_2}}{1+i\,\ov{\si_2}})$,
  thus realness is equivalent to
  \beglab{eqrealsi2}
  \si_1 = \ov{\si_1}+\log(1+i\ov{\si_2}), \quad
  \si_2 = \f{\ov{\si_2}}{1+i\,\ov{\si_2}},
  \quad \text{with}\ens |\si_2| =
  \left| \f{\ov{\si_2}}{1+i\,\ov{\si_2}} \right|
  < \tf12 e^{-2\al_\pi}.
  \edla
  The second equation in~\eqref{eqrealsi2} is equivalent to
  $|\si_2+i|^2=1$, so we write $\si_2 = -i(1-e^{i\th})$ with $\th\in\R$.
  %
  % \begla
  % %
  % \si_2 = -i(1-e^{i\th}) \quad \text{with}\ens \th \in \R.
  % %
  % \edla
  %
Since $|1+i\,\ov{\si_2}|=1$, the third condition in~\eqref{eqrealsi2}
is equivalent to $|\si_2|^2 < \f14 e^{-4\al_\pi}$,
or $\cos\th>1-\f18 e^{-4\al_\pi}$, we thus parametrize~$\si_2$ by
\begla
\si_2 = -i(1-e^{i\th}) \quad \text{with}\ens
\th \in (-\th_*,\th_*),
\quad \text{where} \ens
\th_* \defeq \arccos( 1-\tf18 e^{-4\al_\pi} ).
\edla
The first equation in~\eqref{eqrealsi2} is then equivalent to
$-2i\IM\si_1 = -\log (1+i\,\ov{\si_2}) = i\th$,
we thus must parametrize~$\si_1$ as
\begla
\si_1 = a - i\,\f\th2
\quad \text{with} \ens
a\in\R.
\edla

We see that, with these values of~$\si_1$ and~$\si_2$, the function
  $G^-(z,\si_1,\si_2)$ is analytic in $\{\, z\in\DD_{I^-} \mid \RE z>\f12\ln(2|1-e^{i\th}|)\,\}$;
  it coincides with $G^+(e^{2i\pi}z,\ov {\si_1},\ov {\si_2})$ and is
  thus also analytic in
  $\{\, z\in e^{-2\pi i}\DD_{I^+} \mid \RE
  z>\f12\ln(2|1-e^{i\th}|)\,\}$,
  and real-valued for $\arg z = -\pi$.

%%%%%%%%%%%%%%%
%%%%%%%%%%%%%%%
\iffalse
%%%%%%%%%%%%%%%
%%%%%%%%%%%%%%%

For the direction $\arg z=-\pi$, as discussed in Picture
\ref{figdomainINTERSEC}, we need to restrict $|\si_2|$ to be small
enough so that the intersection is nonempty. Then similar to the proof
of Theorem C(i), further requirements are:
%
\begin{align}
  \ov{\si_1} &=\si_1+\log(1-i\si_2), \label{eq_re1} \\
  \ov{\si_2} &=\f{\si_2}{1-i\si_2}, \label{eq_re2}\\
  |\si_2|&<\f{1}{2}e^{\Re e(2z)},|\ov{\si_2}|<\f{1}{2}e^{\Re e(2z)}.\label{eq_re3}
\end{align}
{
The equation~\eqref{eq_re2} is amount to saying that $|\si_2+i|=1$, so we can set $\si_2+i=ie^{i\th}$ . Further computations about~\eqref{eq_re3} tell us that the requirement $\si_2\ov{\si_2}<\f{1}{4}e^{4\Re e(z)}$ is equivalent to $\cos \th>1-\f{1}{8} e^{4\Re e(z)}$ or
\begin{equation}\label{eq_re4}
-\arccos(1-\tfrac{1}{8} e^{4\Re e(z)})<\th<\arccos(1-\tfrac{1}{8} e^{4\Re e(z)}).
\end{equation}}
In conclusion, when $z\in \DD(I^-,\al)\cap (e^{-2\pi i}\DD(I^+,\al))\cap \R$, if $\si_1$ and $\si_2$ satisfy the following relations
\[\si_1=a +i\tfrac{\th}{2},~\si_2=-i(1-e^{i\th}),~~\textrm{for}~\textrm{all}~ a\in \R\]
with inequality~\eqref{eq_re4} holding, then the Borel sum $$G^+\big (z,a+i\tfrac{\th}{2},-i(1-e^{i\th})\big)=G^-\big(e^{2\pi i}z,a-i\tfrac{\th}{2},i(1-e^{-i\th})\big)$$ is real analytic.

%As a summary, we get the proof of Theorem C(i) and (ii).

%%%%%%%%%%%%%%%
%%%%%%%%%%%%%%%
\fi
%%%%%%%%%%%%%%%
%%%%%%%%%%%%%%%

\begin{Remark}\label{rationalDSL}
  Since rational coefficients are often linked with the enumeration of
  geometric objects, it would be interesting to spell out the
  enumerative meaning of the coefficients of each~$\ti G_k$ or of all
  of~$\ti G$ from the geometrical and non-perturbative topological
  string perspective. Even more interesting would be the understanding
  of the enumerative nature of the connection formula between $G^+$
  and~$G^-$.
\end{Remark}

%%%%%%%%%%%%%%%%%%%%%%%%%%%%%%%%%%%%%%%%%%%%%%
%%%%%%%%%%%%%%%%%%%%%%%%%%%%%%%%%%%%%%%%%%%%%%

%\newpage

\section{Transseries completion for the free energy in the large
  radius limit}\label{LRL}

%%%%%%%%%%%%%%%%%%%%%%%%%%%%%%%%%%%%%%%%%%%%%

In this section we will go from the resurgent properties that we have
proved for the formal series obtained via Alim-Yau-Zhou's double
scaling limit~(\cite{[AYZ]}) to those for the free energy as obtained
via Couso-Santamar\'{i}a's large radius limit, thus rigorously proving
several statements conjectured in~\cite{[RCS]}.
Before doing that, we have a remark on the double scaling process.

%%%%%%%%%%%%%%%%%%%%%%%%%%%%%%%%%%%%%%%%%%%%%

\subsection{{A remark on the interpretation of Alim-Yau-Zhou's
  parameter \texorpdfstring{$\varepsilon$}{ε}}} % in \texorpdfstring{\cite{[AYZ]}}{[2]}}}

To capture the terms $a_gC_{zzz}^{2g-2}(S^{zz})^{3g-3}$ of the
coefficients of the total free energy~$\FF$ in~\eqref{eqFCzzzSzz}, Alim-Yau-Zhou's
paper~\cite{[AYZ]} employs the following double scaling:
  \begin{equation}\label{ds2}
  g_s\mapsto \ve^{-1}g_s, \qquad S^{zz}\mapsto \varepsilon^{\frac{2}{3}} S^{zz},
\end{equation}
along with $\la_s^2=g_s^2C_{zzz}^2(S^{zz})^3$ and $\ve\to 0$
(cf.~\eqref{dsl}).
The indeterminate~$\la_s$ in the resulting free energy $\FFs(\la_s)$
of~\eqref{eqdefFFs} is thus essentially $g_s C_{zzz} (S^{zz})^{3/2}$
and the small parameter~$\ve$ is a device used to capture the terms
that are dominant when the nonholomorphic propagator~$S^{zz}$ is
large.
In Couso-Santamar\'{i}a's large radius limit process~(\cite{[RCS]}),
according to~\eqref{eqFFzdSi} only one variable is
rescaled:
\begin{equation}\label{lol}
  S^{zz}=z^2\Si,
\end{equation}
and then one takes the limit $z\to0$, \ie in the $z$-space one goes to the large radius
point, where the Yukawa coupling~$C_{zzz}$ is singular.

The comparison between~\eqref{ds2} and~\eqref{lol} prompted the author
of~\cite{[RCS]} to propose the relation $\varepsilon = z^{-3}$,
however this is quite misleading:
on the one hand it yields a contradiction since $\varepsilon$ and~$z$
could not go simultaneously to~$0$,
on the other hand the large radius limit $z\to0$ is only half of the
process to get the~$a_g$'s, one must still set
$g_s = C_{zzz}^{-1} \,\Si^{-3/2} \la_s$ and send~$\Si$ to~$\infty$.
A better explanation is that Alim-Yau-Zhou's double scaling limit can
be slightly generalized to
\begin{equation}\label{generalizdsl}
g_s = \varepsilon \,C_{zzz}^{\al-1} \,\Si^{-3/2} \la_s,
    \qquad S^{zz}= \varepsilon^{-\frac{2}{3}}\, C_{zzz}^{-2\al/3} \,\Si
\end{equation}
with arbitrary~$\al$ (instead of just taking $\al=0$), which leads to
\begin{equation}   \label{eqnewrelvareps}
  \varepsilon = z^{-3}C_{zzz}^{-\al}.
\end{equation}
In view of the ``large radius'' feature $C_{zzz} \isequiv{z\to0} \ka z^{-3}$ used
by Couso-Santamar\'{i}a, this new relation results in
\begin{equation}
  \varepsilon \isequiv{z\to0} \ka^{-\al} z^{3(\al-1)},
\end{equation}
which is meaningful for any $\al>1$.

%%%%%%%%%%%%%%%%%%%%%%%%%%%%%%%%%%%%%%%%%%%%%%%%%%%%%%%%

\subsection{From the double scaling limit to the large radius limit}

%%%%%%%%%%%%%%%%%%%%%%%%%%%%%%%%%%%%%%%%%%%%%%%%%%%%%%%%

In Section~\ref{resurgentfreeenergy}, 
we have discussed the resurgent properties of
the total free energy $\FFs(\la_s)$ of~\eqref{eqdefFFs} and its
transseries completion
\[
\GG(\la_s,\si_1,\si_2) = \ti G\big(\tfrac{1}{3\la_s^2}, \si_1,\si_2\big)
\]
 of~\eqref{eqreltiGnGGn},
 solutions to the nonlinear ODE deduced from HAE
 via Alim-Yau-Zhou's double scaling limit~(\cite{[AYZ]}) 
 \begin{equation}
\tag{\ref{eq1}}
  \th^2_{\la_s}\FF+(\th_{\la_s}\FF)^2+2\left(1-\f{2}{3\la_s^2}\right)\th_{\la_s}\FF+\f{5}{9}=0,
  \qquad \th_{\la_s}\defeq\la_s\f{\pt}{\pt \la_s}.
\end{equation}
Recall that, in Couso-Santamar\'{i}a's large radius limit process~(\cite{[RCS]}), HAE leads to the $u$-equation instead:
\begin{equation}
  \tag{\ref{uE}}
\partial_u H-\frac{3}{2} g_s^2 u^3\left( \partial_u
  H+\frac{u}{3}\partial_u^2 H+\frac{u}{3}(\partial_u
  H)^2\right)=\frac{1}{2u}+\frac{1}{u^2}.
\end{equation}

Our goal is now to discuss the resurgent properties of
Couso-Santamar\'{i}a's large radius limit free energy $H\zu(g_s,u)$
of~\eqref{eqdefHzugsu} and to employ alien calculus to derive
 the transseries completion
$H\uu(g_s,u,\sigma)$ of~\eqref{eqdefHuugsusi} or, equivalently,
\beglab{eqHuugiveHHu}
\HHu(g_s,u,\si_1,\si_2) = \si_1+H\uu(g_s,u,\si_2),
\edla
which will be the two-parameter transseries
solution to~\eqref{uE}.

The key observation is that the change of variable
\begin{equation}   \label{eqdefphigs}
  \la_s = \phi_{g_s}\!(u) \defeq
  \left(\f{g_s^2u^3}{(1-2g_s^2u^2)^{3/2}}\right)^{1/2}
  \quad \text{(for any parameter $g_s\in\C^*$)}
\end{equation}
empirically discovered in~\cite{[RCS]} (see especially \cite[eqn.~(48)]{[RCS]}) allows one to directly go
from~\eqref{eq1} to~\eqref{uE}, up to adding an elementary function
of~$u$ and~$g_s$.

\begin{Proposition}\label{proFFtoH}
  For any parameter $g_s\in\C^*$, the change of variable and unknown
  \beglab{eq_tu}
    H(u) =\FF\big(\phi_{g_s}\!(u)\big) + R(g_s,u)
    %
    % H(u) =\FF\big(\phi_{g_s}\!(u)\big)+H\zu\conv(g_s,u)+H\zu_1(u)
    %
    \edla
    with~$\phi_{g_s}$ as in~\eqref{eqdefphigs} and
    \beglab{eqdefRgsu}
    R(g_s,u) \defeq \f14\log\Big(\f{u^2}{1-2g_s^2u^2}\Big)
    + \f{(1-2g_s^2u^2)^{3/2}-1}{3g_s^2u^3}    
  \edla
  % %
  % \begin{align}
  %   %
  %   H\zu\conv(g_s,u) & := \f{(1-2g_s^2u^2)^{3/2}-1+3g_s^2u^2}{3g_s^2u^3}-\f{1}{4}\log(1-2g_s^2u^2),\\
  %   %
  %   H\zu_1(u) & :=-\f1u+\f12\log u,
  %   %
  % \end{align}
  %
  makes the two nonlinear ODEs~\eqref{eq1} and~\eqref{uE} equivalent.
\end{Proposition}

\begin{proof}
  As observed at the beginning of Section~\ref{resurgentfreeenergy},
  the change of variable and unknown
  $\FF(\la_s) = g(\f1{3\la_s^2})$
  transforms~\eqref{eq1} into
  \begin{equation} \tag{\ref{eq_ageq2}}
    g''+(g')^2+2g'+\f{5}{36}z_1^{-2}=0,
  \end{equation}
  where we now call $z_1 = \f1{3\la_s^2}$ (instead of~$z$ as in
  Section~\ref{resurgentfreeenergy}) the variable with respect to
  which the unknown $g=g(z_1)$ is expressed.
  Therefore, one just needs to check that the change of variable and unknown
  \begin{align}
\label{keyvarchange}
  z_1 = \f1{3\la_s^2} &= \f1{3\phi_{g_s}\!(u)^2}
  = \f1{3g_s^2u^3}(1-2g_s^2u^2)^{3/2}, \\[1ex]
  H(u) &= g(z_1) + R(g_s,u)
  %
%   \f14\log\Big(\f{u^2}{1-2g_s^2u^2}\Big)
% %
%   + \f{(1-2g_s^2u^2)^{3/2}-1}{3g_s^2u^3}
%
  \end{align}
  makes~\eqref{eq_ageq2} and~\eqref{uE} equivalent.
    This computation is left to the reader.
\end{proof}

\begin{Proposition}   \label{propHzu}
  The large radius perturbative series~$H\zu$ defined in~\cite{[RCS]} is
  \beglab{eqrelHHsHzu}
  H\zu(g_s,u) = R(g_s,u) + \FFs\big( \phi_{g_s}\!(u) \big),
    %
    % \f14\log\Big(\f{u^2}{1-2g_s^2u^2}\Big)
    % %
    % + \f{(1-2g_s^2u^2)^{3/2}-1}{3g_s^2u^3},
  %
  \edla
  where $R(g_s,u)$ is defined by~\eqref{eqdefRgsu} (or rather its
  Taylor expansion \wrt~$g_s^2$) and the second term
  of the \rhs\ is understood as the substitution of the convergent
  series
  \beglab{eqsubstlassqphigsusq}
  \la_s^2 = \phi_{g_s}\!(u)^2 = \sum_{\ell\gqw1}
  \Big( \f{(2\ell-1)!}{2^{\ell-1}(\ell-1)!^2}u^{2\ell+1} \Big)
  g_s^{2\ell}
  \edla
  in the formal power series $\FFs(\la_s) = \sum_{\ell\gqw1} a_{\ell+1} \la_s^{2\ell}$ of~\eqref{eqdefFFs}.
Among the formal series in~$g_s^2$ with $u$-dependent coefficients,
  the solutions to~\eqref{uE} are the formal series
  \begla
  C(g_s)+H\zu(g_s,u)
  \quad \text{with arbitrary formal series}\ens 
  C(g_s)\in \C[[g_s]].
  \edla
\end{Proposition}

From the perspective of perturbative free energy, we thus have a direct relationship
\begin{center}
\usetikzlibrary {shapes.geometric}
\begin{tikzpicture}[fill=blue!20]
\path (-3,0) node(a) [rectangle,rotate=0,draw,fill] {$\FFs(\la_s)$}
(3,0) node(b) [rectangle,rotate=0,draw,fill] {$H\zu(g_s,u)$};
\draw[thick,->] (a) -- (b)node[above,midway]{~};
\end{tikzpicture}
\end{center}
%
% Couso-Santamar\'{i}a (\cite{[RCS]}) argued that one can promote the
% double scaling limit to another one: the large radius limit which is
% also expected to be universal nature.
% In other words, the paper provides a coordinate substitution in $\FFs(\la_s)$
% \begin{equation}\label{keyvarchange}
%   %
%   \blue{ \la_s^2=\f{g_s^2u^3}{(1-2g_s^2u^2)^{3/2}}, }
%   %
%   \qquad
%   \phi_{g_s}\!(u)\defeq\left(\f{g_s^2u^3}{(1-2g_s^2u^2)^{3/2}}\right)^{1/2}
% \end{equation}
%
%
Note that, as mentioned in~\eqref{eqdefHzugsu}, the formula for~$H\zu$
contains a contribution of genus $g=\ell+1=1$, \ie a constant term in~$g_s^2$
(hinted at in~\cite{[RCS]}),
since
\beglab{eqconstanttermRgsu}
R(g_s,u)
%
% \f14\log\Big(\f{u^2}{1-2g_s^2u^2}\Big)
% %
% + \f{(1-2g_s^2u^2)^{3/2}-1}{3g_s^2u^3}
%
= -\f1u+\f12\log u + H\zu\conv(g_s,u)
\quad \text{with $H\zu\conv(g_s,u) = O(g_s^2)$,}
\edla
but by convention the expansion stemming
from~$\FFs$ starts from genus $g=\ell+1=2$ (cf.~\eqref{eqdefFFs}).

% \begin{equation}\label{eq_tu}
% H\zu(g_s,u) =\FFs(\phi_{g_s}\!(u))+H\zu\conv(g_s,u)+H\zu_1(u)
% \end{equation}
% where
% \begin{align}
% H\zu\conv(g_s,u)&=\f{(1-2g_s^2u^2)^{3/2}-1+3g_s^2u^2}{3g_s^2u^3}-\f{1}{4}\log(1-2g_s^2u^2),\\
% \pt_u H\zu_1&=\f{1}{2u}+\f{1}{u^2}.
% \end{align}

\begin{proof}[Proof of Proposition~\ref{propHzu}]
  If one starts with arbitrary $\FF(\la_s)\in\C[[\la_s^2]]$, performs
  the substitution~\eqref{eqsubstlassqphigsusq} and adds $R(g_s,u)$,
  \begin{center}
\usetikzlibrary {shapes.geometric}
\begin{tikzpicture}[fill=blue!20]
\path (-3,0) node(a) [rectangle,rotate=0,draw,fill] {$\FF(\la_s)\in\C[[\la_s^2]]$}
(3,0) node(b) [rectangle,rotate=0,draw,fill] {$H(g_s,u) = R(g_s,u)+\FF\big( \phi_{g_s}\!(u) \big)$};
\draw[thick,->] (a) -- (b)node[above,midway]{~};
\end{tikzpicture}
\end{center}
then the result is a series $H(g_s,u)$ in the
indeterminate~$g_s^2$ with $u$-dependent coefficients.
The computation outlined in the proof of
Proposition~\ref{proFFtoH} shows that if the initial series
$\FF(\la_s)$ solves~\eqref{eq1}, then the resulting series solves~\eqref{uE}.
In particular, the \rhs\ of~\eqref{eqrelHHsHzu} is a formal solution to~\eqref{uE}.

On the other hand, when plugging an arbitrary formal series with
$u$-dependent coefficients
$H = \sum_{k\gqw0} H_k(u) g_s^{2k}$ into~\eqref{uE},
it is easy to see that each term~$H_k$ is determined by the previous
ones up to the addition of an arbitrary complex constant,
thus 
  \begla
  H = C(g_s)+R(g_s,u)+\FFs\big( \phi_{g_s}\!(u) \big)
  \ens \text{with arbitrary}\ens 
  C(g_s)\in \C[[g_s]].
  \edla

  To conclude the proof, we just need to check that among these
  solutions, $H\zu(g_s,u)$ is the one corresponding to the choice
  $C(g_s)\equiv0$.
  To that end, we observe that
  $R(g_s,u)+\FFs\big( \phi_{g_s}\!(u) \big)$ is a formal series
  in~$g_s^2$ all of whose coefficients are polynomials in~$u$ that
  vanish at $u=0$, with the only exception of the constant term
  in~$g_s^2$ that stems from~\eqref{eqconstanttermRgsu}.
  But \cite[eqn.~(47)]{[RCS]} shows that the
  coefficient of $g_s^{2(g-1)}$ in $H\zu(g_s,u)$ must vanish when
  $u=0$ for each $g\gqw2$;
  this requirement shows that $C(g_s)\equiv0$ is the only possibility
  (note that the constant term \wrt~$g_s$ in $H\zu(g_s,u)$,
  corresponding to $g=1$, is a function of~$u$ that is only determined
  up to an additive constant and our choice is only a matter of
  convention).
\end{proof}

%%%%%%%%%%%%%%%%%%%%%%%%%%%%%%%%%%%%%%%%%%%%%%%%
%%%%%%%%%%%%%%%%%%%%%%%%%%%%%%%%%%%%%%%%%%%%%%%%

\subsection{{Resurgent structure of the transseries completion}}   \label{secresurLR}

%%%%%%%%%%%%%%%%%%%%%%%%%%%%%%%%%%%%%%%%%%%%%%%%
%%%%%%%%%%%%%%%%%%%%%%%%%%%%%%%%%%%%%%%%%%%%%%%%

%\red{EXPRESS AT LEAST ONCE THE RESURGENCE VARIABLE~$z$ IN TERMS
  %OF~$g_s$}

%Announced in Section~\ref{secannumthreeZresur}:
%
We know from Section~\ref{resurgentfreeenergy} that $\FFs(\la_s)$ is
resurgent in $z_1 = \f1{3\la_s^2} = \f1{3\phi_{g_s}\!(u)^2}$.
Now, the core of the above relation~\eqref{eqrelHHsHzu} can be viewed
as a tangent-to-identity change of variable \wrt\ the variable
$z_2=\f{1}{3g_s^2u^3}$,
in the sense that~\eqref{keyvarchange} can be rephrased as
\beglab{eqdefphu}
z_1 = z_2 \Big( 1 - \f2{3u} z_2\ii \Big)^{3/2} = z_2 + \ph_u(z_2),
\qquad
\ph_u(z_2) \in \C\{z_2\ii\},
\edla
where $u\in\C^*$ is now treated as a parameter.
We can thus obtain the resurgence in~$z_2$ of $H\zu(g_s,u)$ from
general resurgence theory,
and alien calculus then produces a transseries completion that formally
solves the $u$-equation~\eqref{uE}:

\begin{thmA1}
  \emph{(i)}   The large radius limit free energy $H\zu(g_s,u)$
  in~\eqref{eqdefHzugsu} with $u\in\C^\ast$
treated as parameter is a divergent simple $2\Z$-resurgent series \wrt\ the variable
$z_2=\f{1}{3g^2_su^3}$,
and thus a divergent simple $\f2{3 u^{3}}\Z$-resurgent series in the variable~$\f{1}{g_s^{2}}$.

  \smallskip

  \emph{(ii)} On $H\zu(g_s,u)$ viewed as a resurgent series
  in~$z_2$, the actions of the algebra automorphisms
  $\exp(\si e^{-2z_2}\De_2)$ and
  $(\DDE_{\R_{\gqw 0}}^+)^\si=\exp(\si \DDE_{\R_{\gqw 0}})$
  of $\ti\RT_{2\Z}[[\si,e^{-2z_2}]]$ coincide,
  and $H\zu(g_s,u)$ can be embedded in a two-parameter transseries
\begin{equation}  \label{eq_uts}
  \HHu(g_s,u,\si_1,\si_2) = % \ti H(z_2,u,\si_1,\si_2)\\ =
  \si_1+H\zu(g_s,u)+\sum_{n\gqw1} \si_2^n e^{-\f{2n}{3g_s^2u^3}}H\nuu(g_s,u)
\end{equation}
  solution to~\eqref{uE} defined by
  \beglab{eqdefHHu}
  \HHu \defeq \big( \DDE_{\R_{\gqw 0}}^+ \big)^{-i\si_2}\big[\si_1+ H\zu(g_s,u)\big].
  \edla
The transseries~$\HHu$ is related to the transseries~$\ti G$ defined
by~\eqref{eqGzsisi} by 
\beglab{eqrelHHutiG}
\HHu(g_s,u,\si_1,\si_2) =
\ti G\big(z_2 + \ph_u(z_2),\si_1,-\si_2\big) + R(g_s,u)
\quad \text{with} \ens z_2=\f{1}{3g^2_su^3}.
\edla

  \smallskip

  \emph{(iii)}
  One also has, in terms of the transseries $\GG(\la_s,\si_1,\si_2)$ of~\eqref{tssss}
  (formal integral of the double scaling limit HAE, as in Theorem~A(ii)),
\beglab{eqHHuGG}
\HHu(g_s,u,\si_1,\si_2)
 = \GG\big(\phi_{g_s}\!(u),\si_1,-\si_2) + R(g_s,u).
\edla

  \smallskip

  \emph{(iv)}
For each $n\gqw1$, the $n$th component of the
transseries~\eqref{eq_uts} involves
\begin{equation}\label{eq_tun}
H\nuu(g_s,u)=-\f{1}{n}e^{\f{2n}{u}}+\sls_{g=1}^\infty g_s^{2g}
e^{\f{2n}{u}}u^g Pol_n(u,2g),
\end{equation}
which is a simple $2\Z$-resurgent series in~$z_2$ where, for
each $g\gqw 1$, $Pol_n(u,2g)$ is a polynomial in~$u$
of degree~$2g$ with rational coefficients.
\end{thmA1}

\begin{Remark}
(1) Here, for sake of simplicity, we stick to $2\Z$-resurgence in the
variable $z_2=\f{1}{3g^2_su^3}$ for fixed $u\in\C^*$
and the operator
$\DDE_{\R_{\gqw 0}}^+ =  \exp\!\big( \sum_{k=1}^{\infty}
  e^{-2kz}\De_{2k} \big)$
acts on the corresponding algebra of transseries.
This is trivially equivalent to
%
% $2 u^{-3}\Z$-resurgence in the variable $z=\f{1}{3}g_s^{-2}$, or
%
$\f2{3 u^{3}}\Z$-resurgence in the variable~$\f{1}{g_s^{2}}$
and can be viewed as an elementary instance of duality between
equational resurgence and parametric resurgence first introduced
in~\cite{JEapplic}.
%
% we will give the action of the corresponding alien
% derivations $\De_{2ku^{-3}}$ in Remark~\ref{remEquivResVarzztwo}.

(2) The coefficients in Theorem~A'(iv) are rational, a phenomenon
parallel to Lemma~\ref{lemrqtionalGk}.
\end{Remark}

Note that we choose to use $-i\si_2$ as power in~\eqref{eqdefHHu} rather
than $i\si_2$ as in~\eqref{eqGzsisi} (and consequently need to
change~$\si_2$ into~$-\si_2$ when going from~$\ti G$
to~$\HHu$ in~\eqref{eqrelHHutiG}) just to align with the corresponding
formulas in~\cite{[RCS]}.

\begin{proof}[Proof of Theorem~A']
  (i)
  For the sake of clarity we use different notation for the series
  according as they are expressed in the variable~$g_s^2$ or in the
  variable~$z_2$:
  \beglab{eqdeftiRu}
  \ti H\zu(z_2,u) \defeq H\zu(g_s,u) \in\C[[z_2\ii]], \quad
  \ti R_u(z_2) \defeq R(g_s,u) \in \C\{z_2\ii\} 
   \edla
  with $z_2 = \f{1}{3g^2_su^3}$, where $u\in\C^*$ is treated as a fixed parameter.
  In view of~\eqref{eqreltigFFs},
  the first statement of Proposition~\ref{propHzu} thus amounts to
  \beglab{eqtiHztig}
  \ti H\zu(z_2,u) = \ti g\big(z_2 + \ph_u(z_2)\big) + \ti R_u(z_2),
  \edla
  where~$\ph_u$ stems from~\eqref{eqdefphu}:
  \beglab{eqphuTaylor}
  \ph_u(z_2) = z_2\big( (1-\tfrac{2}{3u}z_2^{-1})^{3/2} - 1 \big) =
  \sum_{n=1}^\infty\binom{\f{3}{2}}{n}\Big(\!-\f{2}{3u}\Big)^{n}
  z_2^{-(n-1)}\in \C\{z_2^{-1}\}.
  \edla
  In~\eqref{eqtiHztig} we have
  $\ti R_u, \ph_u \in \C\{z_2\ii\} \subset \Td\RT_{2\Z}^\simp$ and,
  according to Proposition~\ref{th_omrs}, $\ti g\in
  \Td\RT_{2\Z}^\simp$.
  We can thus apply Theorem~\ref{thmad} to $\ti H\zu = \ti g\circ(id +
  \ph_u) + \ti R_u$:
  according to~\eqref{nolinearoperator}, $\ti H\zu \in
  \Td\RT_{2\Z}^\simp$.
  Moreover, $\ti H\zu$ is divergent because~$\ti g$ is divergent.
  \medskip
  
  (ii)
  Theorem~\ref{thmad} also entails, according to~\eqref{eq_nlad1},
  that
  \begla
  \De_\om \ti H\zu = e^{-\om\ph_u}\cdot (\De_\om\ti g)\circ(id+\ph_u)
  \quad
  \text{for every $\om\in 2\Z^*$}
\edla
(since $\ph_u$ is convergent and thus $\De_\om\ph_u=0$) and,
consequently,
\begla
  \big(\DDE_{\R_{\gqw 0}}^+\big)^\si \ti H\zu = \big(\DDE_{\R_{\gqw 0}}^+\big)^\si\ti g\circ(id+\ph_u)
  + \ti R_u
  \quad \text{for any $\si\in\C$.}
\edla
Point~(ii) of Theorem~A' thus follows from Proposition~\ref{propkey0}, which says
that
  \begla
  \exp(i\si_2 e^{-2z_1} \De_2)(\si_1+\ti g) =
  \big(\DDE_{\R_{\gqw 0}}^+\big)^{i\si_2}(\si_1+\ti g) =
  \ti G(z_1,\si_1,\si_2).  
  \edla
  
In particular, we get simple $2\Z$-resurgent series
$\ti H\nuu(z_2,u)=H\nuu(g_s,u)$, $n\gqw1$, as components of the
transseries
\beglab{eqHHuG}
\ti\HHu_{\mid (\si_1,\si_2)}
\defeq \big( \DDE_{\R_{\gqw 0}}^+ \big)^{-i\si_2}\big[\si_1+ \ti H\zu\big]
 = \ti G\circ(id+\ph_u)_{\mid
  (\si_1,-\si_2)} + \ti R_u
\edla
(where the notation $K_{\mid (\si_1,\pm\si_2)}$ indicates that the
arguments $(\si_1,\si_2)$ of~$K$ must be replaced with $(\si_1,\pm\si_2)$).
Since neither~$\ph_u$ nor~$\ti R_u$ depend on the transseries
parameter~$\si_2$ (only~$\ti G$ does), the coefficient of $\si_2^n$
in~\eqref{eqHHuG} is
\beglab{eqHnuuGnztwo}
e^{-2n z_2} \ti H\nuu(z_2,u) =
(-1)^n e^{-2n (z_2+\ph_u(z_2))} \ti G_n\big(z_2+\ph_u(z_2)\big)
\quad \text{for each $n\gqw1$,}
\edla
where the components~$\ti G_n$ of~$\ti G$ are the simple
$2\Z$-resurgent series of Proposition~\ref{propkey0}.

\medskip

(iii)
We now return to the variable~$g_s^2$ and focus on the coefficients of
the expansion of $H\nuu$ in powers of
this indeterminate:
\begla
H\nuu(g_s,u) = \sum_{g\gqw0} H\nuu_g(u) \, g_s^{2g}.
\edla
We first rephrase~\eqref{eqHHuG} % \eqref{eqHnuuGnztwo}
by using the transseries~$\GG$ of~\eqref{tssss}
and~\eqref{eqreltiGnGGn} % series~$\GG_n$
\[
  \GG(\la_s,\si_1,\si_2) = \ti G(z_1 = \tfrac{1}{3\la_s^2}, \si_1,\si_2)
      = \si_1 + \sum_{n\gqw0} \si_2^n \, e^{-\f{2n}{3\la_s^{2}}} \, \GG_n(\la_s)
      %
      % \quad \text{with}\ens z_1 = \tfrac{1}{3\la_s^2},
      %
\]
      where, according to Lemma~\ref{lemrqtionalGk} and~\eqref{eqidentifytiGm},
\begla
\GG_n(\la_s) = \sum_{k\gqw0} \GG_{n,k} \la_s^{2k} \in \Q[[\la_s^2]]
\quad \text{with}\ens \GG_{n,0}=\f{(-1)^{n-1}}n.
\edla
When returning to the indeterminate~$g_s^2$, we must replace the
change of variable $z_2=(id+\ph_u)(z_1)$ by the change of variable
$\la_s^2 = \phi_{g_s}\!(u)^2$ and~\eqref{eqHHuG} thus becomes~\eqref{eqHHuGG}.

\medskip

(iv)
We just obtained
\begla
\HHu(g_s,u,\si_1,\si_2) = \si_1 + \sum_{n\gqw0} (-\si_2)^n
e^{-\f{2n}{3\phi_{g_s}\!(u)^{2}}} \GG_n\big( \phi_{g_s}\!(u) \big) + R(g_s,u).
\edla
When extracting the coefficient of $\si_2^n$ for any $n\gqw1$ in this relation, we must
take care of the discrepancy between
$e^{-\f{2n}{3\phi_{g_s}\!(u)^{2}}}$ and $e^{-\f{2n}{3g_s^2u^3}}$.
Since
\begin{align*}
\f1{3\phi_{g_s}\!(u)^{2}} &- \f1{3g_s^2u^3} = -\f1u + g_s^2 u \,c_-\!(g_s^2u^2)
&&\text{with}
&&c_-\!(t) \in \Q[[t]],\\[1.5ex]
\phi_{g_s}\!(u)^2 &= g_s^2u^3 \big( 1 + g_s^2 u^2 c_+\!(g_s^2u^2) \big)
&&\text{with}
&&c_+\!(t) \in \Q[[t]],
\end{align*}
we get
\begin{align*}
H\nuu &=
(-1)^n \, e^{-2n(-\f1u + g_s^2u\,c_-\!)} \, \GG_n\big( \phi_{g_s}\!(u)^2 \big)
\\[1ex] &= e^{\f{2n}u} \Big( \sum_{\ell\gqw0} \f{(-2n)^\ell}{\ell!}
          g_s^{2\ell} u^\ell c_-^\ell \Big)
          \Big( \sum_{k\gqw0} (-1)^n\GG_{n,k} \,g_s^{2k} u^{3k} (1 +
          g_s^2 u^2 c_+\!)^k \Big)
  \\[1ex] &= e^{\f{2n}u} \Big( -\f1n + \sum_{g\gqw1} g_s^{2g}
            \sum_{\ell+k+r=g} \f{(-2n)^\ell}{\ell!} (-1)^n\GG_{n,k}
            \,c_{\ell,k,r} \, u^{\ell+2r+3k} \Big),
\end{align*}
where the rational coefficients $c_{\ell,k,r}$ are defined by the
generating series
\begla
c_-\!(t)^\ell \big(1 + t c_+\!(t)\big)^k = \sum_{r\gqw0} c_{\ell,k,r} t^r.
\edla
This matches the description of~$H\nuu$ announced in~\eqref{eq_tun}.
\end{proof}

\begin{Remark}
We recover the same family of polynomials with rational coefficients
as in Couso-Santamar\'{i}a's article.
For instance, for the first two nontrivial poynomials associated with
$n=1$, our computations give
\begla
Pol_1(u,2) = \f{5u^2}{12}+1,
\quad
Pol_1(u,4) =
  -\f{25u^4}{288}+\f{5u^3}{4}-\f{5u^2}{12}+\f{u}{3}-\f{1}{2},
  \edla
  in acccordance with \cite[eqns.~(55)-(56)]{[RCS]}.
  \end{Remark}

%\begin{Remark}
  %
  % \emph{(1)}
  %
  % When calculating the transseries solution of equation~\eqref{eq_tu},
  % we take $u$ to be a parameter with no need to require $1-2\tau_s/u$
  % to be constant as in \cite{[RCS]}.
%
% \emph{(2)}
%
% In Theorem A'(ii), we calculate the action of the operator
% $(\DDE_{\R_{\gqw 0}}^+)^{-i\si}$ to align with the corresponding
% formulas in \cite{[RCS]}, while in Theorem A(ii), we compute the
% action of the operator $(\DDE_{\R_{\gqw 0}}^+)^{i\si}$ in line of
% Section \ref{bridgetransseries}. The only distinction between them
% lies in a sign discrepancy as indicated by their superscripts, which
% is only a matter of convention.
%
%\end{Remark}

  % In \S \ref{bridgetransseries}, we have already observed that the
  % bridge equation provided us the Stokes automorphisms, and it is the
  % same situation here.  The resurgent property of the series
  % $\Td{g}\circ (z_2+\Td{a}(z_2,u))$, as demonstrated in the proof,
  % plays a pivotal role. Moreover, we can extend our analysis in \S 3.2

  \medskip
  
We now establish the Bridge Equation and compute the Stokes phenomena
for $\HHu(g_s,u,\si_1,\si_2)$, our transseries solution to~\eqref{uE}.

\begin{thmA2}
\emph{(i)} With respect to the resurgence variable $z_2=\f{1}{3g_s^2u^3}$, we have
$\De_\om\HHu=0$ for all $\om\in 2\Z^*\setminus\{-2,2\}$, and
\begin{align}
  \De_2 \HHu(g_s,u,\si_1,\si_2) &= i \,e^{2z_2}\f{\pt}{\pt \si_2} \HHu(g_s,u,\si_1,\si_2)\label{bridge1}  \\
  \De_{-2}\HHu(g_s,u,\si_1,\si_2) &=
       i\, e^{-2z_2}\Big(\si_2\f{\pt}{\pt\si_1}\HHu(g_s,u,\si_1,\si_2)
        -\si_2^2\f{\pt}{\pt\si_2}\HHu(g_s,u,\si_1,\si_2)\Big).\label{bridge2}
\end{align}

\smallskip

\emph{(ii)} The action of the symbolic Stokes automorphism on~$\HHu$
is given by
\begin{align}
  \DDE_{\R_{\gqw 0}}^+ \HHu(g_s,u,\si_1,\si_2) &= \HHu(g_s,u,\si_1,\si_2+i)\label{stk7},  \\
  \DDE_{\R_{\lqw 0}}^+ \HHu(g_s,u,\si_1,\si_2) &=
                                                 \HHu(g_s,u,\si_1+\log(1+i\si_2),\f{\si_2}{1+i\si_2}\big)\label{stk8}.
\end{align}
\end{thmA2}

\begin{proof}
  (i)
  Treating $u\in\C^*$ as a parameter and switching to $\ti\HHu(z_2,u,\si_1,\si_2)$
  explicitly viewed as a transseries in the variable
  $z_2=\f{1}{3g_s^2u^3}$, we have seen in~\eqref{eqHHuG} that
  \begla
\ti\HHu_{\mid (\si_1,\si_2)}
 = \ti G_{\mid (\si_1,-\si_2)}\circ(id+\ph_u) + \ti R_u.
\edla
The Alien Calculus rule~\eqref{eq_nlad1} thus yields
  \begla
\De_\om\ti\HHu_{\mid (\si_1,\si_2)}
 = e^{-\om\ph_u} (\De_\om\ti G)_{\mid (\si_1,-\si_2)}\circ(id+\ph_u) + \ti R_u.
  \edla
Plugging there the formula for $\De_\om\ti G$ obtained in
Proposition~\ref{propbridge}, we get
$\De_\om\ti\HHu=0$ for all $\om\in 2\Z^*\setminus\{-2,2\}$,
\begla
\De_2\ti\HHu =- i \, e^{-2\ph_u} \, e^{2(z_2+\ph_u)} \Big(\f{\pt}{\pt \si_2} \ti G\Big)_{\mid
  (\si_1,-\si_2)}\circ(id+\ph_u)
= i \, e^{2z_2} \f{\pt}{\pt \si_2} \ti\HHu
\edla
and
\begin{multline}
  \De_{-2}\ti\HHu =- i \, e^{2\ph_u} \, e^{-2(z_2+\ph_u)}
  \Big(\si_2\f{\pt}{\pt \si_1}\ti G-\si_2^2\f{\pt}{\pt\si_2}\ti G\Big)_{\mid
  (\si_1,-\si_2)}\circ(id+\ph_u) \\[1ex]
= - i \, e^{-2z_2} \Big(-\si_2\f{\pt}{\pt \si_1}\ti\HHu+\si_2^2\f{\pt}{\pt\si_2}\ti\HHu\Big).
\end{multline}

\medskip

(ii)
Similarly, since $\ph_u(z_2)\in\C\{z_2\ii\}$, Alien Calculus yields
\begla
\DDE_d^+ \ti\HHu_{\mid (\si_1,\si_2)} =
\big(\DDE_d^+ \ti G\big)_{\mid (\si_1,-\si_2)}\circ(id+\ph_u) + \ti R_u
\edla
for $d=\R_{\lqw 0}$ or~$\R_{\gqw 0}$, where the latter case requires
the same care as in Section~\ref{secsymbolSt}.2 (see especially
Lemma~\ref{lemcAsubalgforDDElqw}).
In view of~\eqref{stk3}, we get
\begin{multline*}
\DDE_{\R_{\gqw 0}}^+\ti\HHu =
\big(\DDE_{\R_{\gqw 0}}^+\ti G\big)_{\mid (\si_1,-\si_2)}\circ(id+\ph_u) + \ti R_u
= \ti G\big(z_2+\ph_u,\si_1,\si_2-i\big)_{\mid (\si_1,-\si_2)}+ \ti
R_u \\[1ex]
=\ti G\big(z_2+\ph_u,\si_1,-\si_2-i\big)+ \ti R_u
= \ti\HHu(z_2,\si_1,\si_2+i),
\end{multline*}
while~\eqref{stk4} yields
\begin{multline*}
\DDE_{\R_{\lqw 0}}^+\ti\HHu =
\big(\DDE_{\R_{\lqw 0}}^+\ti G\big)_{\mid
  (\si_1,-\si_2)}\circ(id+\ph_u) + \ti R_u \\[1ex]
= \ti G\big(z_2+\ph_u, \si_1+\log(1-i\si_2),\f{\si_2}{1-i\si_2}\big)_{\mid (\si_1,-\si_2)}+ \ti
R_u \\[1ex]
=\ti G\big(z_2+\ph_u,\si_1+\log(1+i\si_2),-\f{\si_2}{1+i\si_2}\big)+ \ti R_u
= \ti\HHu(z_2,\si_1+\log(1+i\si_2),\f{\si_2}{1+i\si_2}).
\end{multline*}
\end{proof}

%%%%%%%%%%%%%%%%%%%%%%%%%%%%%%%%%%%%%%%%%%%%%%%%
%%%%%%%%%%%%%%%%%%%%%%%%%%%%%%%%%%%%%%%%%%%%%%%%

\subsection{Summability of the large radius expansions, real analytic
  solutions and rationality of coefficients}

 \label{realLR}

%%%%%%%%%%%%%%%%%%%%%%%%%%%%%%%%%%%%%%%%%%%%%%%%
%%%%%%%%%%%%%%%%%%%%%%%%%%%%%%%%%%%%%%%%%%%%%%%%

% \begin{equation}\label{trs3}
% \HHu(z_2,u,\si_1,\si_2)=H(z_2,u,\si_1,\si_2)+\Td{r}_0(z_2,u)+\Td{r}_1(z_2,u)+H\zu_1(u),
% \end{equation}
% where $H(z_2,u,\si_1,\si_2)=G(z_2+\Td{a}(z_2,u),\si_1,\si_2).$
%
% Since $\Td{r}_0,\Td{r}_1\in \C\{ z_2^{-1} \}$, their Borel sums are
% just themselves, and in the following we only need to determine the
% Borel sum of $H(z_2,u,\si_1,\si_2 )$.

We now deduce from the previous section summability results for the
formal series $H\nuu(g_s,u)$ \wrt~$z_2$.

\begin{thmB1}
  \emph{(i)}
  For every $u\in \C^*$,
  the perturbative solution $H\zu(g_s,u)$ to~\eqref{uE} is $1$-summable in the
  directions of $(-2\pi,0)$ \wrt\ the variable
  $z_2=\f{1}{3g_s^2u^3}$ and
  each $H\nuu(g_s,u)$, $n\gqw1$, is $1$-summable \wrt~$z_2$ in the directions of both
  \begla
    I^+ = (-\pi,0) \enspace\text{and}\enspace
    I^- = (0,\pi).
  \edla
  There exist sectorial neighbourhoods of infinity $\DD_{I^+}'(u)$
  and $\DD_{I^-}'(u)$ of opening~$2\pi$, with $\DD_{I^\pm}'(u)$ centred on
  $\arg z_2 = \pm\f\pi2$, such that, for each choice of sign and each
  $(\si_1,\si_2)\in\C^2$, the series of functions
  \begin{equation}\label{eqbspart}
    \HHu_\pm(g_s,u,\si_1,\si_2) \defeq
    \si_1+\sls_{n\gqw 0}\si_2^n\,
    e^{-\f{2n}{3g_s^{2}u^3}} \AS^{I^\pm} H\nuu(g_s,u) 
  \end{equation}
is convergent in the domain
\begin{equation}   \label{eqdefDDppmsi2}
  \DD'^\pm(\si_2)\defeq \big\{ (g_s,u) \mid
  \f{1}{3g_s^2u^3} \in \DD_{I^\pm}'(u) \;\;\text{and}\;\;
  \Re e \Big[ \f{( 1 - 2 g_s^2 u^2)^{3/2}}{g_s^2u^3} \Big]
  > \f32\ln|2\si_2| \big\}
  \end{equation}
and defines an analytic solution\footnote{%
  For each choice of sign, the condition $\f{1}{3g_s^2u^3}
  \in
  \DD'_{I^\pm}(u)$ defines a sectorial neighbourhood of~$0$ of
  opening~$\pi$ in the Riemann surface of the logarithm \wrt\ the
  variable~$g_s$, centred on the ray $\arg g_s=-\f32\arg u\mp\f\pi4$.
}
to the HAE~\eqref{uE}.
\smallskip

\emph{(ii)} The large radius analytic solutions~$\HHu_\pm$ that we
just obtained are related to the double scaling limit analytic solutions of
Theorem~B(i) by the formulas
\beglab{eqHHyASIpmGG}
\HHu_\pm(g_s,u,\si_1,\si_2) =
\AS^{I^\pm}\GG\big( \phi_{g_s}(u), \si_1, -\si_2 \big) + R(g_s,u)
\edla
with $\phi_g$ and~$R$ as in~\eqref{eqdefphigs} and~\eqref{eqdefRgsu}.
\smallskip

\emph{(iii)} Near the direction $\arg z_2=0$ (\ie
$\arg(g_su^{3/2})=0$), the connection between
the families of solutions~$\HHu_+$ and~$\HHu_-$
is given by
\begla
\HHu_+(g_s,u,\si_1,\si_2) =
\HHu_-(g_s,u,\si_1,\si_2+i)
\edla
for $(g_s,u) \in \DD'^+(\si_2)\cap\DD'^-(\si_2+i).$
\smallskip

\emph{(iv)} Near the direction $\arg z_2=-\pi$ (\ie $\arg(g_su^{3/2})=\f\pi2$),
when $|\si_2|<1$ is small enough,
the connection formula is
\begin{equation}
  \HHu_+(e^{-i\pi}g_s,u,\si_1,\si_2) =
  \HHu_-\Big(g_s,\si_1+\log(1-i\si_2),\f{\si_2}{1-i\si_2}\Big)
\end{equation}
for
$(g_s,u) \in \DD'^-(\f{\si_2}{1-i\si_2})\cap(e^{-2\pi
  i}\DD'^+(\si_2))$.
\end{thmB1}

%%%%%%%%%%%%%%%%%%%%%%%%%%%%%%%%%%%%%%%%%%%%%%%%
%%%%%%%%%%%%%%%%%%%%%%%%%%%%%%%%%%%%%%%%%%%%%%%%

\begin{proof}
  For the sake of clarity let us use the notation
  $\ti H\nuu(z_2,u)=H\nuu(g_s,u)$ for the components of the
  transseries~\eqref{eq_uts} expressed in the resurgence variable
  $z_2=\f{1}{3g_s^2u^3}$, as in the proof of Theorem~A'.
According to~\eqref{eqtiHztig} and~\eqref{eqHnuuGnztwo}, we have
\begla
\ti H\zu = \ti g\circ(id+\ph_u) + \ti R_u,
\quad
\ti H\nuu = (-1)^n e^{-2n\ph_u} \, \ti G_n\circ(id+\ph_u)
\quad \text{for $n\gqw1$,}
\edla
where $\ti R_u$ and $\ph_u(z_2)$ are convergent series in~$z_2\ii$,
both of them convergent for $|z_2| > \f2{3|u|}$
(\ie $|g_s^2u^2|<\f12$)
according to~\eqref{eqdefRgsu}, \eqref{eqdeftiRu} and~\eqref{eqphuTaylor}.

By Remark~\ref{remCV}, we can view~$\ti R_u$ and~$\ph_u$ as formal series that are
$1$-summable in the directions of any interval~$I$.
Moreover, Theorem~\ref{thmStab} entails that, for any $\ti\psi \in \C\oplus\NN(I)$,
the composite formal series $\Td{\psi}\circ (id+\ph_u)$ is 1-summable in the directions of~$I$, with
$\AS^I(\Td{\psi}\circ (id +\ph_u))=(\AS^I\Td{\psi})\circ(id+\ph_u)$.
We can apply this to $\ti g=\ti G_0$ or $\ti G_n$ with $n\ge1$ thanks
to Propositions~\ref{propseries} and~\ref{propabs}, according to which 
\begla
\ti g\in\ti\NN((-2\pi,0),\be_0), \quad
\ti G_n\in\C\oplus\ti\NN(I^\pm, \al) \ens\text{for $n\gqw1$}
\edla
with some locally bounded functions $\be_0\col (-2\pi,0)\to\R_{\gqw
  0}$ and $\al\col I^+\cup I^- \to\R_{\gqw 0}$.
This shows the summability of $H\nuu$ \wrt\ $z_2=\f{1}{3g_s^2u^3}$ for
all $n\gqw0$.

We can get quantitative information from \cite[Theorem~5.55]{[M.S.]}:
since the Borel transform of~$\ph_u$ is
\begla
\BB\ph_u =
-\tfrac{1}{u}\de+\sum_{n=0}^\infty\f{\binom{\f{3}{2}}{n+2}(-\f{2}{3u})^{n+2}}{n!}\ze^n
=-\tfrac{1}{u}\de+\wh\ph_u(\ze),
\edla
we see that the entire function~$\wh\ph_u$ satisfies
$|\wh\ph_u(\ze)| \le \f1{6|u|^{2}} e^{|\f{2}{3u}|\cdot|\ze|}$
and \cite[eqn.~(5.71)]{[M.S.]} yields
\begla
\ti g\circ(id+\ph_u) \in\ti\NN((-2\pi,0),\be_0+\tf2{|u|}), \quad
\ti G_n\circ(id+\ph_u) \in\C\oplus\ti\NN(I^\pm, \al+\tf2{|u|}) \ens\text{for $n\gqw1$.}
\edla
We thus define
\begla
\DD'_{I^\pm}(u) \defeq \DD\big(I^\pm, \al+\tf2{|u|} \big)
\edla
with the notation~\eqref{eqDDIaltiC}.
For $z_2=\f{1}{3g_s^2u^3}\in\DD'_{I^\pm}(u)$, we get
\begla
e^{-\f{2n}{3g_s^{2}u^3}} \AS^{I^\pm} H\nuu(g_s,u) =
(-1)^n e^{-2nz_1} \AS^{I^\pm}\ti G_n(z_1)
\quad \text{with} \ens z_1 = z_2+\ph_u(z_2)
\edla
for all $n\gqw0$ (recall that $\al\gqw\be_0$).
The convergence of the series~\eqref{eqbspart} in the domain~\eqref{eqdefDDppmsi2} is then a
direct consequence of the corresponding convergence statement in
Proposition~\ref{propabs},
the result being
\beglab{eqrelHHuGpm}
\HHu_\pm(g_s,u,\si_1,\si_2) = G^\pm(z_1,\si_1,-\si_2)
+ R(g_s,u)
\quad \text{for}\ens (g_s,u) \in \DD'^\pm(\si_2)
\edla
still with notation $z_1 = (id+\ph_u)\big(\f{1}{3g_s^2u^3}\big)$.
This yields Point~(i) of Theorem~B'.

We just obtained the relation~\eqref{eqrelHHuGpm} between the
Borel-Laplace sums (in~$z_2$) of the large radius limit
transseries~$\HHu_\pm$ and the Borel-Laplace sums (in~$z_1$) of the
transseries~$G^\pm$; the latter ones are themselves related to the double
scaling limit solutions $\AS^{I^\pm}\GG$ by
  \begla
  G^\pm(z_1 = \f1{3\la_s^2},\si_1,\si_2) =
  \AS^{I^\pm} \GG(\la_s,\si_1,\si_2)
  \edla
  (cf. Section~\ref{secsummaSt}).
Point~(ii) follows.  

In view of~\eqref{eqrelHHuGpm},
the statements~(iii) and~(iv) of Theorem~B' are consequences
of~\eqref{eq_right} and~\eqref{eq_left_inv};
here is, for instance, the derivation of~(iv):
\begin{multline*}
\HHu_+(e^{-i\pi}g_s,u,\si_1,\si_2) = G^+(z_1,\si_1,-\si_2) + R(g_s,u)
\\[1ex]
= G^-\Big(z,\si_1+\log(1-i\si_2),\f{-\si_2}{1-i\si_2}\Big) + R(g_s,u)
\\[1ex]
= \HHu_-\Big(g_s,u,\si_1+\log(1-i\si_2),\f{\si_2}{1-i\si_2}\Big).
  \end{multline*}
Of course, one could as well obtain the connection formulas directly
from Theorem~A''.
\end{proof}

\begin{Remark}
  Given arbitrary $\de\in(0,\f\pi2)$, we may replace~$I^+$ and~$I^-$ by the
  smaller intervals
    \begla
    I^+_\de = [-\pi+\de,-\de] \enspace\text{and}\enspace
    I^-_\de = [\de,\pi-\de]
  \edla
  and restrict our attention to
  the domain $0<|u| < \big(\sup_{I^\pm_\de}\al\big)\ii$.
This way, we observe that $\DD'^\pm(\si_2)$ is never empty, because
$\DD'_{I^\pm}(u)$ then contains
$\DD'_{I^\pm_\de}(u) \defeq \DD(I^\pm_\de,\f3{|u|}) = |u|\ii\DD(I^\pm_\de,3)$,
hence $\DD'^\pm(\si_2)$ contains
\begla
\DD_\de'^\pm(\si_2) \defeq
  \big\{ (g_s,u) \mid
  \f{1}{g_s^2u^2} \in \DD(I^\pm_\de,9) \;\;\text{and}\;\;
  \Re e \Big[ \f{( 1 - 2 g_s^2 u^2)^{3/2}}{g_s^2u^3} \Big]
  > \f32\ln|2\si_2| \big\}.
\edla
This also allows one to work with~$u$ fixed for summability purposes
\wrt~$g_s^{-2}$, or with~$g_s$ fixed when thinking of~$u$ as the
variable in the large radius limit HAE~\eqref{uE}.
\end{Remark}

We are now ready to distinguish real-analytic solutions to the large
radius limit HAE~\eqref{uE} among the Borel-Laplace sums of the
transseries solution that we have studied in this section.

\begin{thmC1}\label{ess}
\emph{(i)}
  For any $a,b\in \R$, the particular solution
\begin{equation}
\HHu_+\big(g_s,u,a,b-\tfrac{i}{2} \big)=\HHu_-\big(g_s,u,a,b+\tfrac{i}{2}\big)
\end{equation}
is analytic in the domain
\begla
\big\{ (g_s,u) \mid
  \f{1}{3g_s^2u^2} \in \DD'_{I^+}(u)\cup \DD'_{I^-}(u) \;\;\text{and}\;\;
\Re e \Big[ \f{( 1 - 2 g_s^2 u^2)^{3/2}}{g_s^2u^3} \Big]
> \f34\ln(1+4 b^2) \,\big\}
\edla
and it is real-valued in restriction to all $(g_s,u)$ such that $u\in\R^*$
and $\arg(g_s^2u^3)=0$.
\smallskip

\emph{(ii)}
There exists $0<\th_*<\f\pi4$ such that, for any $a\in \R$
and $\th\in(-\th_*,\th_*)$,
the particular solution
\begin{equation}\label{ran2}
\HHu_+\big(e^{-i\pi}g_s,u,a+i\tfrac{\th}{2}, -i(1-e^{-i\th})\big)
  =\HHu_-\big(g_s,u,a-i\tfrac{\th}{2}, i(1-e^{i\th})\big) 
  \end{equation}
  is analytic in the domain
\begla
\big\{ (g_s,u) \mid
  \f{1}{3g_s^2u^2} \in \DD'_{I^+}(u)\cup \DD'_{I^-}(u) \;\;\text{and}\;\;
\Re e \Big[ \f{( 1 - 2 g_s^2 u^2)^{3/2}}{g_s^2u^3} \Big]
> \f32\ln(2|1-e^{i\th}|) \,\big\}
\edla
and it is real-valued in restriction to all $(g_s,u)$ such that $u\in\R^*$
and $\arg(g_s^2u^3)=\pi$.
\end{thmC1}

\begin{proof}
  We could of course derive these properties directly from
  Theorem~B'(iii) and~(iv)
  %,
  but we prefer to use Theorem~C(i) and~(ii)
  and the real-valued solutions along the rays $\{\arg z_1 = 0\}$
    and $\{\arg z_1 = -\pi\}$ obtained there.
      We observe that, in the relation~\eqref{eqHHyASIpmGG}, if $u\in\R^*$, then:
  \begin{itemize}
  \item
The change of variable $z_1=(id+\ph_u)(z_2)$
    (which corresponds to the change $\la_s=\phi_{g_s}(u)$)
    maps any real~$z_2$ with $|z_2|>\f2{3|u|}$ to a real~$z_1$ with
    same argument.
  \item
    The term $R(g_s,u)$ is real when $g_s^2u^3\in\R$ with $|g_s^2u^2|<\f12$.
\end{itemize}
The conclusion thus follows.
\end{proof}

\begin{Remark}\label{rationalLRL}
  What we said in Remark~\ref{rationalDSL} concerning the enumerative
  properties of transseries objects also applies to $Pol_n(u,2g)$ and
  $H\nuu(g_s,u)$ for each~$n$, the whole transseries
  $H\uu(g_s,u,\sigma)$ and the connection formula between~$\HHu_+$
  and~$\HHu_-$.
\end{Remark}

%%%%%%%%%%%%%%%%%%%%%%%%%%%%%%%%%%%%%%%%%%%%%%%%%%%%%%%%%%%
%%%%%%%%%%%%%%%%%%%%%%%%%%%%%%%%%%%%%%%%%%%%%%%%%%%%%%%%%%%

\appendix

\section{Appendix: Remark on $\AS^I\col\tNNe(I,\al)\to\gO\big(\DD(I,\al)\big)$}

\label{appRemNext}

%%%%%%%%%%%%%%%%%%%%%%%%%%%%%%%%%%%%%%%%%%%%%%%%%%%%%%%%%%%
%%%%%%%%%%%%%%%%%%%%%%%%%%%%%%%%%%%%%%%%%%%%%%%%%%%%%%%%%%%

In Section~\ref{secBLsum}, we considered the space of all finite sums of the
form~\eqref{eqtiphsumnujCzii}, which in fact is nothing but
\(
  \sum_{\mu\in\C} z^{-\mu}\C[[z\ii]]
\),
and defined a subspace $\tNNe(I,\al)$.

Let us consider the set of all possible exponents modulo~$\Z$ and use the
notation
\[ % begin{gather*}
  \mu\in\C \mapsto [\mu] = \mu+\Z \in \C/\Z,
\]
%
%  \\[1ex]
%
and, given $\goa\in\C/\Z$, % \text{given $\goa\in\C/\Z$,}\ens
\[
  z^{-\{\goa\}}\C((z\ii)) \defeq
  \bigcup_{\mu\in\goa} z^{-\mu}\C[[z\ii]] = z^{-\mu_0}\C((z\ii))
  \ens\text{for any $\mu_0\in\goa$}
\] % end{gather*}
(where, as usual, we denote by $\C((z\ii))$ the space of formal Laurent
series in the indeterminate~$z\ii$).
Then
\begin{gather}
  \label{eqexplqinnondir}
  \sum_{\mu\in\C} z^{-\mu}\C[[z\ii]] = \bigoplus_{\goa\in\C/\Z} z^{-\{\goa\}}\C((z\ii))
  \\[1ex]
  \notag
  \tNNe(I,\al) = \bigoplus_{\goa\in\C/\Z} z^{-\{\goa\}}\ti\NN(I,\al)
  \quad\text{with}\ens
    z^{-\{\goa\}}\ti\NN(I,\al) \defeq \bigcup_{\mu\in\goa}
    z^{-\mu}\ti\NN(I,\al).
\end{gather}
Using the notation
\[
  \ti\ph = \sum_{\goa\in\C/\Z} \tpha % \ti\ph_{\goa}
\]
for the canonical decomposition of an arbitrary element~$\ti\ph$
(with all but a finite number of $\tpha$ equal to~$0$),
we indeed have
\[
  \ti\ph\in \tNNe(I,\al)
  \ens\Longleftrightarrow\ens
  \forall \goa\in\C/\Z, \; \exists \nu \in \goa \; \text{ such that} \;
  z^{\nu}\tpha \in \ti\NN(I,\al).
\]

%\newpage
\smallskip

\noindent \textbf{Relation with the representation~\eqref{eqtiphsumnujCzii} of an arbitrary~$\ti\ph$}

\smallskip

\noindent
Suppose that
\beglab{eqreptiph}
  \ti\ph = z^{-\mu_1}\ti\psi_1+\cdots+z^{-\mu_N}\ti\psi_N
  \quad \text{for some}\ens N\gqw1,\ens\mu_j\in\C,\ens\ti\psi_j\in
  \C[[z\ii]].  
\edla
Then one can check that, for every $\goa\in\C/\Z$,
\beglab{eqtiphgoa}
  \tpha = \sum_{j\in\{1,\ldots,N\}\;\text{s.t.}\; \mu_j\in\goa}
  z^{-\mu_j}\ti\psi_j.
\edla
Moreover, for every $\goa\in\C/\Z$, % there is a unique
we can pick $\nua\in\C$
% with maximal real part
such that
\beglab{eqchoicenua}
  \mu_j\in\goa \Imp
  \De_j \defeq \mu_j-\nua \in \Z_{\gqw1},
\edla
whence
\[
  z^{\snua} \tpha = 
  \sum_{j\in\{1,\ldots,N\}\;\text{s.t.}\; \mu_j\in\goa}
  z^{-\De_j}\ti\psi_j.
\]
%
% (with the convention $\nua=\infty$ and $z^{-\nua}=0$ if there is no~$j$
%  such that $\mu_j\in\goa$).
  
\bigskip

\noindent \textbf{Definition of
  $\AS^I\col\tNNe(I,\al)\to\gO\big(\DD(I,\al)\big)$}

\smallskip

\noindent
It follows that, for any representation~\eqref{eqreptiph} of 
$\ti\ph\in\tNNe(I,\al)$
with $\ti\psi_1, \ldots, \ti\psi_N\in\ti\NN(I,\al)$
and any choice of
$(\nua)_{\raisebox{-.5ex}{$\scriptstyle \goa\in\C/\Z$}}$ satisfying~\eqref{eqchoicenua},
\begin{multline*}
  \sum_{j\in\{1,\ldots,N\}} z^{-\mu_j}\AS^I\ti\psi_j =
  \sum_{\goa\in\C/\Z} z^{-\snua}
  \left(\sum_{j\in\{1,\ldots,N\}\;\text{s.t.}\; \mu_j\in\goa}
    z^{-\De_j}\AS^I\ti\psi_j \right) \\[1ex]
  = \sum_{\goa\in\C/\Z} z^{-\snua} \AS^I (z^{\snua}\tpha)
\end{multline*}
because $z^{-\De_j}\AS^I\ti\psi_j = \AS^I(z^{-\De_j}\ti\psi_j)$.
Moreover, for a given $\goa\in\C/\Z$, if we consider two different
solutions~$\nuap1$ and~$\nuap2$ of~\eqref{eqchoicenua}, then their difference must be
integer, thus $\nuap2=\nuap1-\De$ with $\De\in\Z_{\gqw1}$
(swapping~$\nuap1$ and~$\nuap2$ if necessary) and
\[
  z^{-\snuap2} \AS^I (z^{\snuap2}\tpha) =
  z^{-\snuap1+\De} \AS^I (z^{\snuap1-\De}\tpha) =
  z^{-\snuap1} \AS^I (z^{\snuap1}\tpha).
\]

\noindent
\textbf{Conclusion:}
The function $\AS^I\ti\ph \defeq \sum_{j\in\{1,\ldots,N\}}
z^{-\mu_j}\AS^I\ti\psi_j$ does not depend on the particular
representation~\eqref{eqreptiph} but only on $\ti\ph\in\tNNe(I,\al)$.
Moreover,
\[
  \AS^I\ti\ph = \sum_{\goa\in\C/\Z} z^{-\snua} \AS^I (z^{\snua}\tpha)
\]
for any $(\nua)_{\raisebox{-.5ex}{$\scriptstyle \goa\in\C/\Z$}}$ such
that
$z^{\snua}\tpha \in \ti\NN(I,\al)$ for each~$\goa$.

%%%%%%%%%%%%%%%%%%%%%%%%%%%%%%%%%%%%%%%%%%%%%%%%
%%%%%%%%%%%%%%%%%%%%%%%%%%%%%%%%%%%%%%%%%%%%%%%%

% \blue{UPDATE THE REFERENCE LIST --- SOME ENTRIES NEED CORRECTION --- COPY
% e.g. \href{https://www.zbmath.org/}{https://www.zbmath.org/} TO AVOID
% MISTAKES --- REFER TO PUBLISHED VERSION RATHER THAN arXiv PREPRINT
% WHENEVER POSSIBLE}

%\bigskip
%\noindent\small School of Mathematical Sciences\\
%\noindent\small Capital Normal University\\
%\noindent\small beijing 210023, P. R. China\\
%\noindent\small E-mail address:  ***@***.com

\end{document}